\documentclass[12pt,letterpaper,oneside]{article}

\usepackage{amsmath}
\usepackage{amsfonts}
\usepackage{times}
\usepackage[dvipsnames]{xcolor}
\pdfoutput=1 
\usepackage{bm}
\usepackage{graphicx}
\usepackage{achemso}
\setkeys{acs}{usetitle = true,etalmode=truncate,maxauthors = 5,biblabel = period}
\usepackage{color}
\usepackage[super]{natbib}
\usepackage{longtable}

\usepackage{srcltx}
\usepackage[normalem]{ulem}
\usepackage[T1]{fontenc}
\usepackage{setspace}
\usepackage[letterpaper,ignoreall,top=2.65cm,bottom=2.7cm,%
left=2.65cm,right=2.65cm,foot=1cm,head=1.5cm]{geometry}

\begin{document}
%%% TITLE PAGE %%%%%%%%%%%%%%%%%%%%%%%%%%%%%%%%%%%%%%%%%%%%%%%%%%%%%%%%%%%%%%%%%%%%%%%%%%%%%%%%%%%%%%%%%%%%%%%%%%%%%%%%%%%
%\tableofcontents

\Large
  \title{The Search for Superconductivity in High Pressure Hydrides} 
  \normalsize
  \author{Tiange Bi, Niloofar Zarifi, Tyson Terpstra, Eva Zurek$^*$\\
  Department of Chemistry    \\
  University at Buffalo          \\
  State University of New York \\
  Buffalo, NY 14260-3000, USA \\[2ex]
  tiangebi@buffalo.edu, nzarifi@buffalo.edu, tysonter@buffalo.edu, $^*$ezurek@buffalo.edu}
  \date{\today}

\maketitle

\section*{Abstract}
The computational and experimental exploration of the phase diagrams of binary hydrides under high pressure has uncovered phases with novel stoichiometries and structures, some which are superconducting at quite high temperatures. Herein we review the plethora of studies that have been undertaken in the last decade on the main group and transition metal hydrides, as well as a few of the rare earth hydrides at pressures attainable in diamond anvil cells. The aggregate of data shows that the propensity for superconductivity is dependent upon the species used to ``dope'' hydrogen, with some of the highest values obtained for elements that belong to the alkaline and rare earth, or the pnictogen and chalcogen families. \\[2ex]

\noindent\textbf{Keywords:} \\
superconductivity, hydrides, high pressure, density functional theory, electronic structure, crystal structure prediction, materials, extreme conditions, metallic hydrogen, Bardeen-Cooper-Schrieffer superconductor

\newpage
%%%%%%%%%%%%%%%%%%%%%%%%%%%%%%%%%%%%%%%%%%%%%%%%%%%%%%%%%%%%%%%%%%%%%%%%%%%%%%%%%%%%%%%%%%%%%%%%%%%%%%%%%%%%%%%%%%%%%%%%%%%
%% Beginning of Document
%%%%%%%%%%%%%%%%%%%%%%%%%%%%%%%%%%%%%%%%%%%%%%%%%%%%%%%%%%%%%%%%%%%%%%%%%%%%%%%%%%%%%%%%%%%%%%%%%%%%%%%%%%%%%%%%%%%%%%%%%%%

%%%%%%%%%%%%%%%%%%%%%%%%%%%%%%%%%%%%%%%%%%%%%%%%%%%%%%%%%%%%%%%%%%%%%%%%%%%%%%%%%%%%%%%%%%%%%%%%%%%%%%%%%%%%%%%%%%%%%%%%%%%%
%%%%%%%%%%%%%%%%%%%%%%%%%%                       High Pressure Review                  %%%%%%%%%%%%%%%%%%%%%%%%%%%%%%%%%%%%%
%%%%%%%%%%%%%%%%%%%%%%%%%%%%%%%%%%%%%%%%%%%%%%%%%%%%%%%%%%%%%%%%%%%%%%%%%%%%%%%%%%%%%%%%%%%%%%%%%%%%%%%%%%%%%%%%%%%%%%%%%%%%

\section{Introduction}
The metalization of hydrogen under pressure was first proposed by J.\ D.\ Bernal, but only later transcribed by Wigner and Huntington in their seminal 1935 paper, which discussed the possibility of hydrogen transforming to an alkali metal-like monoatomic solid at $P>$~25~GPa \cite{Wigner:1935}. In 1968 Ashcroft predicted that this elusive substance had the propensity to be a high temperature Bardeen-Cooper-Schrieffer (BCS) type superconductor based upon its large density of states at the Fermi level, high phonon frequencies (a result of the small atomic mass), and substantial electron--phonon coupling (due to the lack of core electrons and strong covalent bonding) \cite{Ashcroft:1968a}. 
Nearly 40 years later the same considerations led Ashcroft to propose that hydrogen--rich solids, such as the group 14 hydrides, had the potential to become high temperature superconductors under pressure \cite{Ashcroft:2004a, Ashcroft:2004b}. Moreover, he speculated that the addition of a second element to hydrogen could reduce the physical pressure required to metallize the system via doping \cite{Carlsson:1983} or ``chemical precompression'' \cite{Ashcroft:2004a, Ashcroft:2004b}. Herein, we provide a thorough review of the efforts, both experimental and theoretical, undertaken to metallize hydrogen-rich solids under pressure in the search for new superconducting materials. 

Pressure can coerce compounds to assume stoichiometries and geometric arrangements that would not be accessible at atmospheric conditions  \cite{Hemley:2000,Song:2013a,Grochala:2007a,Goncharov:2013b,Bhardwaj:2012a,Dubrovinsky:2013a,McMillan:2013a,Klug:2011a,Naumov:2014a,Zurek:2014i,Zurek:2016b,Hermann-lip}.  Because experimental trial-and-error high pressure syntheses can be expensive to carry out and the results difficult to analyze, it is desirable to predict which elemental combinations and pressures could be used to synthesize compounds with useful properties. However, neither chemical intuition nor data-mining techniques are typically useful for these purposes because they have been developed based upon information gathered at atmospheric conditions. Fortunately, the spectacular advances in computer hardware, coupled with the developments in \emph{a priori} crystal structure prediction (CSP) using Density Functional Theory  (DFT) \cite{Zurek:2014d,woodley:2008a,Schon:2010a,revard,oganov:2010a,Jansen:2015a,uspex5} has led to synergy between experiment and theory in high pressure research \cite{Zurek:2014i,Zhang:2017,Wang:2014a,random1,random3}. Computations are carried out to predict the structures and properties of targets for synthesis, and also to aid in the characterization of phases that have been made in experiment. This synergy has been instrumental in advancing the research carried out on high pressure hydrides.

Ashcroft's original predictions regarding superconductivity in hydrogen--rich systems \cite{Ashcroft:2004a, Ashcroft:2004b} have led to a plethora of theoretical, and some experimental investigations of high pressure hydrides. By now the phase diagrams of most binary hydrides have been explored on a computer. In the early days of CSP it was common to interrogate the stability of a single stoichiometry by carrying out geometry optimizations, as a function of pressure, on simple lattices or crystal structure types that were known for other chemically similar systems. The next major step in CSP was taken when it became standard to employ automated techniques such as random structure searches \cite{random1,random2}, evolutionary/genetic algorithms \cite{Zurek:2011a,gasp,uspex1,maise,Bahmann:2013a,Zunger:2007a,Abraham:2006a,Fadda:2010a,adaptive:2014a}, particle swarm optimization methods \cite{Wang:2010a, Wang:2014a}, basin \cite{Doye:1997a} or minima hopping \cite{Godecker:2004}, metadynamics \cite{Laio:2002a} and simulated annealing \cite{Kirkpatrick:1983} to predict the global minima at a given set of conditions. However, these studies were typically limited in that they investigated the stoichiometries that were known to be stable at atmospheric conditions (e.g.\ H$_2$S). It was eventually realized that under pressure the stable and metastable structures may have very different stoichiometries, and by now it is common practice to use CSP to predict the thermodynamically and dynamically stable phases while varying the hydrogen content (e.g.\ H$_n$S for a range of $n$) as a function of pressure. In fact, there is by now a standard ``recipe'' of how to carry out these studies. However, because of the stochastic nature of CSP searches the only way to be certain that the global minimum has been found is to compute all of the local minima (which is impractical for all but the simplest systems), and because different decisions can be made about which pressures, stoichiometries and unit cell sizes are considered in the CSP searches, it should not be surprising that the findings of two or more investigations on the same system can, at times, yield different results. Moreover, it is not always clear whether or not the phases that are formed in experiment are the global minima or simply metastable. For example, recent work on a sample of compressed hydrogen disulfide has shown that the stoichiometries that are formed, and their $T_c$, depends upon the experimental conditions \cite{Drozdov:2015a}.

Therefore, this review compiles and presents all of the theoretical and much of the experimental data available to date on the high pressure investigations of binary hydrides, with a particular emphasis on the structures that have been proposed as being stable and their propensity for superconductivity (in a few cases the main motivation for the study may have been for other applications such as hydrogen storage \cite{Song:2013a}). The focus is on hydrides of the main group elements, the transition metal elements, and only a few of the rare earths (Sc, Y, La) are discussed in detail, typically at pressures that can be achieved reliably in a diamond anvil cell (DAC), $\sim$400~GPa. The results are organized according to the groups within the periodic table to which the ``dopant'' element added to hydrogen belongs. We note that a number of excellent reviews of the high pressure hydrides have appeared recently, however most of them have focused on particular elements or authors, and none have presented a thorough compilation of the work carried out so far \cite{Zurek:2016d,Shamp:2016,Zhang:2017,Duan:2017a,Wang:2017a,Struzhkin:2015a,random3,Yao-S-review:2018}. 

We hope this review is therefore useful for those interested in comparing the results obtained in different studies of the same set of hydrides, for determining which binary compounds have not yet been intensely studied, and for unveiling chemical trends in the properties and behavior of the binary hydrides under pressure. We also point the reader to a number of reviews that cover advances in high pressure CSP \cite{Zhang:2017,Wang:2014a,random1,random3}, the successes and limitations of DFT calculations in high pressure research \cite{Zurek:2014i}, as well as the methods employed to estimate $T_c$ in BCS-type superconductors \cite{bose2009electron}.

%%%%%%%%%%%%%%%%%%%%%%%%%%%%%%%%%%%%%%%%%%%%%%%%%%%%%%%%%%%%%%%%%%%%%%%%%%%%%%%%%%%%%%%%%%%%%%%%%%%%%%%%%%%%%%%%%%%%%%%%%%%%
%%%%%%%%%%%%%%%%%%%%%%%%%%                       Alkali Metals                         %%%%%%%%%%%%%%%%%%%%%%%%%%%%%%%%%%%%%
%%%%%%%%%%%%%%%%%%%%%%%%%%%%%%%%%%%%%%%%%%%%%%%%%%%%%%%%%%%%%%%%%%%%%%%%%%%%%%%%%%%%%%%%%%%%%%%%%%%%%%%%%%%%%%%%%%%%%%%%%%%%

\section{Group 1: Alkali Metal Hydrides}
%

%\textbf{Lithium, Sodium, Potassium, Rubidium, Cesium}

At atmospheric conditions the alkali metal (M~=~Li, Na, K, Rb, Cs) hydrides adopt an MH stoichiometry and crystallize in the rock-salt ($B1$) structure. The band gaps of these ionic solids are large, ranging from 4-6~eV \cite{Setten:2007a,Lebegue:2003b}. A number of phase changes ($B1 \rightarrow B2$ for M~=~Na, K, Rb, Cs; $B2 \rightarrow$~CrB for M~=~Rb, Cs) occur at progressively lower pressures for the heavier alkali metals \cite{Duclos:1987a,Hochheimer:1985a,Ghandehari:1995a,Ghandehari:1995b}. Computations suggest that under sufficient compression KH will also assume the CrB structure \cite{Ahuja:1998a, Ahuja:1999a}. A hitherto unobserved transition to a $Pnma$ and a $P6_3/mmc$ phase was computationally predicted in CsH \cite{Hooper:2011}, but it has not yet been observed. A pressure induced insulator to metal transition is likely to take place in these systems due to pressure induced broadening of the valence and conduction bands. So far these alkali hydrides have not been metalized, but extrapolation of experimentally determined band gaps and first-principles calculations suggests that band gap closure may occur between 300-1000~GPa  \cite{Lebegue:2003a,Ghandehari:1995c,Ghandehari:1995b}. Because metalization in these systems will likely occur because of pressure induced band overlap, the ``classic'' alkali hydrides are unlikely to have a high density of states (DOS) at the Fermi level ($E_F$), and therefore will not be good candidates for high temperature superconductivity at pressures attainable in a DAC.

The only alkali metal subhydrides that were predicted to become stable under pressure are LiH$_m$ with $m>1$ \cite{Hooper:2012}. They were computed to be thermodynamically stable with respect to decomposition into Li and LiH in a very narrow pressure range ($\sim$50-100~GPa). None of the phases found were good metals, but the band structure of two Li$_5$H compounds that had the lowest enthalpies of formation, $\Delta H_F$, contained two conical bands with a linear dispersion around $E_F$. Thus, even though these phases are not promising for superconductivity their electronic structure features a Dirac cone, hinting that they may have unusual electron transport properties.

\begin{figure}[h!]
\begin{center}
\includegraphics[width=0.9\columnwidth]{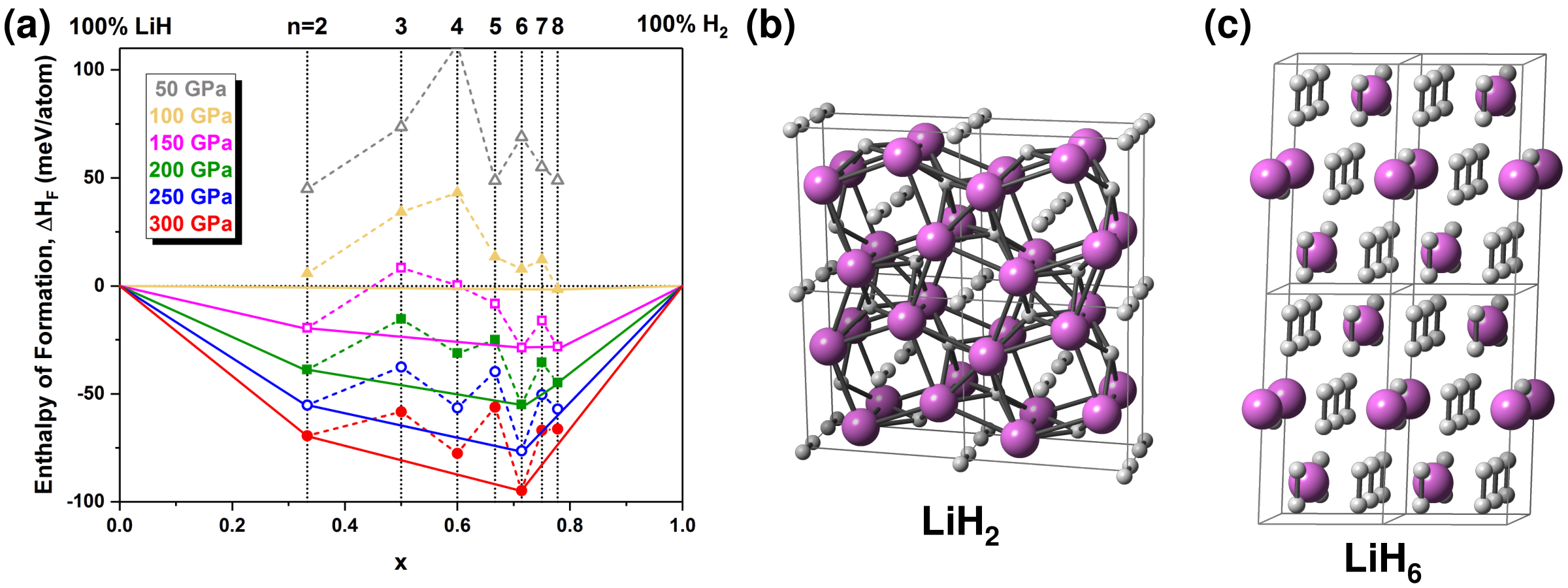}
\end{center}
\caption{(a) Enthalpies of formation, $\Delta H_F$,  with respect to LiH and H$_2$ of the most stable LiH$_n$, $n=1-8$, phases predicted in Ref.\ \cite{Zurek:2009c} between 50-300~GPa. The $x$ axis shows the fraction of H$_2$ in the structures and the solid lines denote the convex hulls. Supercells of the (b) $P4/mbm$-LiH$_2$ and (c) $R\bar{3}m$-LiH$_6$ phases that fell on the hull between 100-300~GPa are shown.}
\label{fig:Group1}
\end{figure}

Evolutionary algorithms coupled with the PBE functional predicted that phases with stoichiometries such as LiH$_2$, LiH$_6$ and LiH$_8$ would become stable at pressures ranging from $\sim$100-300~GPa \cite{Zurek:2009c}. Fig.\ \ref{fig:Group1}(a) plots $\Delta H_F$ for the reaction $\text{LiH}+\left(\frac{n-1}{2}\right)\text{H}_2\rightarrow \text{LiH}_n$ versus the mole fraction of hydrogen in the products. The solid lines denote the convex hulls for the different pressures. Any phase whose $\Delta H_F$ lies on the hull is thermodynamically stable, whereas other phases may be metastable. LiH$_2$, shown in Fig.\ \ref{fig:Group1}(b), was comprised of H$_2$ and H$^-$ units and it was found to metallize via pressure induced broadening and overlap of the H$^-$ donor/impurity band with the H$_2$ $\sigma^*$ anti-bonding band. The DOS at $E_F$ was low, and the $T_c$ of LiH$_2$ was calculated to be 0~K at 150~GPa \cite{Xie:2014a}. 
The most stable LiH$_6$ (shown in Fig.\ \ref{fig:Group1}(b)) and LiH$_8$ phases, on the other hand, only contained molecular hydrogen units that obtained a partial negative charge via electron transfer from the electropositive lithium atom, i.e.\ H$_2^{\delta-}$. These systems were good metals because of the partial filling of the H$_2$ $\sigma^*$ anti-bonding bands even at 1~atm. The high DOS at $E_F$ persisted at pressures where they became stable, wherein DFT calculations predicted $T_c$ values that ranged from $\sim$30-80~K \cite{Xie:2014a}. 

Recently, P\'epin and co-workers succeeded in synthesizing the lithium polyhydrides after squeezing  LiH in a DAC at 300~K above 130~GPa \cite{Pepin:2015a}. Synchrotron infrared (IR) absorption revealed peaks whose frequencies differed significantly from the H-H stretching mode in pure H$_2$ (the H$_2$ vibron), but roughly matched those computed for the LiH$_2$ and LiH$_6$ phases in Ref.\ \cite{Zurek:2009c}. Therefore, P\'epin et al.\ proposed that lithium diffuses into the diamond where it can react with carbon, and this mechanism leads to the formation of an  LiH$_6$ layer at the diamond/sample interface, and an LiH$_2$ layer at the LiH$_6$/LiH interface. Further characterization was not possible, however it was noted that the  sample remained optically transparent until 215~GPa, and the IR measurements did not provide any evidence of metallicity. This suggests that despite the agreement of the measured and computed IR data, the LiH$_6$ phase predicted in Ref.\ \cite{Zurek:2009c} cannot be formed in experiment because it must be metallic as a consequence of the fact that its unit cell contains an odd number of electrons. 

The discrepancy between theory and experiment inspired a theoretical study that employed the nonlocal van der Waals (vdW) including functional of Dion et al.\ (vdW-DF) \cite{Dion:2004}, which incorporates the effects of dispersion self-consistently \cite{Chen:2017a}. It was shown that the inclusion of vdW interactions affects the relative stabilities of the polyhydride phases, so that the LiH$_6$ phase predicted to be stable within PBE in Ref.\ \cite{Zurek:2009c} no longer lay on the convex hull in Ref.\ \cite{Chen:2017a}.  Based on the computed enthalpies, the insulating character observed in experiment, plus a comparison of the experimental and theoretical vibron frequencies as a function of pressure, the authors of Ref.\ \cite{Chen:2017a} concluded that LiH$_2$, LiH$_9$ or a metastable LiH$_7$ phase are the most likely candidates for the species made in experiment. Exploratory calculations indicated that vdW interactions did not influence the structures nor stabilities of the heavier group I polyhydrides.

A theoretical investigation predicting that sodium polyhydrides will become stable above $\sim$25~GPa \cite{Baettig:2011} inspired an experimental study where these phases were synthesized in a DAC above 40~GPa and 2000~K \cite{Struzhkin:2016}. Because the computed X-ray diffraction (XRD) patterns and Raman spectra of the phases predicted as being stable in Ref.\ \cite{Baettig:2011} could not account for the experimental results, further CSP searches were carried out in Ref.\ \cite{Struzhkin:2016} on a larger range of stoichiometries than those originally considered. An NaH$_3$ phase was found to be the lowest point on the convex hull at 50~GPa, and the observables calculated for the NaH$_3$ and NaH$_7$ stoichiometries gave the best agreement with experiment. The NaH$_3$ and NaH$_7$ phases were insulating at 50~GPa, and they were comprised of H$_2$ as well as H$^-$ and H$_3^-$ units, respectively. 

Ref.\ \cite{Zhou:2012a} used CSP to find the most stable structures of KH$_n$ with $n=2,4,6,8$. Calculations were carried out to determine the superconducting properties of a $C2/c$-KH$_6$ phase with a large DOS at $E_F$ that was comprised of H$_2$ units. The $T_c$ was estimated to be 58-70~K at 230~GPa and 46-57~K at 300~GPa. Another study carried out evolutionary searches on KH$_n$ with $n=2-13$ and found that phases with the KH$_6$ stoichiometry did not lie on the convex hull up to pressures of 250~GPa \cite{Hooper:2012a}. Instead, the systems that were thermodynamically and dynamically stable contained H$^-$ and H$_3^-$ units, and they were unlikely to be good superconductors.

Finally, CSP has been employed to predict the structures of the polyhydrides of rubidium \cite{Hooper:2011a} and cesium \cite{Shamp:2012}. Above 30~GPa the rubidium polyhydride with the most negative $\Delta H_F$ was RbH$_5$, which was comprised of linear H$_3^-$ and H$_2$ units. The lowest point on the CsH$_n$ convex hulls between 50-100~GPa was CsH$_3$, and five nearly isoenthalpic structures were found. They were all comprised of linear H$_3^-$ building blocks and Cs$^+$ ions whose positions were related to those of the silicon and thorium atoms, respectively, in either $\alpha$- or $\beta$-ThSi$_2$. Metalization of RbH$_5$ and CsH$_3$ was predicted to occur at high pressures due to pressure induced band overlap of the H$_3^-$ non-bonding bands with either the metal $d$-bands or the H$_3^-$ anti-bonding bands, but the DOS at $E_F$ was low suggesting that these systems are not good candidates for superconductivity.

Most of the thermodynamically and dynamically stable polyhydrides of the alkali metals that were predicted via CSP contained H$^-$ and H$_3^-$ units rendering them insulating at low pressures. Metalization could be induced via pressure induced broadening and overlap of the H$^-$ donor/impurity band with the H$_2$ $\sigma^*$ anti-bonding bands, or the H$_3^-$ non-bonding and anti-bonding bands. As a result these phases had a low DOS at $E_F$, suggesting that their $T_c$ values were likely to be low, like what was found for LiH$_2$ \cite{Xie:2014a}. Phases whose hydrogenic lattices only contained H$_2^{\delta-}$ molecules had a substantial DOS at $E_F$ suggesting they could have a high $T_c$. Calculations predicted $T_c$ values ranging from $\sim$30-80~K for LiH$_6$, LiH$_8$, and KH$_6$. However, it is likely that these phases are metastable.  %) 

%%%%%%%%%%%%%%%%%%%%%%%%%%%%%%%%%%%%%%%%%%%%%%%%%%%%%%%%%%%%%%%%%%%%%%%%%%%%%%%%%%%%%%%%%%%%%%%%%%%%%%%%%%%%%%%%%%%%%%%%%%%%
%%%%%%%%%%%%%%%%%%%%%%%%%%                       Alkaline Metals                       %%%%%%%%%%%%%%%%%%%%%%%%%%%%%%%%%%%%%
%%%%%%%%%%%%%%%%%%%%%%%%%%%%%%%%%%%%%%%%%%%%%%%%%%%%%%%%%%%%%%%%%%%%%%%%%%%%%%%%%%%%%%%%%%%%%%%%%%%%%%%%%%%%%%%%%%%%%%%%%%%%

\section{Group 2: Alkaline Earth Metal Hydrides}

Whereas BeH$_2$ and MgH$_2$ undergo a unique sequence of phase transitions under pressure \cite{Wang:2014,Zhang:2007,Vajeeston:2004a,Vajeeston:2006a, Vajeeston:2002a}, CaH$_2$ \cite{Tse:2007b,Li:2007}, SrH$_2$ \cite{Smith:2009a} and BaH$_2$ \cite{Tse:2009a, Chen:2010b} all undergo the same structural changes, but they occur at lower pressures for the heavier alkaline earth metals. Despite their large band gaps at atmospheric conditions, these phases are computed, within PBE, to metallize at pressures attainable in a DAC. Band gap closure occurs at higher pressures for the lighter systems  \cite{Zhang:2010b}.  However, the DOS at $E_F$ for the metallic phases is low, and DFT calculations predict small to moderate $T_c$ values, for example 38~K \cite{Wang:2014} and 32-44~K \cite{Yu:2014} for BeH$_2$ at 250~GPa, 16-23~K for MgH$_2$ at 180~GPa \cite{Lonie:2012}, and only a few mK for BaH$_2$ at 60~GPa  \cite{Tse:2009a}.

\begin{figure}[h!]
\begin{center}
\includegraphics[width=0.8\columnwidth]{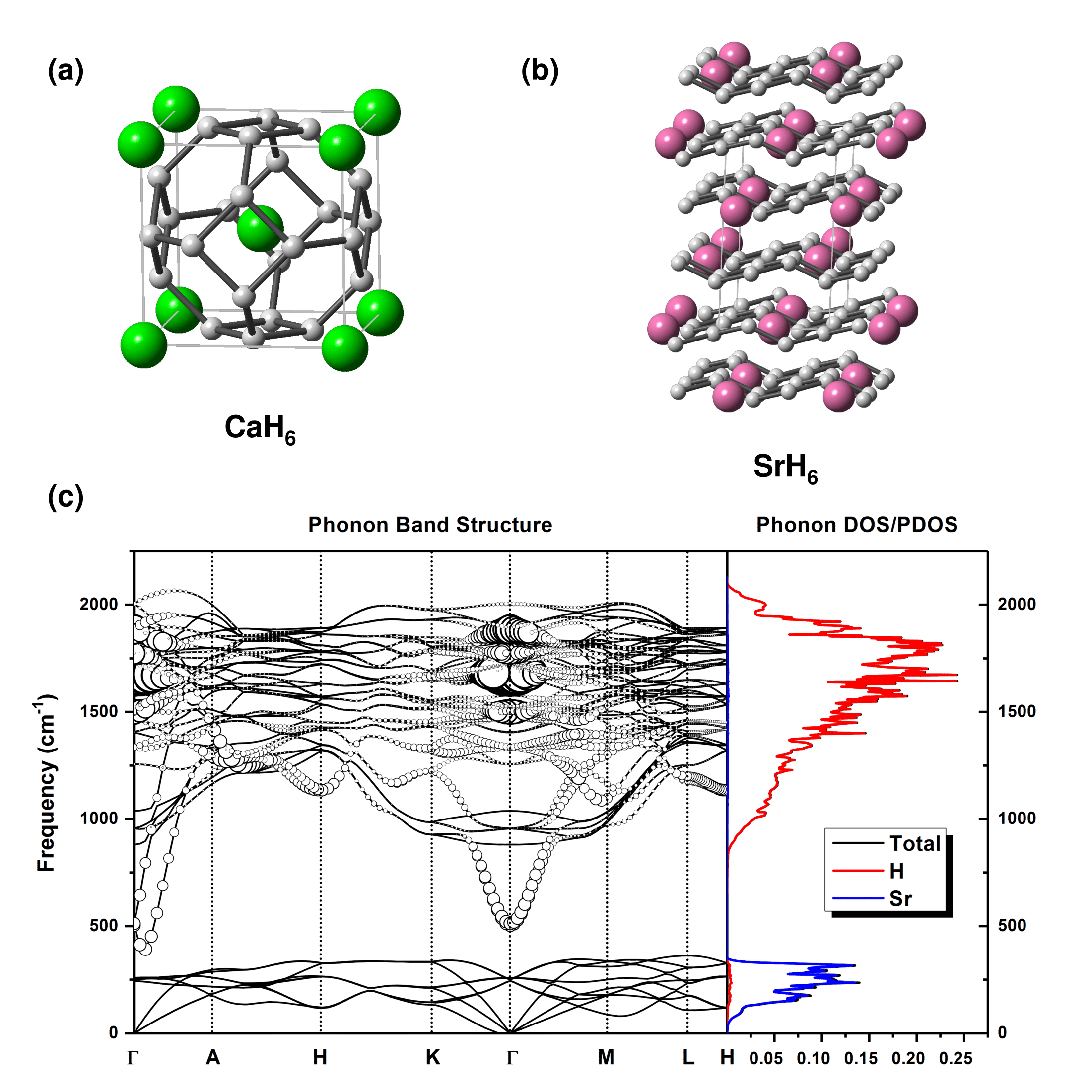}
\end{center}
\caption{Unit cells of (a) the sodalite-like $Im\bar{3}m$-CaH$_6$ \cite{Wang:2012}, and (b) $R\bar{3}m$-SrH$_6$ \cite{Hooper:2013,Wang:2015a} phases predicted to be stable above 150~GPa and 250~GPa, respectively. (c) The calculated phonon band structure along with the phonon projected density of states (PDOS) plot of $R\bar{3}m$-SrH$_6$ at 250~GPa. Circles in the phonon band structure represent the phonon line-width. Because of the separation of the hydrogen-based and strontium-based vibrations, the PDOS overlaps with the total DOS. The calculated values of $\lambda=1.10$, and $\omega_\text{log}=1358$~K gave $T_c$ values of 108~K/156~K as estimated using the modified McMillan/Eliashberg equation for $\mu^*=0.10$.}
\label{fig:Group2}
\end{figure}

CSP techniques have been used to predict the most stable structures of the polyhydrides of magnesium \cite{Lonie:2012,Feng:2015a}, calcium \cite{Wang:2012}, strontium \cite{Hooper:2013,Wang:2015a} and barium \cite{Hooper:2012b}, i.e.\ MH$_n$ with $n>2$, under pressure. Similar to what was observed for the alkali metal polyhydrides, thermodynamic stability was achieved at the lowest pressures for the heaviest systems: whereas BeH$_2$ was the only hydride of beryllium that was stable below 200~GPa, the polyhydrides of barium were predicted to form above 20~GPa \cite{Hooper:2012b}. For comparison LiH$_n$ \cite{Zurek:2009c} and CsH$_n$ \cite{Shamp:2012} with $n>1$ were predicted to stabilize by $\sim$120~GPa and 3~GPa, respectively, as a result of the lower ionization potentials of Li and Cs as compared to those of Be and Ba. Whereas the hydrogen content in the stoichiometry that had the most negative $\Delta H_F$ always increased with increasing pressure for the alkaline earth polyhydrides, this was not always the case for the alkali metal polyhydrides. Another difference, that is important for the $T_c$s of these phases, is that whereas the hydrogenic sublattices of the alkali metal polyhydrides only contained discrete hydrogenic motifs (H$^-$, H$_2$, H$_2^{\delta-}$, and H$_3^-$), a few of the stable alkaline earth polyhydrides were comprised of extended hydrogenic lattices such as clathrate-like structures or one-dimensional chains, as shown in Fig.\ \ref{fig:Group2}(a) and Fig.\ \ref{fig:Group2}(b). It turns out that these structural motifs are linked to high temperature superconductivity. 

In 2012 Wang et al.\ predicted that the CaH$_6$ phase illustrated in Fig.\ \ref{fig:Group2}(a) would become thermodynamically and dynamically stable above 150~GPa \cite{Wang:2012}. The calcium atoms in this $Im\bar{3}m$ symmetry phase are found in a body-centered arrangement and the hydrogen atoms encapsulate them in a sodalite-like clathrate cage. DFT calculations indicated a weak bonding interaction between adjacent hydrogen atoms whose H-H distances measured 1.24~\AA{}. A particularly large electron-phonon-couping (EPC, or $\lambda$) parameter of 2.69 was found, with the largest contribution towards $\lambda$ arising from the modes that resulted from the vibrations of the atoms in the H$_4$ faces comprising the clathrate-like lattice. The estimated $T_c$ at 150~GPa ranged from 220-235~K for a Coulomb pseudopotential ($\mu^*$) value of 0.13-0.1, and $T_c$ decreased under pressure. Because of the large value of $T_c$ predicted for this phase DFT calculations were employed to study MgH$_6$ \cite{Feng:2015a}. It was shown that this structure, which is isotypic to $Im\bar{3}m$-CaH$_6$, became stable with respect to decomposition into MgH$_2$ and H$_2$ above 263~GPa when the zero-point-energy (ZPE) was included. However, because CSP searches were not carried out as a function of stoichiometry at this pressure, it is not clear if this MgH$_6$ phase comprises the convex hull, which has been calculated elsewhere \cite{Lonie:2012}. The electron localization function (ELF) indicated bonding interactions within the hydrogenic framework in MgH$_6$, and a large EPC parameter yielded a $T_c$ ranging from 263-271~K between 300-400~GPa \cite{Feng:2015a}. 

The reason why $Im\bar{3}m$-MgH$_6$ and CaH$_6$ have such a high $T_c$ can be traced back to their large DOS at $E_F$,  which is derived primarily from hydrogen-like states, and the pronounced impact on the electronic structure that results from the motions of the atoms comprising their hydrogenic lattices. As has been recently pointed out, perturbing the hydrogen atoms within quasi-molecular units does not have such a large impact on the electronic structure of the high pressure phases of the hydrides \cite{Shamp:2016,Zhang:2017} yielding lower $\lambda$, and concomitantly $T_c$ values. Therefore, systems with extended hydrogenic lattices are more likely to become superconducting at higher temperatures as compared to those containing H$^-$, H$_2^{\delta-}$ or H$_3^-$ units, for example.

Another phase related to the $Im\bar{3}m$-CaH$_6$ structure is $R\bar{3}m$-SrH$_6$, which becomes stable near $\sim$250~GPa \cite{Hooper:2013,Wang:2015a}. The latter can be derived from the former by elongating four out of the six closest metal-metal contacts and distorting the face that bisects them so it is no longer an ideal hexagon. The elongation of some of the H-H contacts leads to the formation of a hydrogenic lattice comprised of one-dimensional helical hydrogenic chains, as illustrated in Fig.\ \ref{fig:Group2}(b). Fig.\ \ref{fig:Group2}(c) provides the results of our calculations of the phonon band structure and phonon projected density of states of $R\bar{3}m$-SrH$_6$. The EPC of $R\bar{3}m$-SrH$_6$ is lower than what was reported for $Im\bar{3}m$-CaH$_6$ and MgH$_6$, $\lambda=1.10$ at 250~GPa, and so is the estimated $T_c$.

CSP searches were carried out on the polyhydrides of barium up to 150~GPa \cite{Hooper:2012b}, but none of the phases found were isotypic with either $Im\bar{3}m$-CaH$_6$ or $R\bar{3}m$-SrH$_6$. Perhaps higher pressures would yield geometries with motifs that are conducive towards high temperature superconductivity within BaH$_n$.

%%%%%%%%%%%%%%%%%%%%%%%%%%%%%%%%%%%%%%%%%%%%%%%%%%%%%%%%%%%%%%%%%%%%%%%%%%%%%%%%%%%%%%%%%%%%%%%%%%%%%%%%%%%%%%%%%%%%%%%%%%%%
%%%%%%%%%%%%%%%%%%%%%%%%%%                       Transition Metals                     %%%%%%%%%%%%%%%%%%%%%%%%%%%%%%%%%%%%%
%%%%%%%%%%%%%%%%%%%%%%%%%%%%%%%%%%%%%%%%%%%%%%%%%%%%%%%%%%%%%%%%%%%%%%%%%%%%%%%%%%%%%%%%%%%%%%%%%%%%%%%%%%%%%%%%%%%%%%%%%%%%

\section{$d$-Block Elements}

\subsection{Group 3: Scandium, Yttrium, Lanthanum} 

Scandium, yttrium and lanthanum belong to the rare earth (RE) elements, and their chemical behavior resembles that observed in the lanthanides. So, whereas at 1~atm most of the $d$-block elements form hydrides where the H/M ratio is less than 1, Sc, Y and La can form dihydrides that assume a CaF$_2$ type crystal structure where hydrogen fills the tetrahedral holes of a face centered cubic (fcc) lattice, and trihydrides of these elements can be made under pressure. ScH$_3$ and YH$_3$ assume a hexagonal closed packed (hcp) structure. Increasing the pressure leads to a structural transition to an fcc lattice, and this transformation has been studied via DFT calculations  \cite{Pakornchote:2016,Almeida:2009}. The superconducting behavior of ScH$_3$ \cite{Kim:2010a,Wei:2016a}, YH$_3$ \cite{Kim:2010a,Kim:2009} and LaH$_3$ \cite{Kim:2010a} in the fcc structure has been estimated using the Allen-Dynes approximation. The highest $T_c$ values calculated, 20, 40, and 18~K, respectively, were attained at the pressure where the fcc structure first became stable \cite{Kim:2010a}. The $T_c$ was found to decrease with increasing pressure, but a secondary superconducting regime was observed in YH$_3$ above 50~GPa \cite{Kim:2010a}. Later work wherein the Eliashberg formalism was employed to estimate the $T_c$ yielded maximum values of 19.3~K for ScH$_3$ at 18~GPa, and 22.5~K for LaH$_3$ at 11~GPa \cite{Durajski:2014a}.   Another theoretical study found that the  $T_c$ of ScH$_2$ rises steeply under pressure reaching a maximum value of 38~K at 30~GPa, and then it plateaus near 31~K until at least 80~GPa \cite{Wei:2016a}. 

CSP studies have been employed to explore the phase diagram of ScH$_{n}$ ($n = 1-3$) up to 500~GPa \cite{Ye:2015a}. The monohydride was found to assume several phases that were calculated as being more stable at 1 atm than the previously suggested rock-salt structure, which became preferred at 10~GPa, followed by a transition to a $Cmcm$ phase at 265~GPa. ScH$_2$ was found to transform from a CaF$_{2}$-type structure to one with $C2/m$ symmetry at 65~GPa. At 72~GPa, decomposition into ScH and ScH$_3$ was computed to be enthalpically preferred, but higher pressures resulted in the stabilization of ScH$_2$. ScH$_3$ was computed to undergo the following set of  transitions: $P6_3 \rightarrow Fm\bar{3}m \rightarrow P6_{3}/mmc \rightarrow Cmcm$ at 29, 360, and 483~GPa respectively. 
Theoretical calculations have also been undertaken to predict the structure of YH$_3$  at pressures where experimental data is not available \cite{Yao:2010,Liu:2017-Y}. Even though the estimated Debye temperature for a $Cmcm$ phase at 225~GPa was high, the DOS at $E_F$ suggested that the predicted phase would be superconducting only at low temperatures \cite{Yao:2010}. Recently, YH$_3$ compounds with the $P2_1/m$ and $I4/mmm$ spacegroups were predicted to be stable, and their $T_c$ values were estimated as being 19~K and 9~K at 200~GPa, respectively \cite{Liu:2017-Y}. 

\begin{figure}[h!]
\begin{center}
\includegraphics[width=1\columnwidth]{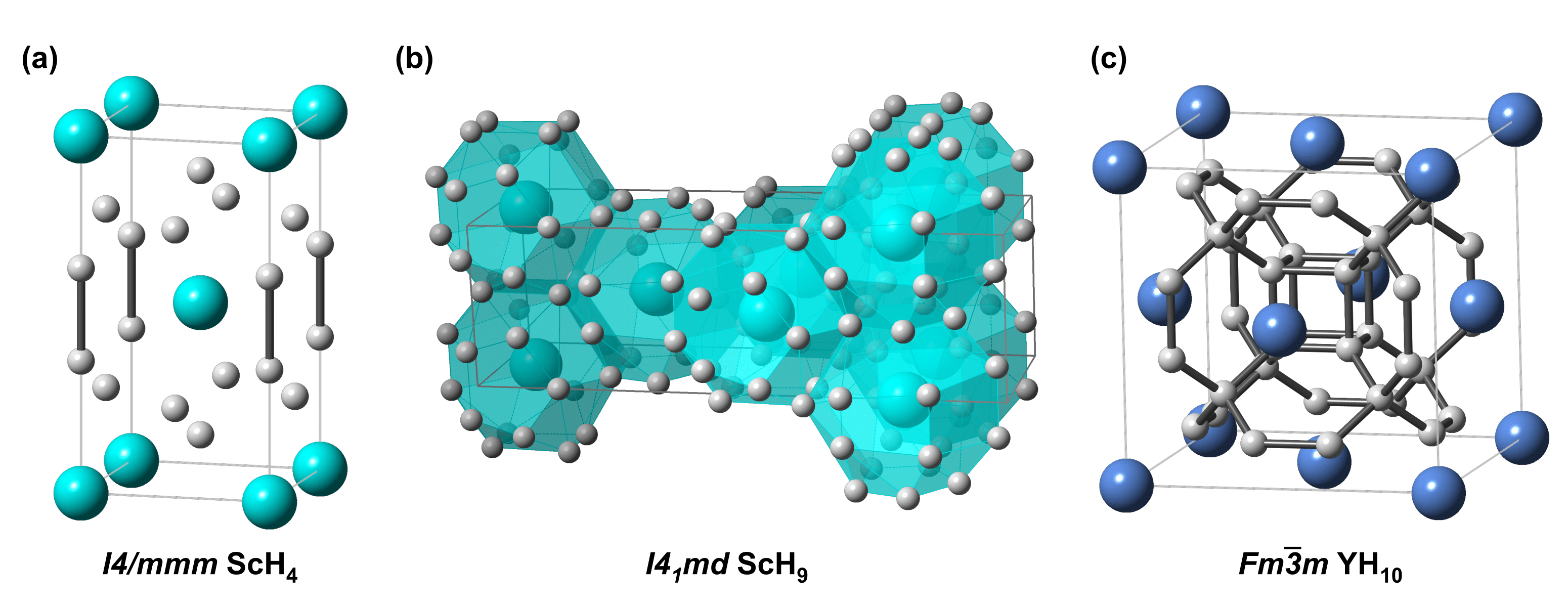}
\end{center}
\caption{Unit cell of (a) $I4/mmm$-ScH$_4$ \cite{Abe:Sc-2017,Qian:Sc-2017,Peng:Sc-2017,Zurek:2018b} (which is isotypic with YH$_4$ \cite{Li:2015a,Liu:2017-La-Y,Liu:2017-Y} and LaH$_4$ \cite{Liu:2017-La-Y}), (b) $I4_1md$-ScH$_9$ \cite{Zurek:2018b}, and (c) $Fm\bar{3}m$-YH$_{10}$ \cite{Li:2015a,Liu:2017-La-Y,Liu:2017-Y} (which is isotypic with the recently predicted \cite{Liu:2017-La-Y} and synthesized  \cite{Geballe:2018a} LaH$_{10}$ phase).}
\label{fig:Group3}
\end{figure}

A number of theoretical studies have recently appeared that investigated the higher hydrides of scandium with $n>3$ \cite{Abe:Sc-2017,Qian:Sc-2017,Peng:Sc-2017,Zurek:2018b}. Abe predicted the following stable phases: $I4/mmm$-ScH$_4$ above 160~GPa, $P6_3/mmc$ ScH$_6$ from 135-265~GPa, and above 265~GPa an $Im\bar{3}m$-ScH$_6$ structure isotypic with the CaH$_6$ phase show in Fig.\ \ref{fig:Group2}(a) that possesses [4$^6$6$^{8}$] polyhedra \cite{Abe:Sc-2017}. The ScH$_4$ structure contained H$^-$ as well as H$_2^{\delta-}$ units as shown in Fig.\ \ref{fig:Group3}(a), and it is isotypic with previously predicted phases for CaH$_4$ \cite{Wang:2012} and SrH$_4$ \cite{Hooper:2013,Wang:2015a}. In addition to these structures the following phases have been predicted above 300~GPa: $Immm$-ScH$_8$, which was found by Qian et al.\ \cite{Qian:Sc-2017}, as well as $P6_3/mmc$-ScH$_9$, $Cmcm$-ScH$_{10}$, and $C2/c$-ScH$_{12}$, which were found by Peng and co-workers \cite{Peng:Sc-2017}. Finally, Ye et al.\ showed that the $I4_1md$-ScH$_9$ phase illustrated in Fig.\ \ref{fig:Group3}(b) and $Immm$-ScH$_{12}$ had somewhat lower enthalpies than the previously proposed structures \cite{Zurek:2018b}. In addition two new stable phases, $Cmcm$-ScH$_6$ and $Cmcm$-ScH$_7$, were predicted. It was therefore proposed that a large number of high hydrides of scandium could be synthesized above 150~GPa, and some of them were computed to have $T_c$ values as high as $\sim$200~K \cite{Peng:Sc-2017,Zurek:2018b}. 

Three manuscripts have explored superconductivity in higher hydrides of yttrium, YH$_n$, $n\ge4$ \cite{Li:2015a,Liu:2017-La-Y,Liu:2017-Y}. CSP techniques found that (in addition to YH$_3$)  YH$_4$ and YH$_6$ phases, which are isotypic to $I4/mmm$-ScH$_4$ and $Im\bar{3}m$-ScH$_6$, were thermodynamically and dynamically stable under pressure. At 120~GPa, the $T_c$ was estimated to be 84-95~K for the former and 251-264~K for the latter \cite{Li:2015a}. Even though the H-H distances at 120~GPa in $Im\bar{3}m$-YH$_6$ were somewhat long, 1.31~\AA{}, the ELF revealed covalent bonding interactions between the hydrogens.  
Another hydride of yttrium that was predicted to be superconducting at high temperatures was the sodalite-like YH$_{10}$ phase illustrated in Fig.\ \ref{fig:Group3}(c) whose $T_c$ was estimated as being 305-326~K at 250~GPa via the Eliashberg equations \cite{Liu:2017-La-Y}.  

Computational explorations of the hydrogen-rich phase diagram of lanthanum also led to the prediction of phases with the propensity for high temperature superconductivity \cite{Liu:2017-La-Y}. At 150~GPa LaH$_n$ with $n=2-5,8,10$ were found to be stable.  LaH$_8$ was comprised of an extended hydrogenic lattice and at 300~GPa it's $T_c$ was estimated as being 114-131~K. LaH$_{10}$ adopted a sodalite-like structure (isotypic to the YH$_{10}$ phase show in Fig.\ \ref{fig:Group3}(c)) wherein the La atoms were arranged on an fcc lattice. This phase contained [4$^6$6$^{12}$] polyhedra, and numerically solving the Eliashberg equations yielded a $T_c$ of 257-274~K at 250~GPa for LaH$_{10}$. Its $T_c$ was found to decrease with increasing pressure. Remarkably, a superhydride of lanthanum consistent with the theoretically predicted structure for LaH$_{10}$ was recently synthesized at 170~GPa \cite{Geballe:2018a}. Decompression led to a $Fm\bar{3}m \rightarrow R\bar{3}m \rightarrow C2/m$ phase transformation.

Recently, an extensive theoretical investigation was carried out on the high hydrides of the REs under pressure \cite{Peng:Sc-2017}. CSP calculations were performed to determine the most stable structures of the hydrides of Sc, Y, La, Ce and Pr and their convex hulls were obtained. It was assumed that the the polyhydrides of Nd, Pm, Sm, Eu, Gd, Tb, Dy, Ho, Er, Tm, Yb and Lu would be isotypic with phases that were found via CSP for the other REs. Therefore, the convex hulls for the hydrides of these twelve REs were generated by carrying out geometry optimizations on the predicted REH$_n$ ($n=3,4,6,9,10$) species. The candidate phases considered were: $Fm\bar{3}m$, $Cmcm$, and $R\bar{3}m$ REH$_3$, $I4/mmm$ REH$_4$, $Im\bar{3}m$, $R\bar{3}c$, and $C2/m$ REH$_6$, $P6_3/mmc$, $F\bar{4}3d$, and $P6_3m$ REH$_9$, and $Fm\bar{3}m$, $R\bar{3}m$, and $Cmcm$ REH$_{10}$. The stable REH$_6$, REH$_9$ and REH$_{10}$ phases resembled clathrate structures (examples of two of these, $Im\bar{3}m$-MH$_6$ and $Fm\bar{3}m$-MH$_{10}$, are shown in Fig.\ \ref{fig:Group2}(a) and Fig.\ \ref{fig:Group3}(c), respectively) with H$_{24}$, H$_{29}$ and H$_{32}$ cages surrounding the metal atoms, and H-H distances of 1-1.2~\AA{}. The $T_c$ of some of these phases was estimated via solving the Eliashberg equations. ScH$_6$, ScH$_9$, YH$_6$, YH$_9$, YH$_{10}$, and LaH$_{6}$ had quite high $T_c$s, with predicted values of up to 303~K at 400~GPa for YH$_{10}$. The $T_c$ of LaH$_9$, CeH$_9$, CeH$_{10}$ and PrH$_9$ were significantly lower, $<$56~K, because they contained heavier elements.

\subsection{Group 4: Titanium, Zirconium, Hafnium}
\begin{figure}[h!]
\begin{center}
\includegraphics[width=0.7\columnwidth]{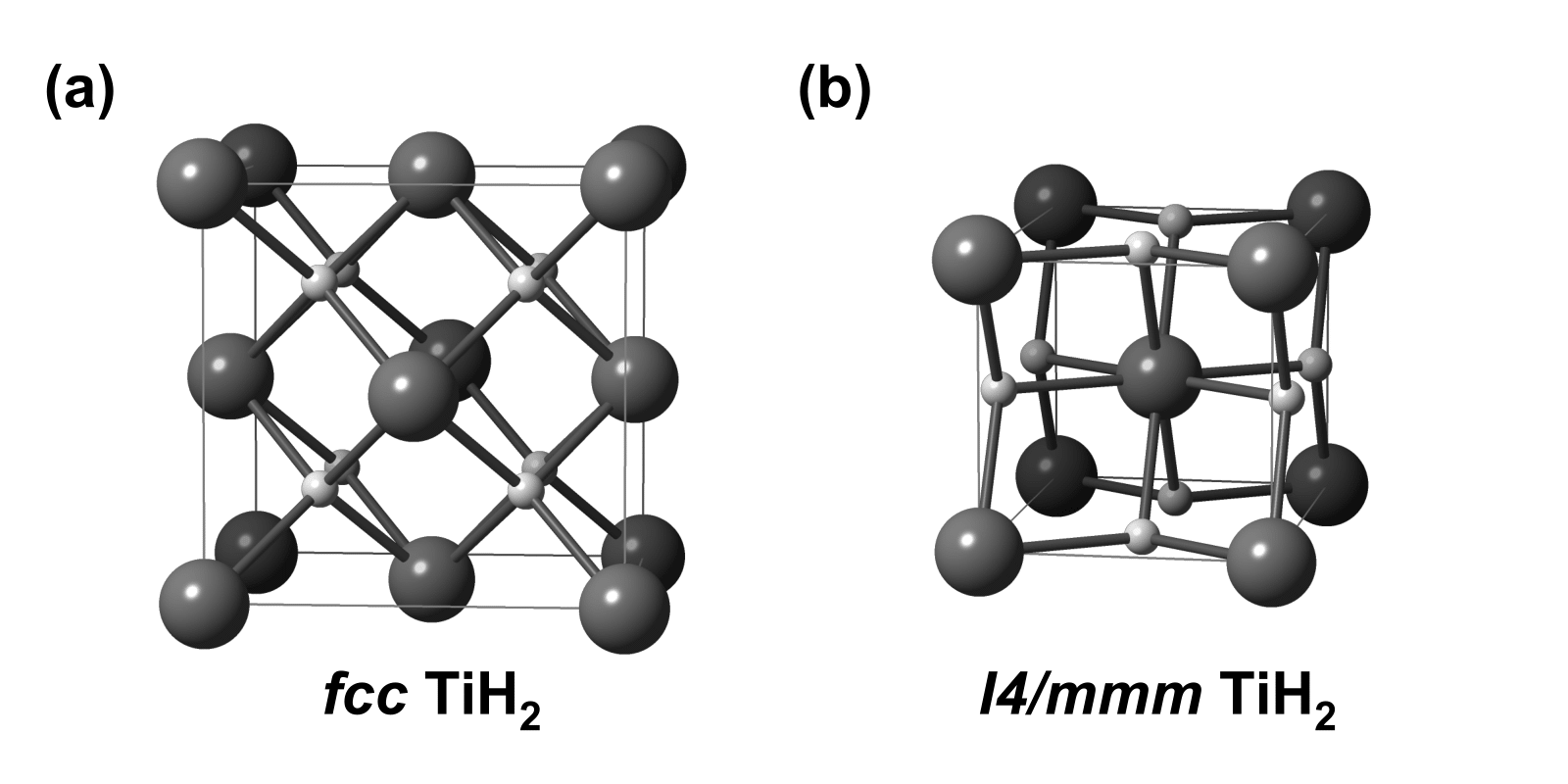}
\end{center}
\caption{Two structures often adopted by metal hydrides: (a) the CaF$_2$ (fcc) structure, and (b) the $I4/mmm$ structure. For example, TiH$_2$ assumes the fcc arrangement at ambient conditions, and the $I4/mmm$ configuration at lower temperatures.}
\label{fig:group4}
\end{figure}

The superconducting properties of TiD$_{0.74}$ have been measured under pressure, and it was shown that $T_c$ varied from 4.17-4.43~K between 14-30~GPa \cite{Bashkin:1998a}, somewhat lower than the value of 5.0~K obtained when a metastable form of this system was quenched to atmospheric pressures. 
As is common for many transition metal dihydrides, TiH$_2$ crystallizes in the CaF$_{2}$ (fcc) structure illustrated in Fig.\ \ref{fig:group4}(a) at room temperature. At lower temperatures a transition to the $I4/mmm$ structure in Fig.\ \ref{fig:group4}(b) occurs. A recent DFT study computed a $T_c$ of 6.7~K and 2~mK for the high and low temperature phases, respectively \cite{Shanavas:2016a}. Experiments revealed that the fcc$\rightarrow I4/mmm$ transition also occurs at room temperature and 0.6~GPa, and suggested that this phase remains stable up to 90~GPa \cite{Kalita:2010}. CSP techniques, on the other hand, predicted the following sequence of transitions:  $I4/mmm \rightarrow P4/nmm \rightarrow P2_1/m$ at 63 and 294~GPa, respectively \cite{Gao:2013}. The computed XRD patterns of the $P4/nmm$ structure were found to be in better agreement with experimental results up to 90~GPa than those of $I4/mmm$.

A number of phases have been reported in the Zr/H phase diagram at ambient pressure, including two non-stoichiometric compounds: fcc $\delta$-ZrH$_x$ with $1.4<x<1.7$, and an $I4/mmm$ symmetry $\epsilon$-ZrH$_x$ structure, as well as a stoichiometric $P4_{2}/n$-ZrH phase \cite{Maimaitiyili:2016}. At ambient conditions ZrH$_2$ assumes the same $I4/mmm$ symmetry structure as TiH$_2$ (see Fig.\ \ref{fig:group4}(b)), and DFT calculations suggested that it would undergo an $I4/mmm\rightarrow P4/nmm$ transition at 103~GPa \cite{Huang:2014}. Phases with Zr$_2$H, ZrH, ZrH$_{1.5}$, and ZrH$_2$ stoichiometries have been studied theoretically \cite{Zhu:2010}, and CSP techniques were employed to find stable phases with the stoichiometry Zr$_x$H$_y$ ($x=1, y=1-8; x=2, y=3, 5$) up to 150~GPa \cite{Li:2017}. For pressures up to 100~GPa ZrH, ZrH$_2$ and ZrH$_3$ were the only thermodynamically stable phases that were identified, whereas at 150~GPa ZrH$_6$ also emerged as a stable phase. The following structures were found to be dynamically stable and metallic at the pressures in the parentheses: $P4_2/mmc$-ZrH (0~GPa), $Cmcm$-ZrH (120~GPa), $R\bar{3}m$-ZrH (150~GPa), $I4/mmm$-ZrH$_2$ (50~GPa), $Pm\bar{3}n$-ZrH$_3$ (50~GPa), and $Cmc2_1$-ZrH$_6$ (140~GPa). The only phase that exhibited a significant EPC, wherein the motions of the Zr atoms contributed towards 82\% of the total $\lambda$, was $Cmcm$-ZrH, whose $T_c$ was estimated as being 10.6~K at 120~GPa.

The phases adopted by HfH$_2$ under pressure, as well as their electronic structure and properties, were recently explored theoretically \cite{Liu:2015}. The $I4/mmm$ phase was calculated as being more stable than the CaF$_2$ structure at atmospheric conditions (both structure types are shown in Fig.\ \ref{fig:group4}). A transition to a $Cmma$ structure at 180~GPa was predicted, followed by a transformation to a $P2_1/m$ phase at 250~GPa. The $T_c$ was estimated via the Allen-Dynes modified McMillan equation as being 47-193~mK at 1~atm, 5.99-8.16~K at 180~GPa and 10.62-12.8~K at 260~GPa for the aforementioned phases. For each phase $T_c$ was found to decrease as the pressure increased. 

At the time of writing this review the chemistry of the hydrides of titanium and of hafnium with unique stoichiometries had not yet been investigated.

\subsection{Group 5: Vanadium, Niobium, Tantalum} 

A number of molecular hydrides of the group 5 metals, including VH$_2$(H$_2$), NbH$_4$ and TaH$_4$ have been synthesized using laser ablation, and studied via molecular quantum mechanical calculations \cite{Wang:2011}. 

In the solid state experiments have shown that VH$_2$ adopts the CaF$_2$ structure, and CSP investigations have also found this to be the most stable phase at 1~atm \cite{Chen:2014}. A further transition to a $Pnma$ phase was predicted at 50~GPa, and the $T_c$ was estimated as being 0.5~K and 4~K for VH$_2$ at 0 and 60~GPa, respectively. This finding is in line with experiments that did not show any hints of superconductivity for VH$_n$, $n<1.93$, above 1.5~K \cite{Ohlendorf:1979a}.  To the best of our knowledge, the structures and properties of vanadium hydrides with $n>2$ have not yet been explored computationally.

\begin{figure}[h!]
\begin{center}
\includegraphics[width=0.7\columnwidth]{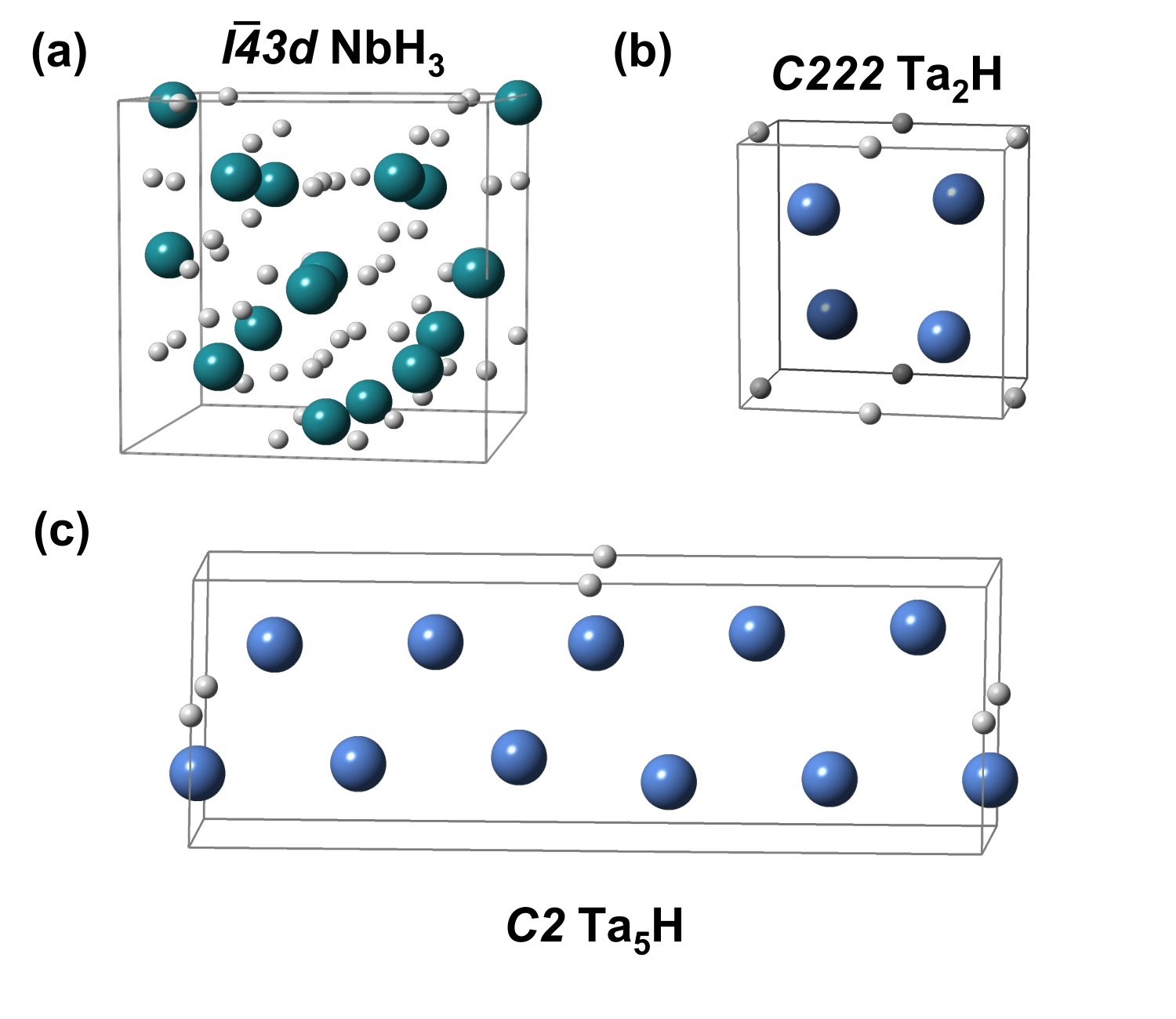}
\end{center}
\caption{Hydrides of group 5 transition metals that have been synthesized and studied theoretically: (a) $I\bar{4}3d$-NbH$_3$ \cite{Gao:2013a,Liu:2017}, (b) $C222$-Ta$_2$H, and (c) $C2$-Ta$_5$H \cite{Zhuang:2017a}.}
\label{fig:group5}
\end{figure}

At ambient temperature and pressure phases with the stoichiometries NbH$_x$, $x\le0.9$, are known, and various forms of NbH$_x$, $x<1$, have been proposed to exist at different temperatures. NbH$_2$ has been synthesized at 2~atm in the CaF$_2$ structure shown in Fig.\ \ref{fig:group4}(a). DFT calculations found this to be the most stable phase at atmospheric pressures, and a transition to a $P6_3mc$ phase was predicted to occur at 45~GPa \cite{Chen:2014}. The $T_c$ of NbH$_2$ was estimated as being 1.5~K and 0.5~K at 0 and 60~GPa, respectively. A comprehensive theoretical investigation of NbH$_n$ ($n=0.75, 1-6$) up to 400~GPa has been carried out \cite{Gao:2013a}. At 1~atm and 10~GPa NbH$_{0.75}$, NbH and NbH$_2$ were computed to be thermodynamically stable. At 50~GPa the NbH$_3$ stoichiometry joined them on the convex hull. By 400~GPa species with $n=1-4$ were found to be stable. $Cccm$-NbH (1~atm), $Fm\bar{3}m$-NbH$_2$ (50~GPa), $I\bar{4}3d$-NbH$_3$ (100~GPa), $I4/mmm$-NbH$_4$ (300~GPa) and $Cmmm$-NbH$_6$ (400~GPa) were good metals at the pressures given in the parentheses, suggesting that all of them have the potential to be superconductors. Due to the computational expense involved the $T_c$ of only a few phases could be estimated via the Allen-Dynes modified McMillan equation. The $T_c$ values were calculated to be: 1.5-2.4~K for NbH at 1~atm, 1.5-2.6~K and 0.7-1.5~K for NbH$_2$ at 1~atm and 50~GPa, respectively, and 38-47~K for NbH$_4$ at 300~GPa. The reason for the higher $T_c$ obtained for NbH$_4$ can be traced back to the larger $\lambda$ and average logarithmic frequency ($\omega_\text{log}$), which is a result of the presence of a larger mole ratio of hydrogen as compared to the other phases. Within the framework of strongly-coupled Eliashberg theory, the $T_c$ of NbH$_4$ was calculated as being somewhat higher, 49.6~K \cite{Durajski:2014b}. A recent combined experimental and theoretical study showed that NbH$_{2.5}$ could be synthesized below 46~GPa, and above 56~GPa the NbH$_3$ phase illustrated in Fig.\ \ref{fig:group5}(a) was made \cite{Liu:2017}. The formation of some of the phases that were experimentally observed could only be explained when finite temperature contributions to the free energy were considered.

A number of Ta$_x$H$_y$ compounds with varying stoichiometries including Ta$_2$H and TaH$_2$ have been synthesized at ambient conditions. A review of the experimental studies, coupled with the results of DFT calculations that investigated the structures and electronic structures of the hydrides of tantalum up to 300~GPa is given in Ref.\ \cite{Zhuang:2017a}. This study showed that 
the XRD patterns for the theoretically predicted ambient pressure phases of Ta$_2$H ($C222$) and Ta$_5$H ($C2$), shown in Fig.\ \ref{fig:group5}(b,c), agreed well with those determined experimentally. The computed enthalpy of a $P6_3mc$ symmetry TaH$_2$ phase was somewhat lower than that of the experimentally found $Fm\bar{3}m$ (CaF$_2$ type, see Fig.\ \ref{fig:group4}b) structure. At 1~atm only TaH and TaH$_2$ were thermodynamically stable. By 50~GPa TaH$_3$ and TaH$_4$ joined them on the convex hull, and TaH$_6$ became stable by 270~GPa. Molecular H$_2$ units were absent in the most stable structures, which were all metallic. Phonon calculations verified dynamic stability for $Pnma$-TaH$_2$, $R\bar{3}m$-TaH$_4$ and $Fdd2$-TaH$_6$ at 200, 250 and 300~GPa, respectively. At these pressures the $T_c$s were predicted to be 5.4-7.1~K, 23.9-31~K, and 124.2-135.8~K, respectively. Recently, a new dihydride of tantalum that had an hcp metal lattice was synthesized under pressure \cite{Kuzovnikov:2017a}.

\subsection{Group 6: Chromium, Molybdenum, Tungsten} 
\begin{figure}[h!]
\begin{center}
\includegraphics[width=0.5\columnwidth]{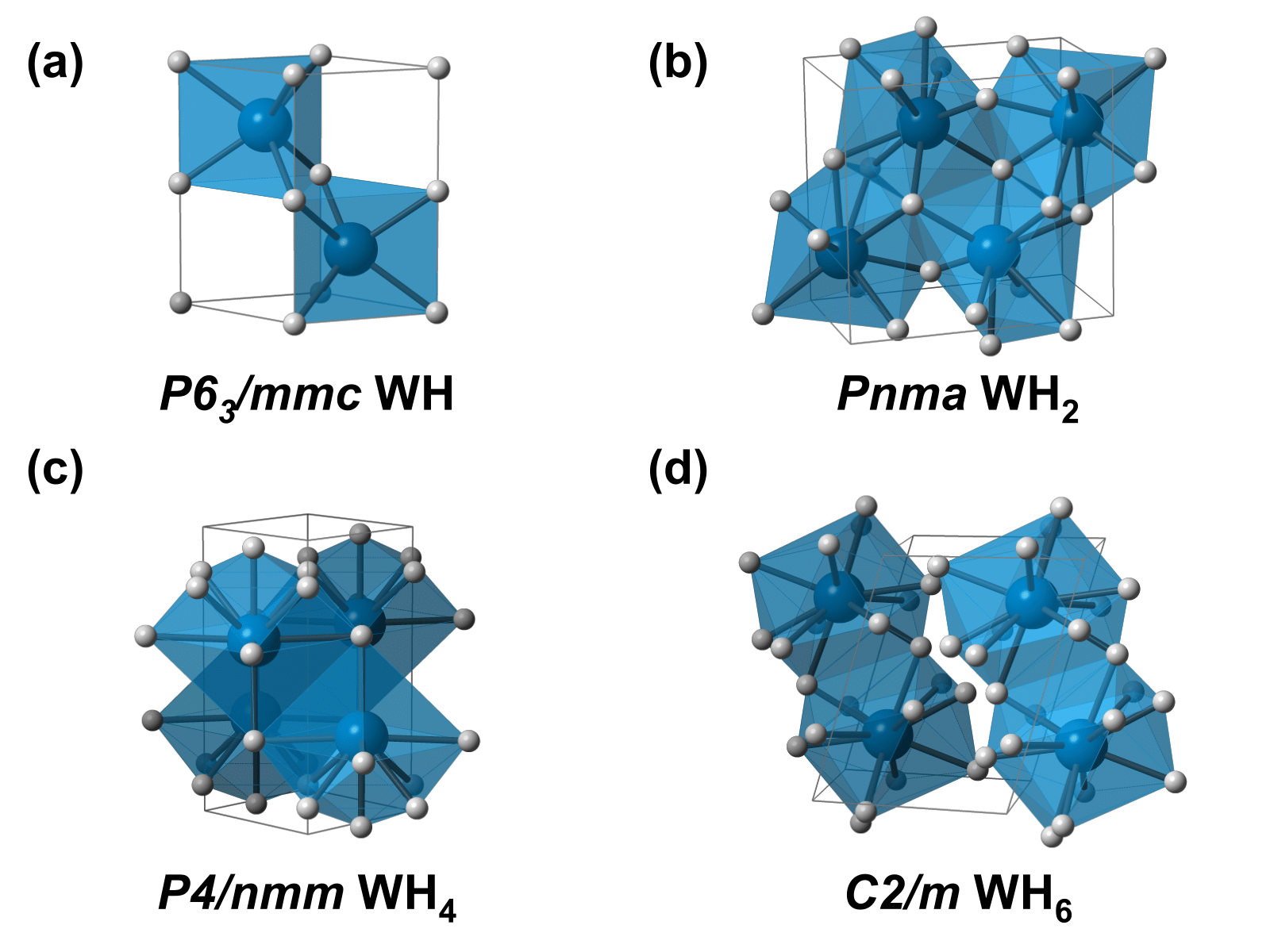}
\end{center}
\caption{The hydrides of tungsten studied theoretically and experimentally in Ref.\ \cite{Zaleski:2012a}. WH has been synthesized in the anti-NiAs ($P6_3/mmc$ symmetry) structure under pressure \cite{Zaleski:2012a, Scheler:2013b}.}
\label{fig:tungsten}
\end{figure}

In the solid state at 1~atm CrH, MoH and WH crystallize in an anti-NiAs structure, shown in Fig.\ \ref{fig:tungsten}(a), wherein the hydrogen atoms are found in the interstitial sites. Less is known about a cubic polymorph of CrH, because of the difficulties of preparing it reproducibly \cite{Snavely:1949,Pozniak:2001}. The dihydride and trihydride of chromium have been made, but they have not yet been structurally characterized. Molecular hydrides of the group 6 transition metals have been synthesized with a wide range of stoichiometries, including species with high hydrogen content such as (H$_2$)$_2$CrH$_2$ \cite{Wang:2003}, MoH$_6$ \cite{Wang:2005}, WH$_6$ \cite{Wang:2002}, and WH$_4$(H$_2$)$_4$ \cite{Wang:2008a}. In fact, quantum chemical calculations showed that the formation of MH$_{12}$ (M=Cr, Mo, W) from MH$_6$ and 3H$_2$ molecules is energetically favorable \cite{Gagliardi:2004}. The existence of these molecules has led to the speculation that high pressure could potentially be used to stabilize high hydrides of the group 6 transition metals, despite the fact that such species are not known at ambient conditions.

Recently, CSP has been employed to predict the most stable structures of compounds with the Cr$_x$H$_y$ stoichiometry up to pressures of 300~GPa \cite{Yu:2015}. Whereas at 1~atm CrH was the only thermodynamically stable phase found, pressure promoted the stabilization of hydrogen rich phases. When ZPE effects were included Cr$_2$H$_3$, CrH$_2$, CrH$_3$, CrH$_4$, and CrH$_8$ were found to lie on the convex hull at some pressure. All of these structures contained the common feature that the metal sublattices were hexagonal close-packed, and hydrogen atoms were found in the octahedral or tetrahedral sites. The EPC parameter was calculated for CrH and CrH$_3$ as representative structures for these phases, and the $T_c$ was estimated using the Allen-Dynes equation. The $T_c$ of CrH was calculated as being 10.6~K at 0~GPa, and it was found to decrease with increasing pressure to 3.1~K at 200~GPa. Because of the increased hydrogen content, the EPC and $\omega_{log}$ of CrH$_3$ was calculated as being larger than that of CrH, with a concomitantly larger $T_c$ of 37.1~K at 81~GPa. Again, $T_c$ decreased under pressure, dropping to $\sim$28~K at 200~GPa. 

Recent experiments uncovered the crystal structures assumed when Mo was subject to hydrogen pressures up to 30~GPa \cite{Kuzovnikov:2017}. At 4~GPa a phase transformation from a bcc to an hcp metal hydride occurred. The H:Mo ratio increased continuously with pressure and reached a saturation limit of 1.35:1 at about 15~GPa. First principles calculations have shown that phases with MoH and MoH$_2$ stoichiometries are dynamically stable from 0-100~GPa \cite{Feng:2016}. Whereas $P6_3/mmc$-MoH was found on both the 20~GPa and 100~GPa convex hulls, MoH$_2$ only became thermodynamically stable by 100~GPa, and MoH$_3$ did not lie on the convex hull between 0-100~GPa. At 2~GPa the most stable MoH$_2$ phase assumed the $P6_3mc$ spacegroup, and a $Pnma$ phase became preferred past 24~GPa. In all of the stable structures Mo atoms were found in the hexagonal sites and H atoms in the octahedral and tetrahedral sites. Even though both MoH and MoH$_2$ were metallic at 100~GPa, the DOS at $E_F$ was dominated by metal $d$-states, suggesting that their $T_c$s will be lower than that of elemental Mo.

Because the solubility of hydrogen in tungsten is low, this metal is often employed as a gasket material to seal hydrogen in DACs at high pressure. However, during an experimental study of a mixture of H$_2$ and SiH$_4$ under pressure, diffraction measurements showed evidence for the serendipitous formation of a tungsten hydride \cite{Strobel:2009a}. This, along with the synthesis of a number of molecular hydrides of tungsten in low temperature matrices (see Ref.\ \cite{Wang:2008a} and Refs.\ within) inspired the computational search for high hydrides of tungsten in the solid state at 1~atm and under pressure \cite{Labet:2012e,Zaleski:2012a}. Some of the phases predicted are illustrated in Fig.\ \ref{fig:tungsten}. CSP studies found WH and WH$_4$ phases that lay on the convex hull at 25~GPa, where they were joined by WH$_2$ by 100~GPa, and WH$_6$ by 150~GPa \cite{Zaleski:2012a}. Moreover, computational experiments hinted that WH$_6$ has the ability to polymerize at ambient conditions \cite{Labet:2012e}. Phonon calculations confirmed that all of these phases were dynamically stable at 150~GPa. Only WH and WH$_2$ were found to be good metals, whereas the Fermi level of WH$_4$ and WH$_6$ lay in a pseudogap. Unfortunately, attempts to synthesize hydrides beyond WH, which was found to crystallize in an anti-NiAs structure, were not successful. 
A more recent computational and experimental study confirmed the synthesis of WH by 25~GPa \cite{Scheler:2013b}. However, it was shown that between 25 and 50~GPa the equation of states data was best explained by a combination of WH and WH$_2$, suggesting that the maximum H:W ratio attained in the synthesis is $\sim 1\frac{1}{3}$. The superconducting properties of the compressed tungsten hydride phases has so far not been explored theoretically or experimentally.

\subsection{Group 7: Manganese, Technetium, Rhenium} 

At the time of writing this manuscript, very little had been reported about the high pressure structures of the group 7 hydrides, and their propensity for superconductivity. On the other hand, molecular hydrides of these elements such as MnH$_2$ and ReH$_4$ have been made via laser ablation, and their electronic structure has been studied via first-principles calculations \cite{Wang:2003a}. 

MnH$_x$ phases with $x<1$ are known to adopt structures that are based on fcc, hcp and double hexagonal closed packed (dhcp) metal lattices, and their phase diagrams have been studied as a function of H/Mn content up to 1000~$^\circ$C and 7.6~GPa \cite{Antonov:1996,Fukai:2002}. 

Below 2~GPa the H:Tc ratio in hydrides of technetium is less than 1, and the $T_c$ of the hydrogenated system is lower than that of the pure metal \cite{Antonov:1989,Spitsyn:1982}. Theoretical CSP investigations found that at 50~GPa $P6_3/mmc$-TcH is the only thermodynamically stable phase, but by 150~GPa $I4/mmm$-TcH$_2$ and $Pnma$-TcH$_3$ also lie on the convex hull \cite{Li:2016}. By 200~GPa TcH$_2$ is no longer thermodynamically stable. However, these phases, and others, are metastable across a broad pressure range. The Allen-Dynes modified McMillan equation was employed to estimate the following $T_c$s: 5.4-10.7~K for $I4/mmm$-TcH$_2$ between 100-200~GPa, 8.6~K for $Cmcm$-TcH$_2$ at 300~GPa, and 9.9~K for $P4_2/mmc$-TcH$_3$ at 300~GPa.

An experimental study showed that above 50~GPa silane decomposes and the released H$_2$ reacts with  metals in the DAC to form metal hydrides \cite{Degtyareva:2009a}.  At 50~GPa the diffraction pattern of the rhenium metal indicated that its volume had expanded, presumably due to the uptake of hydrogen. Based upon the volume it was proposed that a compound with ReH$_{0.39}$ stoichiometry had formed, in agreement with previous studies \cite{Atou:1995}, and this species was found to be stable up to at least 108~GPa. An ReH$_{0.5}$ stoichiometry was synthesized at 15~GPa in the layered anti-CdI$_2$ structure, and heating this compound promoted a phase transition to the NiAs structure type wherein the hydrogen content was increased to ReH$_{0.85}$ \cite{Scheler:2011b}.

\subsection{Group 8: Iron, Ruthenium, Osmium} 

The Earth's core is composed primarily of iron alloyed with nickel and light elements. However, because seismic models suggest that the density of the Earth's core is several percent lower than estimates made for iron-nickel alloys, it has been proposed that iron hydrides may be important constituents of the core. A number of experiments have shown that pressure dramatically increases the solubility of hydrogen in iron \cite{Fukai:1982} yielding Fe:H ratios approaching 1:1 \cite{Badding:1991} that assume a number of potential structure types including dhcp, hcp and fcc \cite{Yamakata:1992,Antonov:2002,Narygina:2011}. DFT calculations on FeH up to 130~GPa suggested the following sequence of structural phase transitions: $\text{dhcp}\rightarrow \text{hcp} \rightarrow \text{fcc}$ \cite{Isaev:2007}. 

The propensity for the stabilization of novel stoichiometries under pressure has inspired theoretical and experimental exploration of iron hydrides with non-classical compositions.
Evolutionary structure searches coupled with DFT calculations were employed to predict the most stable Fe$_x$H$_y$ ($x=1-4$, $y=1-4$) structures at pressures of 100-400~GPa \cite{Bazhanova:2012a}. Even though FeH was the lowest point on the convex hull within the whole pressure range studied, all of the other stoichiometries either lay on the hull or close to it. For Fe$_x$H with $x\ge1$ the iron atoms in the most stable structures assumed close-packed lattices, and the hydrogen atoms were located in the octahedral voids. At pressures similar to those in the center of the Earth, FeH was found to adopt a rock-salt structure wherein the iron atoms were fcc-packed. At 300 and 400~GPa the preferred FeH$_3$ geometries were predicted to assume the Cu$_3$Au structure (spacegroup $Pm\bar{3}m$), and the Cr$_3$Si ($Pm\bar{3}n$) structure types, respectively. At both of these pressures the most stable FeH$_4$ structure adopted $P2_1/m$ symmetry. A theoretical investigation focused on the FeH$_4$ stoichiometry from 80-400~GPa \cite{Li:2017a}. CSP suggested that the following pressure induced phase transitions would occur in this phase: $P2_13 \rightarrow Imma \rightarrow P2_1/m$ at 109 and 242~GPa, respectively. Only $Imma$-FeH$_4$ was found to be a metal with the HSE hybrid functional. Its $T_c$ was estimated as being 1.7~K using the Allen-Dynes modified McMillan equation, where 75\% of the EPC originated from the motions of the hydrogen atoms. Recent evolutionary structure searches have predicted hitherto unknown phases to be stable above 150~GPa: $P4/mmm$-Fe$_3$H$_5$, $Immm$-Fe$_3$H$_{13}$, $I4/mmm$-FeH$_5$, and $Cmmm$-FeH$_6$ whose $T_c$ was estimated to be 43~K at 150~GPa \cite{Kvashnin:2018a}.

\begin{figure}[h!]
\begin{center}
\includegraphics[width=0.95\columnwidth]{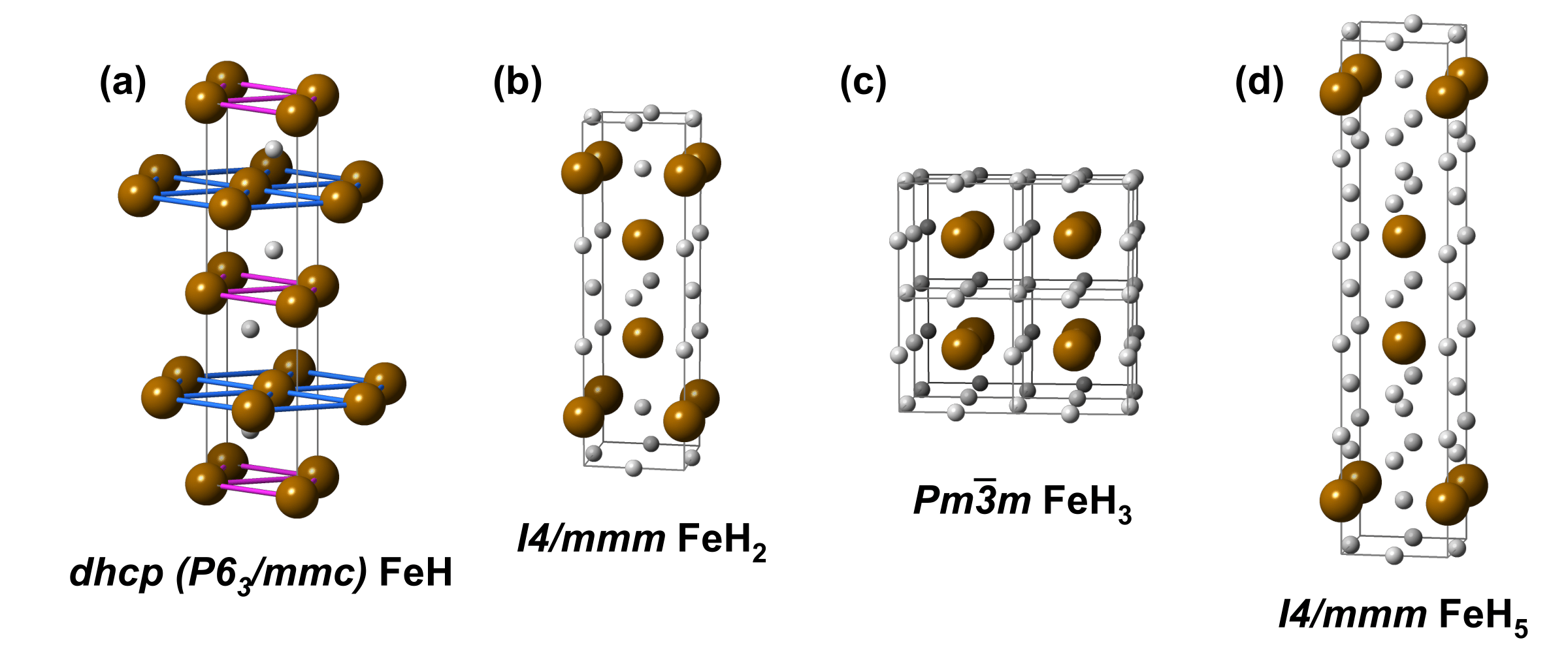}
\end{center}
\caption{Iron hydride phases that have been synthesized under pressure (a) dhcp-FeH \cite{Pepin:2014}, (b) $I4/mmm$-FeH$_2$ \cite{Pepin:2014}, (c) $Pm\bar{3}m$-FeH$_3$ \cite{Pepin:2014}, and (d) $I4/mmm$-FeH$_5$ \cite{Pepin:2017a}.}
\label{fig:iron}
\end{figure}

The work of Bazhanova et al.\ \cite{Bazhanova:2012a} inspired a combined experimental/theoretical investigation  wherein laser heating of a DAC was employed to synthesize higher hydrides of iron \cite{Pepin:2014}. In the experiment the dhcp-FeH structure illustrated in Fig.\ \ref{fig:iron}(a), which was calculated to undergo a ferromagnetic (FM) to nonmagnetic (NM) transition at 45~GPa, was observed. At 67~GPa hydrogen uptake occurred leading to an $I4/mmm$-FeH$_{\sim2}$ phase with FM order. A recent study showed that the positions of the hydrogen atoms in the most stable dihydride phase, illustrated in Fig.\ \ref{fig:iron}(b), differed from those proposed in the original study \cite{Kvashnin:2018a}. The NM $Pm\bar{3}m$-FeH$_3$ phase shown in Fig.\ \ref{fig:iron}(c), which was previously predicted by CSP in Ref.\ \cite{Bazhanova:2012a}, formed at 86~GPa. Calculations showed that all of the synthesized phases were metallic, hinting that they could potentially be superconducting. The FM to NM transition of the hcp and dhcp phases of FeH under pressure have been studied via DFT calculations \cite{Tsumuraya:2012}. 

Recently, the FeH$_5$ phase illustrated in Fig.\ \ref{fig:iron}(d) was synthesized after laser heating in a DAC above 130~GPa \cite{Pepin:2017a}. This phase contained FeH$_3$ units separated by slabs of atomic hydrogen wherein the H-H distances resembled those that would be found in bulk atomic hydrogen. Calculations predicted this phase to be superconducting below $\sim$50~K around 150~GPa \cite{Majumdar:2017a,Kvashnin:2018a}.

Experiments carried out to 9~GPa have not provided evidence for the formation of hydrides of ruthenium. However, CSP studies suggested that a monohydride in the fcc structure (also known as the NaCl structure with $Fm\bar{3}m$ symmetry) would become stable above 10~GPa \cite{Gao:2012}. In a later study of RuH$_{n}$ ($n = 1-8$) from 50-300~GPa three stoichiometries were found to lie on the convex hull \cite{Liu:2015c}. $Fm\bar{3}m$-RuH (NaCl type) became stable above 7~GPa and remained the preferred monohydride until 300~GPa. RuH$_3$ emerged as a stable structure at 66~GPa, assuming the $Pm\bar{3}m$ structure illustrated in Fig.\ \ref{fig:iron}(c) below 120~GPa, and a $Pm\bar{3}n$ symmetry structure at higher pressures. Another stable phase, RuH$_6$, adopted the $Pm$ spacegroup at 19.5~GPa and transitioned to an $Imma$ symmetry phase at 95~GPa. By 100~GPa, however, the decomposition of RuH$_6$ to RuH$_3$ and H$_2$ was found to be thermodynamically preferred. Whereas the mono and trihydride only contained atomistic hydrogen, RuH$_6$ was also comprised of  molecular H$_2$ units. RuH and RuH$_{3}$ were found to be metallic, whereas RuH$_{6}$ was semi-conducting. The $T_{c}$ for RuH was estimated as being 0.41~K at 100~GPa, and for RuH$_3$ it was found to be 3.57~K and 1.25~K at 100~GPa and 200~GPa, respectively. A subsequent experimental study illustrated that a hydride of ruthenium could be synthesized in a DAC between 14-30~GPa \cite{Kuzovnikov:2016}. The phase was identified as the monohydride wherein the metal lattice had the fcc structure and the hydrogen atoms were located on the octahedral sites, in agreement with the original prediction by Gao \emph{et al.} \cite{Gao:2012}.

Evolutionary structure searches were carried out to identify new phases of OsH$_n$ ($n=1-8$) from 50-300~GPa \cite{Liu:2015b}. Three stable stoichiometries were predicted, namely, OsH ($P\ge94$~GPa), OsH$_{3}$ ($P=140-246$~GPa), and OsH$_{6}$ ($P=38-155$~GPa). OsH and OsH$_{3}$ adopted a single stable phase (with $Fm\bar{3}m$ and $Cmm2$ symmetry, respectively), while OsH$_{6}$ assumed a $P2_1/c$ phase below, and an $Fdd2$ phase above 104~GPa. Whereas OsH and OsH$_3$ were found to be good metals, OsH$_6$ was semi-conducting.
The hydrogenic sublattices of the stable and metastable phases contained some of the following structural motifs: H, H$_2$, as well as linear, bent and triangular H$_3$ units. The estimated $T_{c}$ of OsH was 2.1~K at 100~GPa, which is higher than that of pure osmium metal.

\subsection{Group 9: Cobalt, Rhodium, Iridium}

Recent experiments carried out in a DAC show that cobalt undergoes a two-step hydrogenation process at and above room temperature \cite{Ishimatsu:2012}. Above 2.7~GPa the hcp structure was maintained and hydrogen uptake resulted in a CoH$_{0.6}$ stoichiometry, followed by a transformation to a fcc CoH$_{0.9}$ structure above 4.2~GPa \cite{Ishimatsu:2012}. Between 5-11~GPa the monohydride, FM fcc-CoH, was formed. Recent experiments carried out in a DAC up to 22~GPa led to hydrogen uptake by cobalt at $\sim$4~GPa \cite{Kuzovnikov:2015}. Above this pressure a monohydride with the fcc structure formed, and the hydrogen content could not be increased further. Neutron diffraction experiments have shown that hydrogen occupies octahedral sites in both the hcp \cite{Fedotov:1999} and fcc \cite{Antonov:2005} structures. Despite the fact that higher hydrides of cobalt have not been synthesized in the solid state, the molecular hydrides CoH, CoH$_2$, and CoH$_3$ have been created in an electric field \cite{Bialek:2001}, and studied via unrestricted calculations using the B3LYP functional \cite{Uribe:2010}. Moreover, CSP studies have found that $Fm\bar{3}m$-CoH$_2$, $I4/mmm$-CoH$_2$, and $Pm\bar{3}m$-CoH$_3$ (the latter two structures are isotypic with the FeH$_2$ and FeH$_3$ structures shown in Fig.\ \ref{fig:iron}(b,c)) are stable between 10-42~GPa, 42-300~GPa, and 30-300~GPa, respectively \cite{Wang:2018a}. Even though these ionic solids were metallic, calculations did not reveal superconductivity up to 200~GPa.

Rhodium assumes the fcc structure, with one octahedral and two tetrahedral sites per metal atom. In the monohydride, hydrogen atoms occupy the octahedral sites. Calculations have shown that RhH in the NaCl structure type has a magnetic moment of 0.45~$\mu_B$ at 1~atm \cite{Cui:2009}, and predicted that it would undergo the following structural phase transitions: NaCl$\rightarrow$ zincblende $\rightarrow$CsCl$\rightarrow$NiAs, at 11, 154 and 382~GPa \cite{Sudhapriyanga:2014}. First principles calculations suggested that under mild pressures an RhH$_2$ stoichiometry, which is isotypic with the TiH$_2$ structure shown in Fig.\ \ref{fig:group4}, wherein each tetrahedral site was filled with a hydrogen atom, and each octahedral site was vacant, would be the thermodynamically preferred hydride of rhodium \cite{Li:2011a}. In this same study experiments were carried out in a DAC up to 19~GPa revealing RhH formation above 4~GPa (in agreement with previous experiments \cite{Tkacz:1998}), and RhH$_2$ formation above 8~GPa. Upon decompression dehydrogenation occurred, yielding the pure metal by 3~GPa at room temperature. At low temperatures the dihydride could be quenched to atmospheric pressures. This was the first dihydride of the platinum group metals to be synthesized.

Until 2013, no binary hydride of iridium was known. Experiments up to 125~GPa wherein the metal was compressed in a DAC in a hydrogen medium revealed the formation of a new phase at 55~GPa \cite{Scheler:2013c}. The difference in volume between this phase and that of the pure metal suggested that a hydride with the IrH$_3$ stoichiometry had formed, and XRD patterns showed the metal lattice assumed the same $Pm\bar{3}m$ structure as FeH$_3$ (see Fig.\ \ref{fig:iron}(c)). On decompression, decomposition of the trihydride began at 15-20~GPa, and it was not fully complete by 6~GPa. First-principles calculations verified that the synthesized phase is most likely a trihydride, and the thermodynamic and dynamic stability, as a function of pressure of various structural candidates was explored. The experiments did not support the existence of a $Pnma$ structure that was found to have the lowest enthalpy above 68~GPa, and it was hypothesized that the formation of this phase was hindered by a kinetic barrier. At the same time Zaleski-Ejgierd carried out an independent theoretical study of the hydrides of iridium under pressure \cite{Zaleski:2014}. Various CSP techniques were used to predict the most stable structures for a wide range of hydrogen content at 25 and 125~GPa. At 50~GPa a dynamically stable $P6_3mc$-IrH$_3$ phase, which was semiconducting at 25~GPa, was the lowest point on the convex hull. At low pressures this phase was comprised of molecular IrH$_3$ units. A metallic IrH$_2$ phase had the most negative $\Delta H_F$ at 100-125~GPa. The experimentally observed $Pm\bar{3}m$-IrH$_3$ phase was also found via CSP, but its enthalpy was higher than that of $P6_3mc$-IrH$_3$ throughout the pressure range studied. Interestingly, the $Pnma$-IrH$_3$ phase predicted by Scheler and co-workers \cite{Scheler:2013c} had nearly the same structure as $P6_3mc$-IrH$_3$ found by Zaleski-Ejgierd.

\subsection{Group 10: Nickel, Palladium, Platinum} \label{sec:group10}

Hydrogenation of FM Ni to NiH$_x$ occurs at $\sim$0.8~GPa, and an fcc structure is formed \cite{Ishimatsu:2012}. As $x$ increases the magnetization drops, with the formation of the paramagnetic (PM) phase occurring by $x\sim0.6$ \cite{Bauer:1961a}. The disappearance of the magnetic state upon hydrogenation has been observed in Linear Muffin Tin Orbital (LMTO) \cite{Vargas:1987},  full-potential linear augmented plane-wave (FLAPW) \cite{Ishimatsu:2012}, and pseudopotential plane-wave calculations \cite{San:2006}. The latter suggest that the most stable NiH$_x$ structures up to 210~GPa have a metal fcc lattice with the hydrogen atoms filling the octahedral sites \cite{San:2006}. Moreover, the concentration at which the FM to PM transition is calculated to occur decreases under pressure from NiH$_{0.375}$ at 210~GPa to NiH$_{0.6875}$ at 4~GPa. 
\begin{figure}[h!]
\begin{center}
\includegraphics[width=0.45\columnwidth]{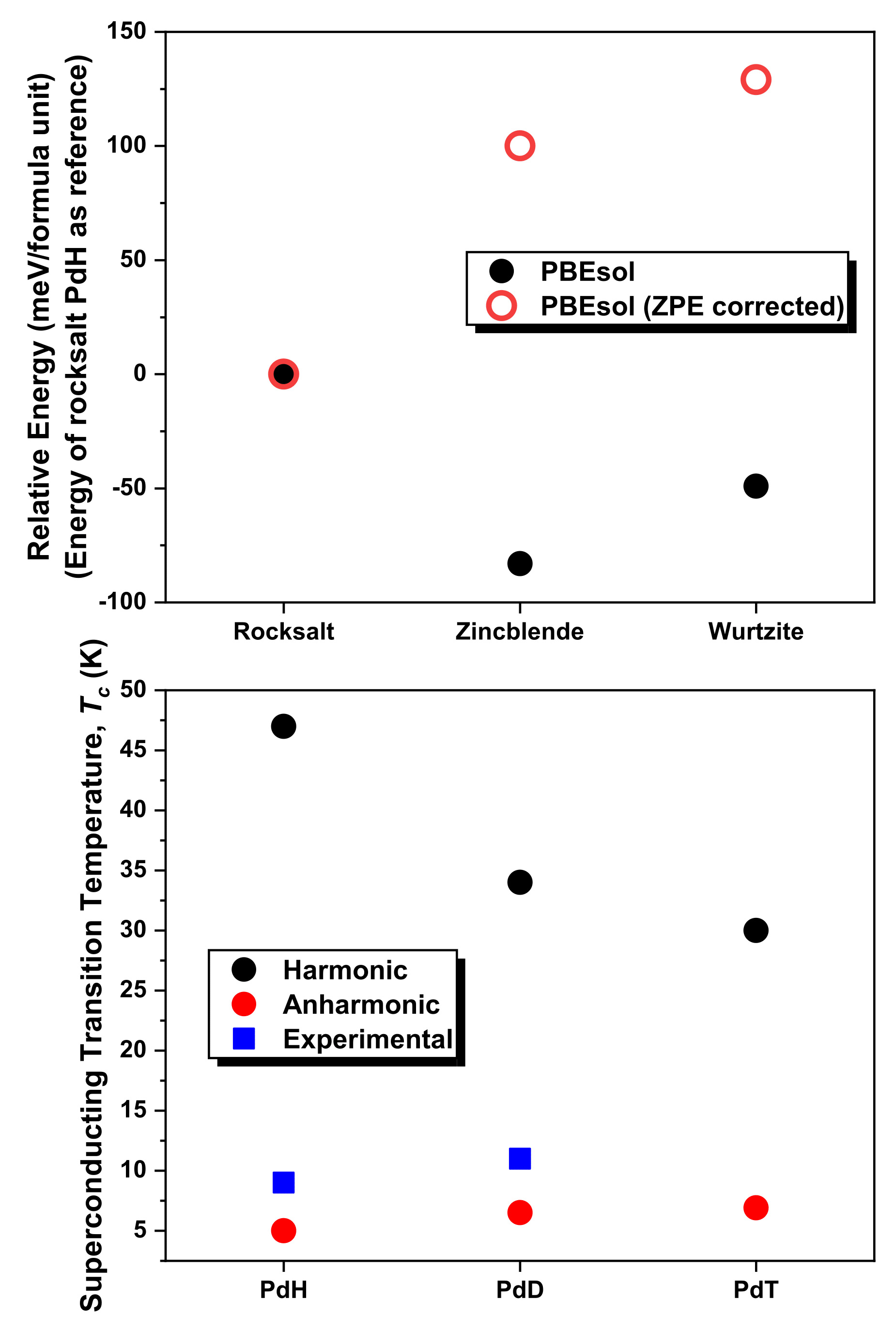}
\end{center}
\caption{(top) The relative energies of different PdH structures at 1~atm with (open circles) and without (closed circles) the ZPE corrections \cite{Houari:2014}. (bottom) The experimental superconducting temperature of PdH and its isotopes at 1~atm\cite{Stritzker:1972}, along with theoretically computed values \cite{Errea:2013}.}
\label{fig:pdh}
\end{figure}

Palladium hydride, one of the first transition metal hydrides to be synthesized, was first made over 150 years ago \cite{Graham:1866}. At atmospheric pressure the metal lattice adopts an fcc structure. DFT calculations on PdH$_x$ with $x\le 1$ that neglected the ZPE showed that hydrogen prefers to occupy the tetrahedral sites as compared to the octahedral ones when the hydrogen content is large \cite{Caputo:2003,Houari:2014}. This contradicts the results of most experimental observations, which are indicative of octahedral site filling \cite{Worsham:1957}. Inclusion of the ZPE was found to have a dramatic impact on the relative enthalpies of the zincblende, wurtzite and rock-salt structures, stabilizing the latter \cite{Houari:2014}, as shown in Fig.\ \ref{fig:pdh}. The Pd-H distances are $\sim$0.25~\AA{} shorter in the NaCl structure, which leads to lower frequency vibrations that affect the total ZPE.  At ambient conditions PdH is superconducting with a $T_c$ of 8-9~K \cite{Schirber:1974}. When hydrogen is replaced with deuterium the $T_c$ increases to 10-11~K, the so-called ``inverse isotope effect'' \cite{Schirber:1974,Stritzker:1972}. The $T_c$ of both phases is reduced under pressure \cite{Skoskiewicz:1974a,Hemmes:1989a}. A tremendous amount of research has been devoted to uncovering the origin of the inverse isotope effect, and recently state-of-the-art first-principles calculations have illustrated that this phenomenon originates from the large anharmonicity in the phonon modes \cite{Errea:2013}, see Fig.\ \ref{fig:pdh}. The superconductivity was shown to be phonon mediated, but neglecting the anharmonic motion leads to a large overestimation of the $T_c$. At pressures of 5~GPa or less phase segregation into PdH and Pd$_3$H$_4$, resulting from the removal of Pd atoms, occurs \cite{Fukai:1994}.

\begin{figure}[ht!]
\begin{center}
\includegraphics[width=0.6\columnwidth]{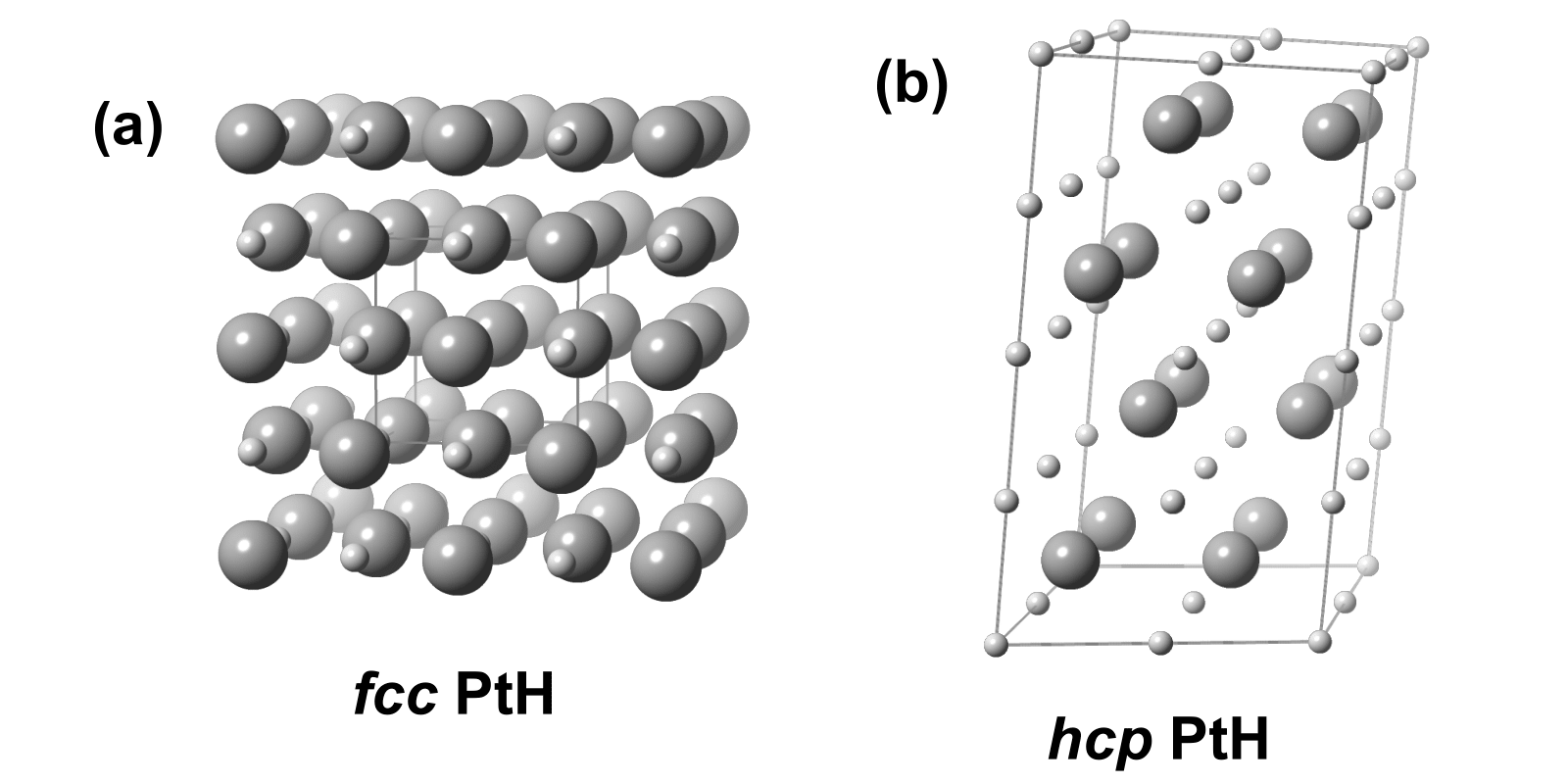}
\end{center}
\caption{Two PtH phases predicted to be stable under pressure using CSP \cite{Zhou:2011a,Kim:2011a,Scheler:2011a}. These fcc (NaCl structure with $Fm\bar{3}m$ symmetry) and hcp ($P6_3/mmc$ symmetry) structures have been predicted or observed in many transition metal monohydrides. Calculations confirmed that both fcc and hcp PtH were superconducting. It has been proposed that one of these phases formed when the platinum comprising the electrodes employed in experiment reacted with the H$_2$ that was liberated from the decomposition of compressed SiH$_4$, and the PtH phase was responsible for the superconductivity observed in Ref.\ \cite{Eremets:2008a}.}
\label{fig:pth}
\end{figure}

The theoretical suggestion that compressed silane may be superconducting at pressures lower than those required to metallize hydrogen \cite{Ashcroft:2004a,Feng:2006a,Pickard:2006a} inspired a number of experiments. One of these concluded that metalization of SiH$_4$ occurs at 50~GPa, and measured a  $T_c<$17~K at 96~GPa and 120~GPa  \cite{Eremets:2008a}. However, discrepancies between theory and experiments (see Sec.\ \ref{sec:tetragen} for further details) coupled with the realization that the platinum electrodes employed to measure the conductivity in the sample may react with hydrogen released from the pressure induced decomposition of SiH$_4$, led to the suggestion that a superconducting hydride of platinum formed under pressure instead \cite{Degtyareva:2009a}. CSP investigations showed that the PtH stoichiometry, which was found to be thermodynamically favored over the elemental phases between 3~GPa \cite{Gao:2012} to 20~GPa \cite{Kim:2011a}, was the most stable point on the Pt/H phase diagram at $\sim$100~GPa \cite{Zhou:2011a}. The two phases that were nearly isoenthalpic at 100~GPa, hcp \cite{Zhou:2011a,Kim:2011a}, and fcc \cite{Kim:2011a} PtH, shown in Fig.\ \ref{fig:pth}, were found to be superconducting. The highest $T_c$ occurred at the onset of dynamic stability, with  fcc-PtH having a slightly larger value than hcp-PtH \cite{Kim:2011a}. The $T_c$ of both phases decreased with the application of pressure \cite{Kim:2011a,Zhou:2011a,Zhang:2011}, approaching 0~K by $\sim$200~GPa. The XRD pattern calculated for hcp-PtH matched well with the experimental results for `silane' \cite{Zhou:2011a}, as did the computed $T_c$ values \cite{Kim:2011a,Zhou:2011a,Scheler:2011a}. Subsequent experiments confirmed the room-temperature synthesis of PtH above 27~GPa, and the formation of an hcp structure above 42~GPa \cite{Scheler:2011a}. However, first principles computations that took anharmonic effects into account found the $T_c$ of hcp-PtH to be $<1$~K at 100~GPa \cite{Errea:2014}. This strong suppression in $\lambda$ and in $T_c$ resulting from anharmonicity led the authors to question whether or not the superconductivity observed in experiment did in fact originate from PtH. 

\subsection{Group 11: Copper, Silver, Gold} 

The first synthesis of copper hydride was by Wurtz in 1844 \cite{Wurtz:1844}, and it is the only coinage metal hydride that has been prepared to date. The protocol yielded a hcp crystal with the CuH stoichiometry that possessed the $P6_3mc$ spacegroup, i.e.\ the wurtzite structure type \cite{Muller:1926,Goedkoop:1955}. This phase is stable below -5$^\circ$C, but at room temperature it decomposes below 8.5~GPa \cite{Tkacz:2004}. 
The synthesis of CuH  above 14.4~GPa, on the other hand, yielded a CuH$_{0.4}$ stoichiometry where the metal lattice is fcc and the hydrogen atoms reside in the interstitial regions ($\gamma$-CuH) \cite{Burtovyy:2004}. A third phase, $\epsilon$-Cu$_2$H, that possesses an anti-CdI$_2$ structure ($P\bar{3}m1$ spacegroup) has been made in a DAC between 18.6-51~GPa  \cite{Donnerer:2013}. The hydrogen atoms within this phase were found to be arranged in layers, instead of filling the void sites randomly. It is not clear why the method of preparation affects which hydride of copper is formed.

Experiments wherein silver and gold were compressed in a hydrogen medium up to 87~GPa and 113~GPa, respectively, did not yield hydrides of these metals at, or above, room temperature \cite{Donnerer:2013}. In another study, AuH was reportedly synthesized by annealing pure gold in a hydrogen atmosphere at $\sim$5~GPa and $\sim$400$^\circ$C \cite{Antonov:1982}. However, its structure is unknown, and the experiment has not been confirmed since. Based on the diffraction pattern, it was suggested that the structure may be related to the $tI2$ phase of mercury \cite{Degtyareva:2015}. Theoretical work has therefore been undertaken to determine if hydrides of the heavier coinage metals could be synthesized. DFT calculations on AgH phases assuming structures that were predicted to be metastable or stable for PtH and RuH wherein different tetrahedral and octahedral sites were populated with hydrogen atoms suggested that AgH would become stable above 180~GPa \cite{Gao:2012}. The most stable structure was fcc with hydrogen occupying the octahedral sites. At 100~GPa AgH was calculated to be a poor metal, suggesting that it is unlikely to be a high temperature superconductor. Another investigation found that dynamic stability within fcc-AgH was achieved by 50~GPa, and at this pressure the system was a semi-conductor precluding it from superconductivity \cite{Kim:2011a}. The theoretical calculations carried out so far do not support the stability of the monohydride of gold at the low pressures employed in Ref.\ \cite{Antonov:1982}. For example, AuH in the fcc structure was computed to become dynamically stable above 220~GPa \cite{Kim:2011a}. And, in a subsequent DFT study none of the structure types that were considered for AuH were found to be thermodynamically stable with respect to the elemental phases at pressures attainable in a DAC \cite{Gao:2012}.

\subsection{Group 12: Zinc, Cadmium, Mercury}

At ambient conditions in the solid state the group 12 metals form dihydride phases. It is believed that HgH$_2$ is a covalent molecular solid \cite{Wang:2005a}, whereas ZnH$_2$ and CdH$_2$ contain hydrogen bridges between the metal atoms. Because solid HgH$_2$ and CdH$_2$ decompose to the elemental phases at low temperatures, they have not been intensely investigated  \cite{Shayesteh:2005}. At the time of writing this review, we were unable to locate any theoretical or experimental studies that had examined the structures and superconducting properties of the hydrides of zinc, cadmium or mercury as a function of pressure.

\section{Group 13: Icosagen Hydrides}

\begin{figure}[h]
\begin{center}
\includegraphics[width=0.6\columnwidth]{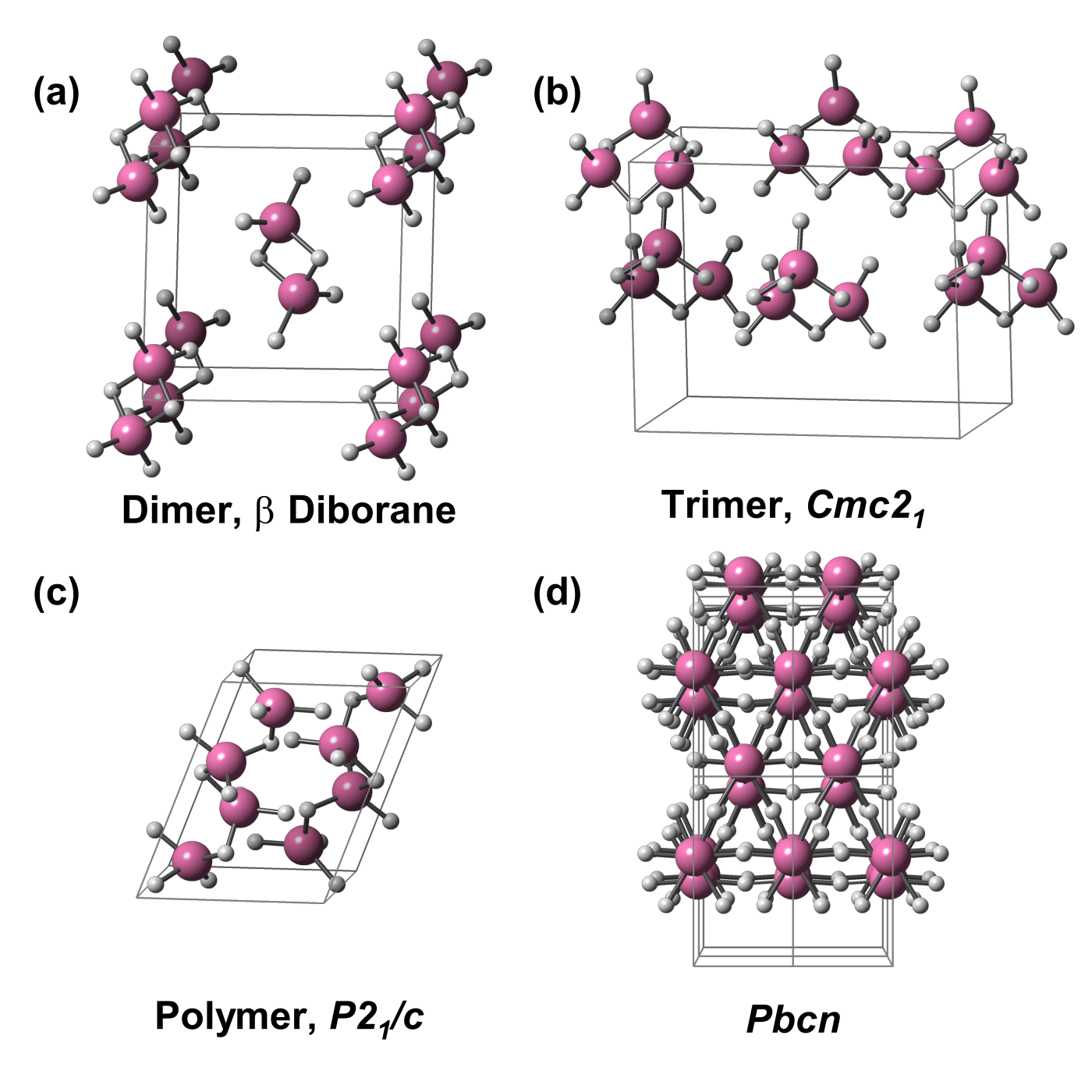}
\end{center}
\caption{(a) The $\beta$-diborane phase that forms above 90~K at atmospheric pressures. Predicted high pressure phases of BH$_3$ including: (b) one that contains B$_3$H$_9$ trimers \cite{Yao:2011a}, (c) one that consists of linear (BH$_3$)$_n$ polymeric chains \cite{Yao:2011a}, and (d) a $Pbcn$ phase whose $T_c$ was estimated as being 125~K at 360~GPa \cite{Abe:2011a}.}
\label{fig:borane}
\end{figure}

\textbf{Boron}\\
Even though the chemistry of boron hydride clusters has been actively researched \cite{Greenwood:1992}, not as much effort has been placed into studying these systems in the solid state and under pressure. One exception is diborane, B$_2$H$_6$. 
At ambient conditions molecular diborane assumes a geometry wherein two hydrogens each bridge two boron atoms, and it is metastable towards hydrogen loss. Crystalline BH$_3$ is unknown, however solid diborane adopts the $\alpha$ phase below 60~K, and annealing above 90~K yields the $\beta$ phase shown in Fig.\ \ref{fig:borane}(a). 
Raman spectroscopy has been employed to study diborane up to 24~GPa \cite{Murli:2009a}. At 4~GPa the system underwent a liquid-solid transition to phase I, followed by a transformation to phase II at 6~GPa, and phase III at 14~GPa. The phase transitions were reversible upon decompression. Infrared (IR) spectroscopy studies up to 50~GPa  provided further evidence for the phase transitions observed via Raman \cite{Song:2009-B}. The IR measurements suggested that the B$_2$H$_6$ molecule remains intact within these phases. Spectroscopic studies also provided evidence for further structural transitions at 42~GPa and 57~GPa \cite{Torabi:2015-B}. Another boron hydride whose high pressure behavior has been scrutinized experimentally is decaborane, B$_{10}$H$_{14}$ \cite{Nakano:2002-B}. It's Raman spectrum did not show any dramatic changes up to 50~GPa. Above this pressure the sample changed color from transparent yellow to orange/red, and Raman spectroscopy suggested that the backbone of the molecule had been perturbed. The sample became black above 100~GPa implying that a transition into a non-molecular phase, which was shown to be semiconducting, had occurred. 

Early theoretical studies concluded that molecular boranes become thermodynamically unstable towards systems comprised of extended bonded networks by $\sim$100-300~GPa, and that a BH$_3$ analogue of AlH$_3$ would become metallic below 30~GPa \cite{Barbee:1997-B}. Twenty years later DFT calculations \cite{Torabi:2013-B,Torabi:2015-B} were carried out to help characterize the phases investigated in Refs.\ \cite{Murli:2009a,Song:2009-B}. The computed IR and Raman spectra of ten candidate phases that contained the molecular diborane unit were  compared with those obtained experimentally \cite{Torabi:2013-B}. This study showed that phase I corresponds to the $\beta$-diborane structure, and the best candidates for phases II and III possessed a $P2_1/c$ symmetry lattice, but with different molecular orientations. Importantly, all of the experimental data was consistent with phases containing B$_2$H$_6$ units, suggesting that transformation to the thermodynamically preferred products, cyclic oligomers and polymer chains, is kinetically hindered. Further theoretical studies concluded that the phases observed at 42~GPa and 57~GPa possessed $P1$ symmetry and were comprised of B$_2$H$_6$ molecules \cite{Torabi:2015-B}. Moreover, a geometry optimization of the proposed phase IV structure showed that near 110~GPa the molecular diborane motifs polymerize forming a phase containing  one-dimensional zig-zag chains of boron atoms. This phase was found to become metallic near 138~GPa within hybrid DFT. Another theoretical investigation carried out at about the same time showed that crystals based upon B$_3$H$_9$ trimers become stabilized with respect to $\beta$-diborane between 4-36~GPa, and at higher pressures linear polymers become preferred \cite{Yao:2011a}. Two of the proposed structures are illustrated in Fig.\ \ref{fig:borane}(b) and Fig.\ \ref{fig:borane}(c). The kinetic barrier for trimer formation was estimated to be large, and it was therefore concluded that it is unlikely that such phases were made in Refs.\ \cite{Murli:2009a,Song:2009-B}. The calculations also showed that the structural interconversion between $\beta$-B$_2$H$_6$ and a $P2_1/c$ symmetry polymeric phase was likely to be facile, suggesting that the latter could be a candidate for the experimentally observed phase III. None of the phases studied were found to be metallic. Another computational study showed that phases with the B$_2$H$_6$ stoichiometry become thermodynamically stable with respect to segregation into the elemental structures above 350~GPa \cite{Abe:2011a}. Two metallic systems, one with $Pbcn$ and the other with $Cmcm$ symmetry were predicted, however the latter was not dynamically stable within the harmonic approximation. At 360~GPa the $T_c$ of the $Pbcn$ phase shown in Fig.\ \ref{fig:borane}(d) was estimated as being 125~K. 

Finally, CSP techniques have been employed to predict the most stable structures with the BH and BH$_3$ stoichiometries up to 300~GPa \cite{Hu:2013-B}, and the B$_4$H$_{10}$, B$_4$H$_8$ and B$_4$H$_6$ stoichiometries between 50-300~GPa \cite{Suarez:2014-B}. In Ref.\ \cite{Hu:2013-B} the BH stoichiometry was found to be stable above 50~GPa, and it was the only boron hydride phase thermodynamically favorable above 153~GPa. Below 175~GPa the preferred BH phases were composed of hydrogen-terminated puckered boron sheets, and a three-dimensional $P6/mmm$ symmetry structure that was metallic with an estimated $T_c$ of 14-21~K at 175~GPa was stable at higher pressures.
Ref.\ \cite{Suarez:2014-B} found that the lowest enthalpy B$_4$H$_{10}$ stoichiometry phases at 50 and 150~GPa consisted of hydrogen-capped boron layers separated by H$_2$ units, suggesting that segregation into the elemental phases is preferred at these pressures. At 300~GPa a three-dimensional network was predicted to be the most stable. With few exceptions, the structures predicted for B$_4$H$_8$ and B$_4$H$_6$ consisted of molecular or polymeric units that did not undergo phase segregation. The reaction $\text{B}_4\text{H}_{10} \rightarrow \text{B}_4\text{H}_{8}+\text{H}_2$ was found to be exothermic at the pressures considered.\\

\noindent\textbf{Aluminum}\\
In the gas phase a number of molecular hydrides of aluminum, including AlH$_n$ ($n=1-3$), Al$_2$H$_4$ and Al$_2$H$_6$ have been formed via laser ablation \cite{Wang:2003-Al}. In the solid state, however, only the AlH$_3$ compound is known, and at atmospheric conditions it is metastable releasing H$_2$ molecules when heated \cite{Sinke:1967-Al}. Solid AlH$_3$ can adopt one of four different modifications depending upon the method of synthesis. According to DFT calculations at 0~K the $\beta$ phase has the lowest energy at atmospheric pressures, and it transitions to the $\alpha^\prime$ and $\alpha$ structures at 2.4 and 4.3~GPa \cite{Vajeeston:2008-Al}. A number of theoretical \cite{Lu:2012-Al,Vajeeston:2008-Al,Feng:2014-Al,Graetz:2006-Al} and experimental \cite{Tkacz:2008-Al,Drozd:2012-Al,Shimura:2010-Al,Graetz:2006-Al,Molodets:2009-Al,Besedin:2011-Al} studies have focused on the structural transitions and properties of AlH$_3$ under pressure. 

\begin{figure}[h!]
\begin{center}
\includegraphics[width=0.6\columnwidth]{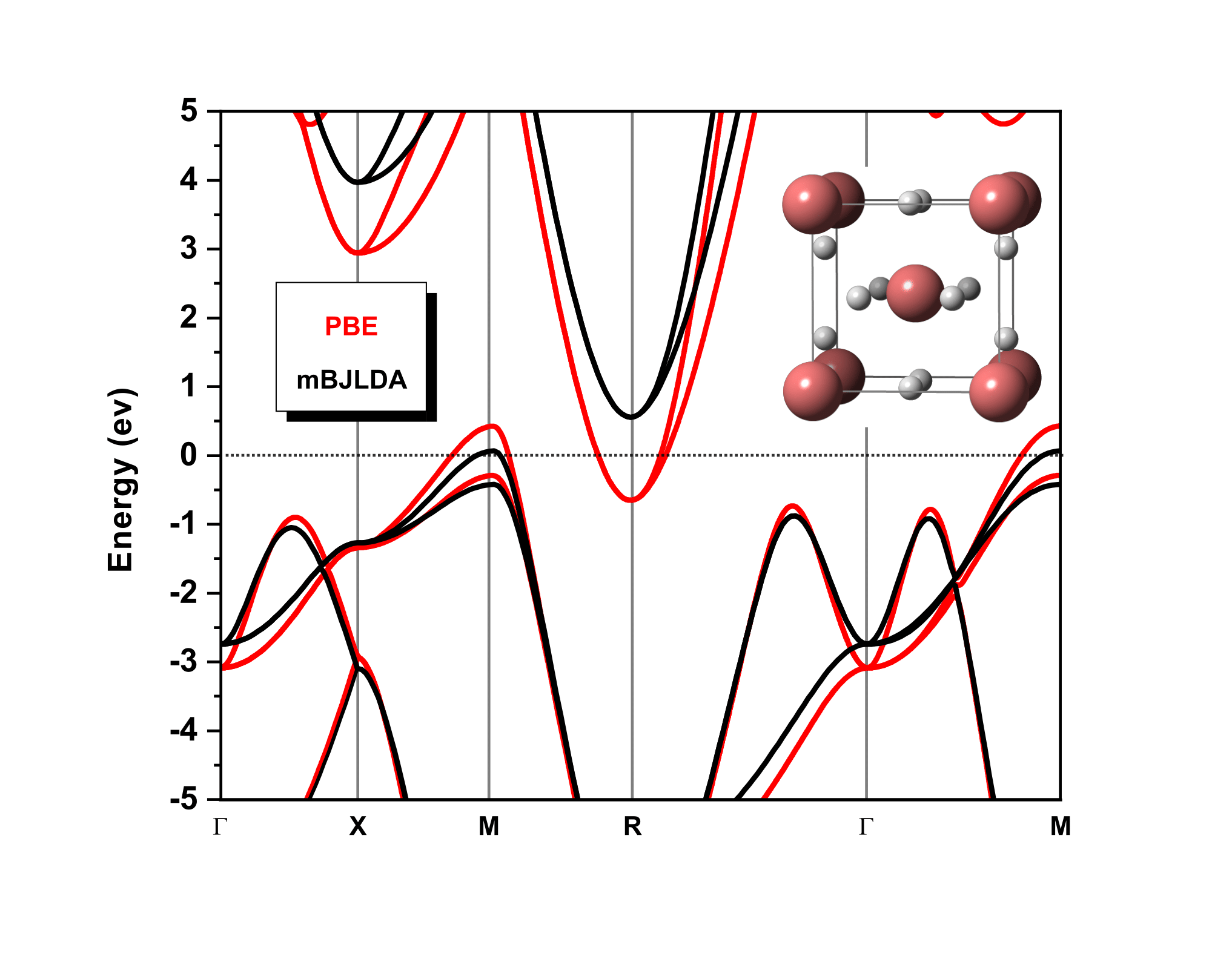}
\end{center}
\caption{The band structure of the $Pm\bar{3}n$-AlH$_3$ phase shown in the inset as calculated with the PBE and TB-mBJLDA functionals at 160~GPa \cite{Shi:2012-Al}.}
\label{fig:aluminum}
\end{figure}

At 1~atm AlH$_3$ has a large band gap, like other ionic solids. However, because this high hydrogen content material should become metallic via pressure induced band broadening, it was suggested that AlH$_3$ might become superconducting when squeezed. \emph{Ab~Initio} random structure searches predicted that the $\alpha$ phase would transition to an insulating layered $Pnma$ structure at 34~GPa \cite{Pickard:2007b}. At 73~GPa a transformation to the semi-metallic $Pm\bar{3}n$ symmetry structure shown in the inset of Fig.\ \ref{fig:aluminum} was found. Both of these phases were stable with respect to dehydrogenation under pressure. A later theoretical study showed that $Pm\bar{3}n$-AlH$_3$ is dynamically stable between 72-106~GPa at 0~K, as well as at 1~atm and $\sim$470~K \cite{Kim:2008-Al}. Within PBE the metallicity of $Pm\bar{3}n$-AlH$_3$ was found to arise from the conduction and valence bands crossing the Fermi level at the $R$ and $M$ points, respectively, as shown by the red bands in Fig.\ \ref{fig:aluminum}. The DOS at $E_F$ decreased with increasing pressure, and GW calculations showed that this phase becomes insulating by 200~GPa because the center of the Al $3s$ band and the H $1s$ band become increasingly separated with pressure \cite{Geshi:2013-Al}. At 72~GPa the $T_c$ of $Pm\bar{3}n$-AlH$_3$ was estimated as being 11~K via the modified McMillan equation, and $T_c$ decreased with increasing pressure, approaching zero by 165~GPa \cite{Wei:2013-Al}. 

The aforementioned predictions inspired a combined theoretical/experimental study \cite{Goncharenko:2008-Al}. $\alpha$-AlH$_3$, which adopts the $R\bar{3}c$ structure, was found to undergo a structural transition to an unknown phase at 63~GPa. At 100~GPa a transition to a phase that could be indexed to either the $Im\bar{3}m$ or the $Pm\bar{3}n$ spacegroup, which differ only in the positions of the hydrogen atoms, occurred. Resistance measurements were indicative of an insulator to metal transition, but superconductivity was not observed up to 165~GPa at temperatures as low as 4~K. DFT calculations showed that the $Pm\bar{3}n$ structure is enthalpically preferred, and this metallic phase was calculated to have a $T_c$ of 24~K at 110~GPa and 6~K at 165~GPa. However, the reason for the discrepancy in the superconducting properties between experiment and theory could not be found. A later computational study revealed that a large fraction of the electron-phonon coupling in this phase arises from modes that are highly anharmonic, and when the anharmonicity was taken into account the electron-phonon coupling, and therefore $T_c$, was greatly diminished \cite{Rousseau:2010-Al,Rousseau:2011-Al}. Another theoretical study illustrated that the band structure near the Fermi level, and in particular bands that give rise to nested pieces of Fermi surface that are integral for the electron-phonon coupling, depend upon the computational method employed \cite{Shi:2012-Al}. In contrast to calculations performed with the PBE functional, both the TB-mBJLDA functional and GW calculations predicted that at 160~GPa AlH$_3$ should be a small gap semi-conductor, as shown by the black bands in Fig.\ \ref{fig:aluminum}. Thus, the disagreement between experiment and theory might be due to a number of factors that were not considered in the original calculations, such as the anharmonicity of the phonon modes or the way in which exact exchange influences the band structure around the Fermi level.

CSP techniques have been employed to predict the structures adopted by AlH$_n$ ($n=5,7,9$) up to 300~GPa \cite{Hou:2015-Al}. The most stable phase below 73~GPa, $P1$-(AlH$_3$)H$_2$, could be thought of as a vdW compound. Pressure was found to induce a transformation to a semiconducting $P1$ symmetry phase, which had the lowest enthalpy of any AlH$_5$ configuration examined below 250~GPa. Above this pressure a metallic $P2_1/m$-AlH$_5$ structure became preferred, and its large electron-phonon coupling resulted in a $T_c$ of 132-146~K. \\

\noindent\textbf{Gallium}\\
A number of molecular hydrides of gallium including GaH, GaH$_2$, GaH$_3$, and Ga$_2$H$_2$ \cite{Himmel:2002-13}, have been prepared in solid noble gas matrices. In addition, wet chemical methods have been employed to synthesize the bridge-bonded molecule digallane, H$_2$Ga($\mu$-H)$_2$GaH$_2$ \cite{Pulham:1991}. This highly reactive compound \cite{Downs:1994-Ga} condenses as a white solid, which may have a polymeric structure, and it decomposes into the elements above 243~K. On the theoretical front, computations have been undertaken to predict the most stable solid state structures of GaH$_3$ between 5-300~GPa \cite{Gao:2011a}. At low pressures the preferred systems contained H$_2$ units, suggesting that they are prone towards decomposition. Above 160~GPa a $Pm\bar{3}n$ phase with monoatomic hydrogen atoms, which is isotypic with the high pressure form of AlH$_3$ shown in the inset of Fig.\ \ref{fig:aluminum}, was computed to become stable with respect to the elemental phases. The $T_c$ of this ionic solid was estimated as being 76-83~K at 160~GPa via the Allen-Dynes modified McMillan equation, and it was found to decrease with increasing pressure. At 120~GPa the $T_c$ of $Pm\bar{3}n$-GaH$_3$ calculated by solving the Eliashberg equations ranged from 90-123~K \cite{Szczesniak:2014}.\\

\noindent\textbf{Indium and Thallium}\\
Despite claims whose origins date back over 60 years, it is unlikely that the trihydrides of indium and thallium were ever successfully synthesized \cite{Downs:1994}. In fact, some studies suggest that InH$_3$ and TlH$_3$ are not stable enough to be isolated in the solid state at ambient pressure and temperature \cite{Hunt:1996-13,Andrews:2004-In}. However, molecular hydrides of these elements have been synthesized in the gas phase at cryogenic temperatures via laser ablation in inert matrices. These include In$_2$H$_2$, InH$_3$,  In$_2$H$_4$, In$_2$H$_6$ \cite{Himmel:2002-13,Andrews:2004-In,Wang:2004-In}, TlH, TlH$_2$, TlH$_3$, Tl$_2$H$_2$ and TlTlH$_2$ \cite{Wang:2004-Tl}. CSP techniques have therefore been used to determine if hydrides of indium could become stable under pressure \cite{Liu:2015a}. InH$_{3}$ and InH$_{5}$ became enthalpically preferred over the elemental phases by 200~GPa. They contained H$_2$ or linear H$_3$ units, wherein charge was donated from the indium to the hydrogen atoms, and at 200~GPa their $T_c$ values were estimated as being 34-40~K and 22-27~K, respectively. With increasing pressure, the $T_c$ of $R\bar{3}$-InH$_3$ decreased slightly. To the best of our knowledge CSP studies on TlH$_n$ have not yet been carried out. But, based upon the trends observed for the icosagen hydrides we suspect that the hydrides of thallium become stable at pressures larger than 200~GPa.

%%%%%%%%%%%%%%%%%%%%%%%%%%%%%%%%%%%%%%%%%%%%%%%%%%%%%%%%%%%%%%%%%%%%%%%%%%%%%%%%%%%%%%%%%%%%%%%%%%%%%%%%%%%%%%%%%%%%%%%%%%%%
%%%%%%%%%%%%%%%%%%%%%%%%%%                       Tetragens                             %%%%%%%%%%%%%%%%%%%%%%%%%%%%%%%%%%%%%
%%%%%%%%%%%%%%%%%%%%%%%%%%%%%%%%%%%%%%%%%%%%%%%%%%%%%%%%%%%%%%%%%%%%%%%%%%%%%%%%%%%%%%%%%%%%%%%%%%%%%%%%%%%%%%%%%%%%%%%%%%%%

\section{Group 14: Tetragen Hydrides} \label{sec:tetragen}

\noindent\textbf{Carbon}\\ 
It is beyond the scope of this review to discuss all of the  work that has been carried out on compressed solids containing hydrocarbon based molecules. Instead, we briefly describe the high pressure behavior of methane and methane/hydrogen mixtures, which have been intensely studied because of their relevance in planetary sciences. Methane has a rich phase diagram \cite{Chen:2011-C}, and various binary molecular compounds with the general formula (CH$_4$)$_n$(H$_2$)$_m$ ($n=1,2$ and $m=1,2,4$) have been characterized spectroscopically up to 30~GPa \cite{Somayazulu:1996}. A wide variety  of CSP techniques have been used to predict the phases methane adopts under pressure, but none of them found any stable metallic structures up to pressures as high as 550~GPa \cite{Martinez:2006-C,Gao:2010-C,Juan:2010-C,Lin:2011-C}. CSP investigations have also suggested that both methane \cite{Gao:2010-C,Liu:2016-C} (95-200~GPa) and a 1:1 mixture of CH$_4$ and H$_2$ \cite{Liu:2014-C} ($P >$230~GPa) become thermodynamically unstable towards decomposition into other hydrocarbon based phases by the pressures given in the parentheses, even though they may remain dynamically stable. The lowest enthalpy systems were found to be large band gap insulators to at least 150~GPa \cite{Liu:2014-C}. These theoretical studies suggest that it is unlikely that a methane-based hydride could exhibit superconductivity at pressures that are currently accessible via static compression. \\

\begin{figure}[h!]
\begin{center}
\includegraphics[width=0.5\columnwidth]{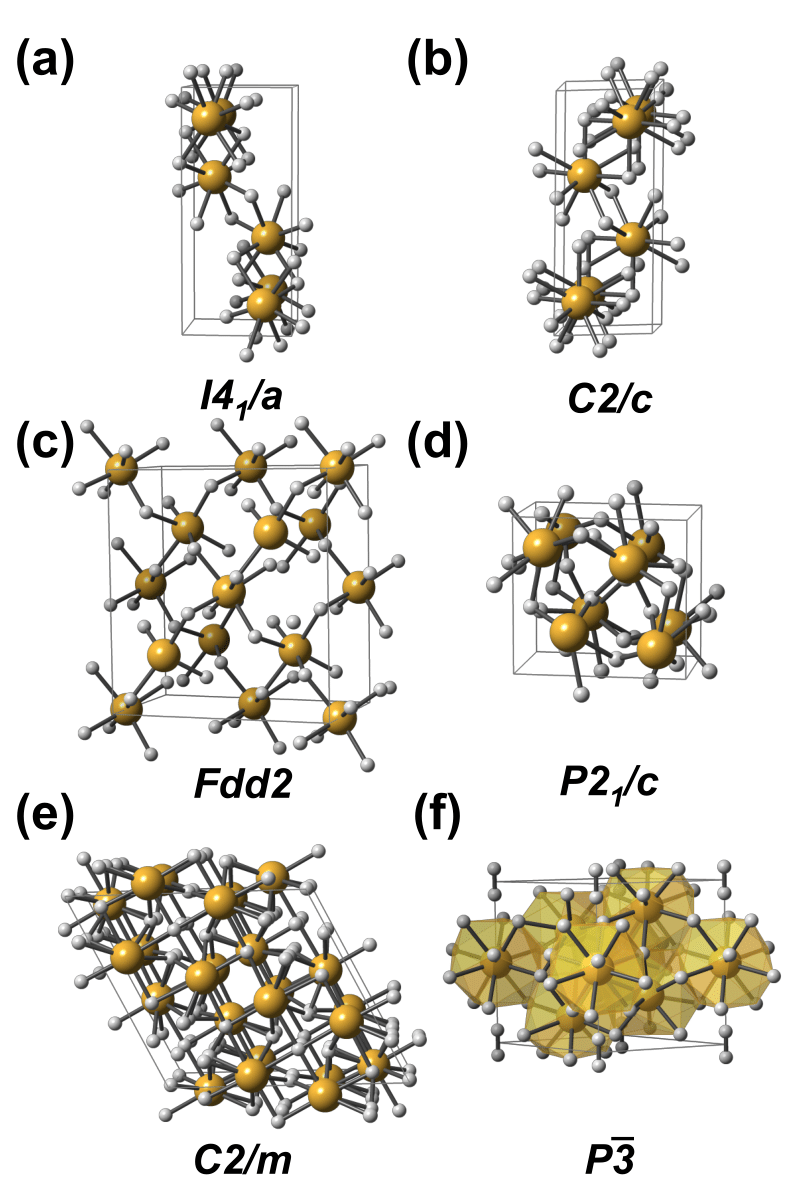}
\end{center}
\caption{Various SiH$_4$ phases that have been investigated under pressure: (a) an $I4_1/a$ symmetry phase that was theoretically predicted  \cite{Pickard:2006a} and experimentally synthesized \cite{Hanfland:2011-Si}, (b) a theoretically predicted $C2/c$ phase \cite{Pickard:2006a}, (c) a theoretically predicted $Fdd2$ phase \cite{Kurzydlowski:2011a}, which is a likely candidate for an experimentally observed polymeric phase, (d) a theoretically predicted $P2_1/c$ phase with $T_c=$~32~K at 400~GPa \cite{Zhang:2015-Si}, (e) a theoretically predicted $C2/m$ phase with $T_c=$106~K at 610~GPa \cite{Zhang:2015-Si}, (f) a theoretically predicted $P\bar{3}$ phase with $T_c=$~32~K at 300~GPa \cite{Cui:2015-Si}.}
\label{fig:silane}
\end{figure}

\noindent\textbf{Silicon} \\
In 2004 Neil Ashcroft proposed that the same attributes that would render metallic hydrogen a high temperature superconductor would be applicable to hydrogen dominant alloys, and in particular those containing a group 14 element such as silicon \cite{Ashcroft:2004a,Ashcroft:2004b}. He also predicted that the group 14 hydrides would become metallic at pressures lower than those required to metallize elemental hydrogen because of ``chemical precompression''. These predictions inspired numerous studies of the hydrides of silicon under pressure. In the first study, carried out in 2006 by Feng et al., DFT calculations were performed on 13 candidate SiH$_4$ structures \cite{Feng:2006a}. The most stable phase underwent pressure induced band gap closure just under 100~GPa, and it was suggested that it might be a high temperature superconductor. A subsequent study by Pickard and Needs, which used random searching instead, found a $I4_1/a$ phase between 50-263~GPa and a $C2/c$ phase at higher pressures -- these phases were more stable than the structures considered by Feng and co-workers \cite{Pickard:2006a}. $I4_1/a$-SiH$_4$, shown in Fig.\ \ref{fig:silane}(a), underwent band gap closure at 200~GPa. $C2/c$-SiH$_4$, shown in Fig.\ \ref{fig:silane}(b), was a good metal, suggestive of high temperature superconductivity. Simulated annealing was also employed to investigate the behavior of silane under pressure \cite{Yao:2007-Si}.  A metallic $C2/c$ symmetry phase, which differed from the one predicted by Pickard and Needs, was singled out for further investigation. Despite the fact that it had an enthalpy higher than that of $I4_1/a$-SiH$_4$, it was found to be dynamically stable between 65-150~GPa. The Allen-Dynes modified McMillan equation yielded estimates of 45-55~K for the $T_c$ of the metastable $C2/c$-SiH$_4$ phase at 125~GPa, whereas $I4_1/a$-SiH$_4$ was not a superconductor at 150~GPa. A combined experimental and theoretical study proposed that SiH$_4$ assumes the $P2_1/c$ symmetry spacegroup between 10-25~GPa \cite{Degtyareva:2007-Si}. Within this pressure range the enthalpy of this phase was found to be lower than any of the previously proposed systems. Experiments showed the emergence of a new phase around 27~GPa, and theoretical work suggested that the structure formed might be the polymeric $Fdd2$ symmetry phase illustrated in Fig.\ \ref{fig:silane}(c).

Meanwhile, optical experiments suggested that silane undergoes an insulator-semiconductor phase transition around 100~GPa, but metalization did not occur below 210~GPa \cite{Sun:2006-Si}.  A comprehensive experimental study of SiH$_4$ showed no evidence for metalization to at least 150~GPa, at which pressure the band gap was estimated as being 0.6-1.8~eV \cite{Strobel:2011-Si}. On the other hand, Raman and infrared spectroscopy detected three phase transitions below 30~GPa, and reflectivity measurements suggested the onset of metalization above 60~GPa \cite{Chen:2008-Si}. Remarkably, resistance measurements showed that silane metalized at 50~GPa and superconductivity with $T_c=17$~K was observed at 96~GPa and 120~GPa \cite{Eremets:2008a}. Based upon XRD, a $P6_3$ symmetry structure was proposed for the superconducting phase. Above 120~GPa an insulating transparent phase, whose diffraction pattern matched with the one obtained for the $I4_1/a$-SiH$_4$ structure of Pickard and Needs, formed. A few years later this polymeric $I4_1/a$-SiH$_4$ phase was synthesized at 124~GPa and 300~K \cite{Hanfland:2011-Si}.

Further theoretical studies found that the superconducting phase  synthesized in Ref.\ \cite{Eremets:2008a} could not possess the proposed $P6_3$ symmetry structure, since it was found to be dynamically unstable, and its enthalpy was significantly higher than that of other alternatives \cite{Chen:2008-2-Si,Kim:2008-Si,Martinez-Canales:2009-Si,Yan:2010-Si}. Computations suggested that metastable structures with $Cmca$ \cite{Chen:2008-2-Si}, $Pbcn$ \cite{Martinez-Canales:2009-Si} or $P4/nbm$ \cite{Kim:2008-Si} symmetries could be candidates for the superconducting phase. Eliashberg theory was employed to calculate the superconducting properties of some of these phases \cite{Wei:2010-Si,Szczkesniak:2013-2-Si,Durajski:2014-Si}. However, none of the proposed structures could fully explain the experimental results. Because of this, it has been suggested that the measured superconductivity originated from unintended reaction products formed from the decomposition of silane under pressure, as described in Sec.\ \ref{sec:group10}.

CSP techniques have recently been employed to explore the phases that silane adopts at significantly higher pressures than those considered previously, some of these are illustrated in Figs.\ \ref{fig:silane}(c-f). One of the phases found possessed  $P2_1/c$ symmetry above 383~GPa, and another $C2/m$ symmetry above 606~GPa with estimated $T_c$ values of 32~K at 400~GPa, and 106~K at 610~GPa, respectively \cite{Zhang:2015-Si}. Another study found a $P\bar{3}$ symmetry phase, which could be described as a polymeric Si-H structure intercalated with H$_2$ units whose $T_c$ was 32~K at 300~GPa, to be the most stable alternative above 241~GPa \cite{Cui:2015-Si}. 

\begin{figure}[h!]
\begin{center}
\includegraphics[width=\columnwidth]{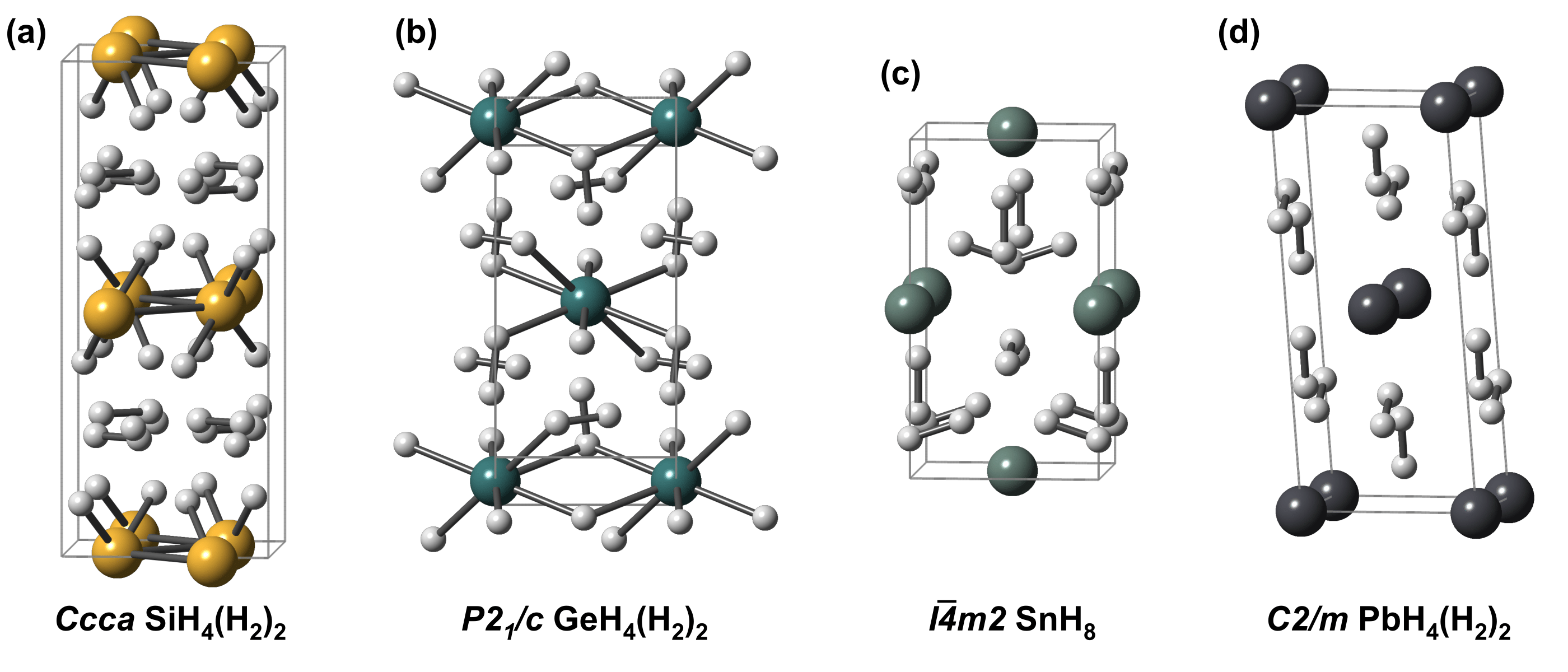}
\end{center}
\caption{Unit cells of superconducting phases predicted for group 14 hydrides with the MH$_4$(H$_2$)$_2$ stoichiometry: (a) $Ccca$-SiH$_4$(H$_2$)$_2$ phase with $T_c=$~98-107~K at 250~GPa \cite{Li:2010-Si}, (b) $P2_{1}/c$-GeH$_4$(H$_2$)$_2$ phase with $T_c=$~76-90~K at 250~GPa \cite{Zhong:2012-Ge}, (c) $I\bar{4}m2$-SnH$_8$ phase with $T_c=$~63-72~K at 250~GPa \cite{Zhang:2015a} and (d) $C2/m$-PbH$_4$(H$_2$)$_2$ phase with $T_c\approx$~107~K at 230~GPa \cite{Cheng:2015}.}
\label{fig:Group14}
\end{figure}

Mixtures of silane and molecular hydrogen have also been intensely investigated both experimentally and theoretically. Spectroscopic evidence for the formation of compounds with the general formula SiH$_4$(H$_2$)$_n$ at 6.5-35~GPa was obtained nearly simultaneously by two different groups \cite{Strobel:2009a,Wang:2009-Si}. First principles calculations \cite{Yao:2010-Si,Yim-2010-Si,Michel:2010-Si,Shanavas:2012-Si} have studied potential candidates for the SiH$_4$(H$_2$)$_2$ phase synthesized by Strobel and co-workers \cite{Strobel:2009a}. CSP at higher pressures predicted that the $Ccca$-SiH$_4$(H$_2$)$_2$ phase illustrated in Fig.\ \ref{fig:Group14}(a) would become stable with respect to the elements above 248~GPa, and its $T_c$ was estimated to be 98-107~K at 250~GPa \cite{Li:2010-Si}. Its superconducting properties have subsequently been examined using the Eliashberg formalism \cite{Szcze:2013-Si,Durajski:2013-Si}. 

The experimental availability of the Si$_2$H$_6$ molecule at standard conditions inspired CSP calculations on this stoichiometry. It was shown that disilane becomes stable with respect to decomposition into the elements at 135~GPa \cite{Jin:2010-Si}. Above this pressure phases with $P\bar{1}$, $Pm\bar{3}m$, and $C2/c$ symmetries with estimated $T_c$ values of 65-76~K at 175~GPa, 139-153~K at 275~GPa and 34-42~K at 300~GPa, respectively, were predicted as being stable. A later study found that disilane is thermodynamically unstable towards decomposition into SiH$_4$ and the elemental phases below 190~GPa \cite{Flores:2012-Si}. Above this pressure a $Cmcm$ symmetry phase was found to have the lowest enthalpy up to 280~GPa, and its $T_c$ was estimated as being 20~K at 100~GPa and 13~K at 220~GPa. \\

\noindent\textbf{Germanium} \\ 
The high pressure behavior of germane, GeH$_4$, has also been intensely investigated. A theoretical study that considered germanium analogues of previously proposed candidate SiH$_4$ and CH$_4$ phases predicted that an insulating fcc structure was preferred below, and a metallic SnF$_4$-like structure was preferred above 72~GPa \cite{Canales:2006a}. Another study employing a similar approach concluded that germane would metallize at a pressure lower than silane \cite{Li:2007-Ge}. However, an evolutionary algorithm based investigation predicted phases whose enthalpies were significantly lower \cite{Gao:2008a}. In this study it was also shown that solid germane is thermodynamically unstable with respect to decomposition into the elemental phases below 196~GPa. Above this pressure a metallic $C2/c$ symmetry phase, which contained H$_2$ motifs with elongated bonds, was stable, and its $T_c$ was estimated as being 64~K at 220~GPa. The superconducting properties of $C2/c$-GeH$_4$ were subsequently analyzed  \cite{Szczesniak:2015-Ge}. 

Even though germane is thermodynamically unstable at atmospheric conditions it does not decompose, implying that  metastable phases may be accessible under pressure. Therefore, a theoretical study was carried out to find the most stable phases containing intact GeH$_4$ units \cite{Zhang:2010-Ge}. The following set of pressure induced transitions were proposed: $P2_1/c\rightarrow Cmmm \rightarrow P2_1/m \rightarrow C2/c$. The metastable $Cmmm$-GeH$_4$ phase was predicted to have a $T_c$ of 40~K at 20~GPa \cite{Zhang:2010-2-Ge}, and its superconducting properties have been studied in greater detail \cite{Szczkesniak:2013-Ge}. Recently, CSP techniques have predicted two hitherto unknown GeH$_4$ phases, one with $Ama2$ symmetry at 250~GPa and one with $C2/c$ symmetry at 500~GPa, to be thermodynamically, mechanically and dynamically stable with estimated $T_c$ values of 47-57 and 70-84~K, respectively \cite{Zhang:2015-Ge}.

Spectroscopic evidence has shown that germane and H$_2$ can form a compound with the approximate GeH$_4$(H$_2$)$_2$ stoichiometry  at pressures of 7.5-27~GPa \cite{Strobel:2010-Ge}. The GeH$_4$ molecules within this phase are rotationally disordered and occupy fcc sites, whereas the H$_2$ molecules fill both octahedral and tetrahedral lattice sites. These experiments inspired theoretical studies on phases with the GeH$_4$(H$_2$)$_2$ stoichiometry up to center of the earth pressures \cite{Zhong:2012-Ge}. Structures with $I\bar{4}m2$ and $Pmn2_1$ symmetry, that differed only in the orientation of the H$_2$ units, were proposed as the most likely candidates for the experimentally observed phases. Below 220~GPa these phases were thermodynamically unstable. Above this pressure a stable, metallic $P2_1/c$ symmetry phase (see Fig.\ \ref{fig:Group14}(b)) with a $T_c$ of 76-90~K at 250~GPa was found. The superconducting properties of $P2_1/c$-GeH$_4$(H$_2$)$_2$ have been investigated \cite{Szcze:2014-Ge}, and its $T_c$ was found to decrease with increasing pressure \cite{Zhong:2013-Ge}. 

Evolutionary algorithms have also been employed to search for the most stable binary compounds of germanium and hydrogen over a wide composition range \cite{Hou:2015-Ge,Esfahani:2017-Ge}. The following stoichiometries were found to lie on the convex hull: Ge$_3$H (which was stable with respect to the elements already by 40~GPa), Ge$_2$H, GeH$_3$, Ge$_3$H$_{11}$ and GeH$_4$. Various GeH$_3$ phases were found to be superconducting with estimated $T_c$ values that exceeded 100~K at 180~GPa  \cite{Abe:2013-Ge}, and  80~K at 300~GPa \cite{Hou:2015-Ge}. The $T_c$s of $C2/m$-GeH$_4$ and $I\bar{4}m2$-Ge$_3$H$_{11}$ were predicted to be 56-67~K at 280~GPa, and 34-43~K at 285~GPa, respectively, and they were found to decrease at higher pressures \cite{Esfahani:2017-Ge}.  \\

\noindent\textbf{Tin} \\
A number of theoretical studies have examined the high pressure behavior of stannane, SnH$_4$, which is an unstable molecule at atmospheric pressures. Simulated annealing calculations predicted a metallic, $P6/mmm$-symmetry structure as having the lowest enthalpy between 70-160~GPa \cite{Tse:2007a}. Soft phonon modes resulting from Fermi surface nesting were found to give rise to a large EPC and a $T_c$ of 80~K at 120~GPa. A couple of years later evolutionary searches uncovered two SnH$_4$ phases with lower enthalpies, an $Ama2$ structure that was stable between 96-180~GPa and a $P6_3/mmc$ symmetry phase that was preferred above 180~GPa \cite{Gao:2010a}. Both of these phases contained hexagonal layers of Sn atoms and H$_2$ units. It was shown that SnH$_4$ becomes thermodynamically stable with respect to decomposition into the elemental species above 96~GPa. The $T_c$ was estimated as being 15-22~K for the $Ama2$ phase at 120~GPa, and 52-62~K for the $P6_3/mmc$ structure at 200~GPa. Computations that considered pressures up to 600~GPa predicted that a $C2/m$ symmetry SnH$_4$ phase will become stable at 400~GPa  \cite{Zhang:2016a}. Its $T_c$ was estimated as being 64-74~K at 500~GPa, and the EPC was found to arise primarily from the vibrational modes of hydrogen. It was shown that at pressures where SnH$_4$ is thermodynamically unstable with respect to the elemental phases, the most stable structures unsurprisingly segregated into single-component slabs or layers, some of which may be kinetically stable \cite{Gonzalez:2010-Sn}. 

Two theoretical studies have considered phases with novel hydrogen to tin ratios. Between 250-350~GPa the $I\bar{4}m2$-SnH$_8$ structure illustrated in Fig.\ \ref{fig:Group14}(c), which contains H$_2$ and slightly bent H$_3$ molecules, was found to lie on the convex hull \cite{Zhang:2015a}. Its  $T_c$ was calculated to be 63-72~K at 250~GPa, and it increased slightly under pressure. In addition to this phase, a later study also identified $C2/m$-SnH$_{12}$ and $C2/m$-SnH$_{14}$, which were predicted to become stable above 250~GPa and 280~GPa, respectively \cite{Esfahani:2016}. A unique motif of linear H$_4^-$ units was observed in SnH$_{12}$, whereas linear H$_3^-$ moieties were found in SnH$_{14}$. The $T_c$ was estimated as being 81~K, 93~K, and 97~K for SnH$_8$ at 220~GPa, SnH$_{12}$ at 250~GPa and SnH$_{14}$ at 300~GPa, respectively. \\

\noindent\textbf{Lead} \\ 
The lead tetrahydride analogue of methane, tetrahedral PbH$_4$, is thermodynamically unstable in the gas phase. PbH$_4$ was first synthesized via laser ablation in a solid hydrogen matrix, along with the Pb$_2$H$_2$ and H(Pb$_2$H$_2$)H molecules \cite{Wang:2003-Pb}. The instability of PbH$_4$ has precluded experimental studies of its solid state structure, and so far the high pressure behavior of this, and other hydrides, of lead have only been investigated theoretically. Calculations showed that solid PbH$_4$ becomes enthalpically favorable with respect to the elemental phases above 132~GPa \cite{Zaleski-Ejgierd:2011a}. Below $\sim$300~GPa the most stable phase was found to have a three-dimensional lattice, whereas the lowest enthalpy phase above this pressure was distinctly layered. Both phases contained H$_2$ molecules whose intermolecular distances were comparable to those within elemental hydrogen, and these hydrogenic sublattices were found to exhibit liquid-like behavior. They were good metals, with a nearly free electron like DOS. CSP techniques have also been employed to investigate the PbH$_4$(H$_2$)$_2$ stoichiometry \cite{Cheng:2015}.  Enthalpically stable structures, which contained H$_2$ molecules that separated the Pb atoms, were predicted above 133~GPa. The $C2/m$ symmetry phase illustrated in Fig.\ \ref{fig:Group14}(d) was found to be a good metal, with an estimated $T_c$ of 107~K at 230~GPa, and the large electron-phonon coupling was primarily due to vibrations associated with the hydrogen atoms.

%%%%%%%%%%%%%%%%%%%%%%%%%%%%%%%%%%%%%%%%%%%%%%%%%%%%%%%%%%%%%%%%%%%%%%%%%%%%%%%%%%%%%%%%%%%%%%%%%%%%%%%%%%%%%%%%%%%%%%%%%%%%
%%%%%%%%%%%%%%%%%%%%%%%%%%                       Pnictogens                            %%%%%%%%%%%%%%%%%%%%%%%%%%%%%%%%%%%%%
%%%%%%%%%%%%%%%%%%%%%%%%%%%%%%%%%%%%%%%%%%%%%%%%%%%%%%%%%%%%%%%%%%%%%%%%%%%%%%%%%%%%%%%%%%%%%%%%%%%%%%%%%%%%%%%%%%%%%%%%%%%%

\section{Group 15: Pnictogen Hydrides} 

\noindent\textbf{Nitrogen} \\ 
At ambient pressure and temperature gas phase NH$_3$ is the only pnictogen hydride that is thermodynamically stable. Six molecular NH$_3$ phases have been studied experimentally including: a low-temperature ordered phase \cite{Hewat:1979}, higher temperature rotationally disordered phases II and III, an orthorhombic phase IV \cite{Loveday:1996,Datchi:2006}, phase V (whose spacegroup is unknown) \cite{Gauthier:1988a}, and phase VI (which may exhibit symmetric hydrogen bonding) \cite{Datchi:2006}. DFT calculations showed that hydrogen bond symmetrization does not occur in ammonia up to at least 300~GPa, but it was pointed out that quantum proton motion may promote symmetrization at lower pressures \cite{Fortes:2003a}. Random searches at 0~K found the previously reported phase I and phase IV structures \cite{Pickard:2008a}. However, above 90~GPa a previously unknown $Pma2$ symmetry phase consisting of alternating layers of NH$_4^+$ and NH$_2^-$ ions, which had a band gap of 3.6~eV at 100~GPa, was found to be the most stable. Experiments provided evidence for the existence of an ionic phase around 150~GPa, and further calculations showed that a $Pca2_1$ symmetry ionic lattice is more stable than $Pma2$ above 176~GPa \cite{Ninet:2014}. The high pressure behavior of mixtures of N$_2$ and H$_2$ \cite{Spaulding:2014a,Goncharov:2015a}, as well as NH$_3$ and H$_2$ \cite{Chidester:2011a} has been studied experimentally. \\

\begin{figure}[h!]
\begin{center}
\includegraphics[width=\columnwidth]{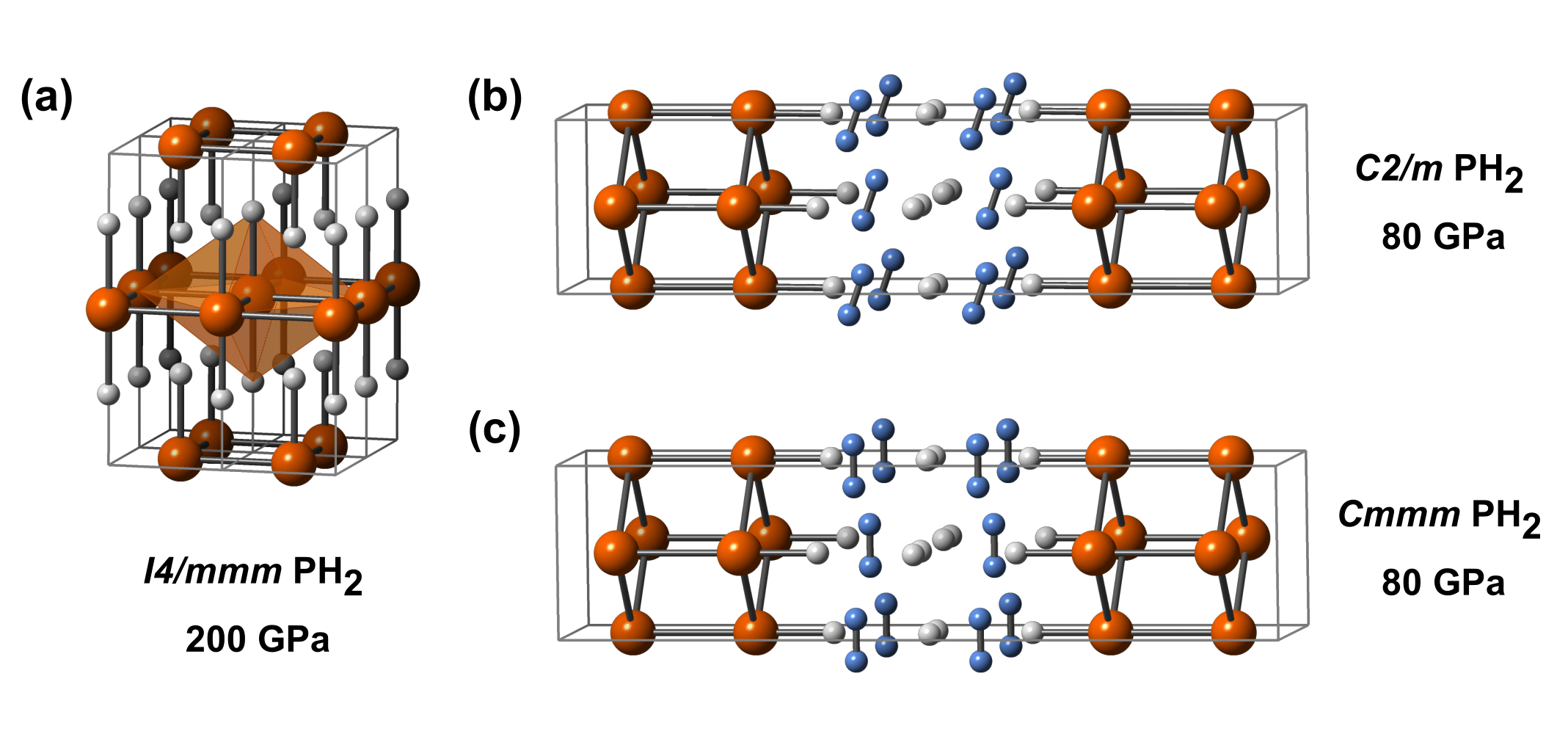}
\end{center}
\caption{Theoretically predicted PH$_2$ phases that are likely contributors to the superconductivity observed in compressed phosphine \cite{Drozdov:2015-P}. (a) $I4/mmm$-PH$_2$ whose $T_c$ was calculated as being $\sim$70~K at 200~GPa \cite{Shamp:2016a,Flores:2016-P}, (b) $C2/m$-PH$_2$ and $Cmmm$-PH$_2$ differ only by a rotation of the H$_2$ molecules colored purple. Their $T_c$ was estimated to be $\sim$40~K at 80~GPa \cite{Bi:2017-P}.}
\label{fig:phosphorus}
\end{figure}

\begin{figure}[h!]
\begin{center}
\includegraphics[width=0.9\columnwidth]{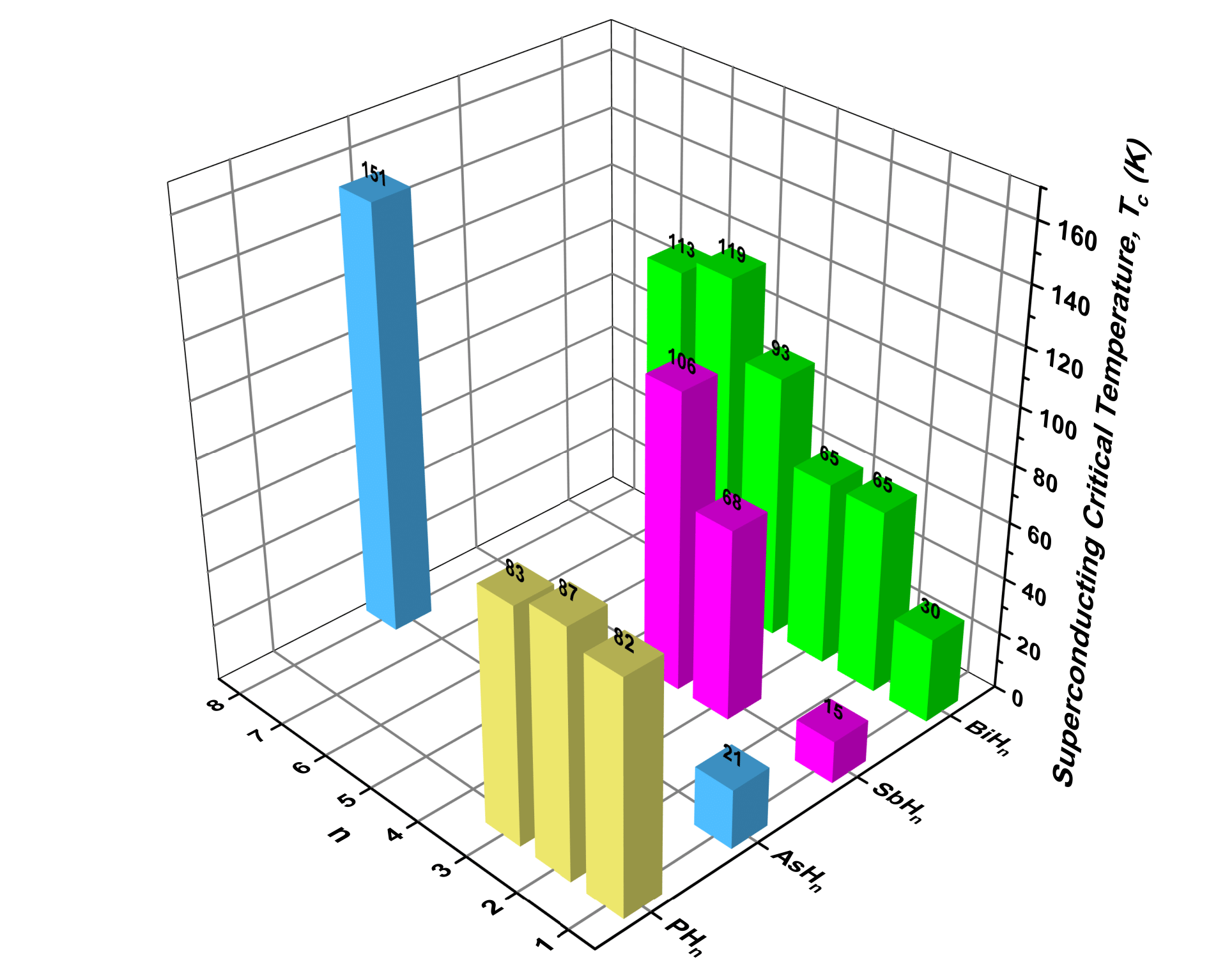}
\end{center}
\caption{Theoretically predicted $T_c$ values for hydrides of group 15 elements, including PH$_n$ (yellow) \cite{Flores:2016-P,Shamp:2016a,Liu:2016-P,Bi:2017-P,Durajski:2016-S-P}, AsH$_n$ (blue) \cite{Fu:2016a}, SbH$_n$ (pink) \cite{YanbinMa:2015b,Fu:2016a,Abe:2015a} and BiH$_n$ (green) \cite{Abe:2015a,YanbinMa:2015a}, $n=1-8$. The $T_c$s provided are the highest ones obtained for the given composition, and may have been calculated at different pressures and using slightly different values of $\mu^*$.}
\label{fig:Group15}
\end{figure}

\noindent\textbf{Phosphorus} \\ 
Recent resistance measurements on phosphine, PH$_3$, which was liquefied and compressed in a DAC revealed a $T_c$ of 30~K at 83~GPa and 103~K at 207~GPa \cite{Drozdov:2015-P}. The lack of experimental structural information motivated a series of theoretical studies that used CSP to identify several candidate structures for the superconducting phases between 100-200~GPa including PH \cite{Flores:2016-P}, PH$_2$ \cite{Shamp:2016a,Flores:2016-P}, and PH$_3$ \cite{Flores:2016-P,Liu:2016-P}. Even though the most stable phases identified were found to be unstable with respect to decomposition into the elements under pressure, they were dynamically stable and superconducting.  The experimental pressure dependence of the $T_c$ agreed most closely with that of the $I4/mmm$ symmetry PH$_2$ structure \cite{Flores:2016-P} illustrated in Fig.\ \ref{fig:phosphorus}(a), but it was concluded that the observed superconductivity is likely due to a mixture of metastable phases that form from the decomposition of phosphine under pressure. Migdal-Eliashberg theory has been employed to study the superconducting properties of one of the predicted phases with the PH$_3$ stoichiometry \cite{Durajski:2016-S-P}. Calculations have also been undertaken to investigate the P/H phase diagram at $P<100$~GPa \cite{Bi:2017-P}. The structure with the most negative $\Delta H_F$ at 80~GPa was a non-metallic (PH$_5$)$_2$ phase whose structure and bonding was analogous to that of diborane. A number of metastable phases that were composed of hydrogen-capped simple cubic like phosphorus layers and mobile molecular H$_2$ layers were found to be superconducting. However, the $T_c$ of the two PH$_2$ phases illustrated in Fig.\ \ref{fig:phosphorus}(b) and Fig.\ \ref{fig:phosphorus}(c) agreed the best with the experimental results.  The maximum $T_c$s calculated for PH, PH$_2$ and PH$_3$ are provided in Fig.\ \ref{fig:Group15}.
\\

\noindent\textbf{Arsenic, Antimony, Bismuth} \\ 
% %
Only a few theoretical studies have investigated the heavier pnictogen polyhydrides, and the highest $T_c$s obtained for each superconducting phase are plotted in Fig.\ \ref{fig:Group15} \cite{Fu:2016a, YanbinMa:2015b, Abe:2015a, YanbinMa:2015a}. Fu and co-workers systematically explored the hydrogen rich phase diagram of the hydrides of P, As, and Sb \cite{Fu:2016a}. The crystalline hydrides of phosphorus were found to be unstable with respect to decomposition into the elements between 100-400~GPa, whereas those of arsenic became stable by 300~GPa, and antimony by 200~GPa. The predicted stable structures included: $Cmcm$-AsH, which adopts a three-dimensional network with five-coordinate As and H atoms; $C2/c$-AsH$_8$, which is formed from irregular AsH$_{16}$ polyhedral motifs connected to one another in a three-dimensional network containing quasi-molecular H$_{2}$ units with a bond length of 0.8-0.9~\AA{}; $Pnma$-SbH, which is composed of Sb-H chain-like motifs where each Sb atom is coordinated to three H atoms; $Pmmn$-SbH$_3$, which is comprised of irregular SbH$_{10}$ and SbH$_{12}$ polyhedra with quasi-molecular H$_{2}$ bridges; and $P6_3/mmc$-SbH$_4$, which is made up of regular SbH$_{14}$ octadecahedra that are connected through shared  corner H atoms in a three-dimensional network forming quasi-molecular H$_{2}$ units. The estimated $T_c$s of AsH$_8$ and SbH$_4$ were $\sim$150~K at 350~GPa and 100~K at 150~GPa, respectively, whereas all other compounds possessed a $T_c$ of $\sim$20~K or lower. The same $P6_3/mmc$-SbH$_4$ phase was predicted in a prior work \cite{YanbinMa:2015b}. Abe and Ashcroft computationally studied the SbH$_2$ and SbH$_3$ stoichiometries, and they found that at 170~GPa $Pnma$-SbH$_3$ was stable and superconducting with a $T_c$ of $\sim$68~K for $\mu^*=0.13$ \cite{Abe:2015a}. The hydrides of bismuth were calculated to become stable with respect to the elemental phases above 105~GPa \cite{YanbinMa:2015a, Abe:2015a}. $P6_3/mmc$-BiH contained monoatomic hydrogen atoms, whereas quasi-molecular H$_2$ units were present within many of the BiH$_n$ ($n=2-6$) phases. In addition, $C2/m$-BiH$_{5}$ was also comprised of linear H$_{3}$ units \cite{YanbinMa:2015a}. $T_c$s ranging from 20-119~K have been calculated for these hydrides, and the highest value obtained was for BiH$_5$ at 300~GPa. 

%%%%%%%%%%%%%%%%%%%%%%%%%%%%%%%%%%%%%%%%%%%%%%%%%%%%%%%%%%%%%%%%%%%%%%%%%%%%%%%%%%%%%%%%%%%%%%%%%%%%%%%%%%%%%%%%%%%%%%%%%%%%
%%%%%%%%%%%%%%%%%%%%%%%%%%                       Chalcogens                            %%%%%%%%%%%%%%%%%%%%%%%%%%%%%%%%%%%%%
%%%%%%%%%%%%%%%%%%%%%%%%%%%%%%%%%%%%%%%%%%%%%%%%%%%%%%%%%%%%%%%%%%%%%%%%%%%%%%%%%%%%%%%%%%%%%%%%%%%%%%%%%%%%%%%%%%%%%%%%%%%%

\section{Group 16: Chalcogen Hydrides}

\noindent\textbf{Oxygen} \\ 
The phase diagram of H$_2$O has been extensively studied because of its relevance towards life on earth, as well as the interiors of icy extraterrestrial objects. Sixteen crystalline phases of H$_2$O have been found experimentally \cite{Petrenko}. Perhaps the most well-known high pressure phase of water is ice X, within which the oxygen atoms are found on a bcc lattice and each hydrogen atom lies midway between two oxygen atoms so that the intra and intermolecular H-O distances become equalized \cite{Loubeyre:1999a}. A number of theoretical studies have probed the structure of ice at center of the earth pressures and higher. A $Pbcm$ structure was predicted to become stable at $\sim$300~GPa \cite{Benoit:1996}. Over a decade later it was shown that it transitions to a $Pbca$ structure at 0.76~TPa, followed by a metallic $Cmcm$ phase at 1.55~TPa \cite{Militzer:2010}. A couple of years later four theoretical studies appeared nearly simultaneously that used different CSP techniques to predict the structures of water at TPa pressures \cite{Wang:2011a,Hermann:2012a,Ji:2011a,McMahon:2011c}. The calculations differed not only in the structure prediction methods that were employed, but also in the pressures at which the searches were carried out, and in the sizes of the unit cells considered. As a result, each study proposed a somewhat different phase diagram. One thing that they all had in common, however, was the discovery of insulating phases that were more stable than the previously proposed metallic $Cmca$-H$_2$O phase. Thus, the pressure at which ice is thought to become metallic was shifted to higher values; somewhere between 2-7~TPa. 

Because H$_2$O and H$_2$ are known to form hydrogen clathrate compounds at high pressures \cite{Vos:1993a}, evolutionary algorithms have been employed to search for hitherto unknown H$_2$O-H$_2$ clathrate-like structures up to 100~GPa \cite{Qian:2014a}. Moreover, CSP methods have found that at planetary pressures non-intuitive reactions and decomposition mechanisms of binary H/O systems can occur. For example, above 5~TPa H$_2$O is predicted to decompose into H$_2$O$_2$ and hydrogen rich phases \cite{Pickard:H2O}, and at 1.4~TPa an H$_4$O structure is preferred over elemental water and hydrogen \cite{Zhang:2013a}. \\

\begin{figure}[h!]
\begin{center}
\includegraphics[width=\columnwidth]{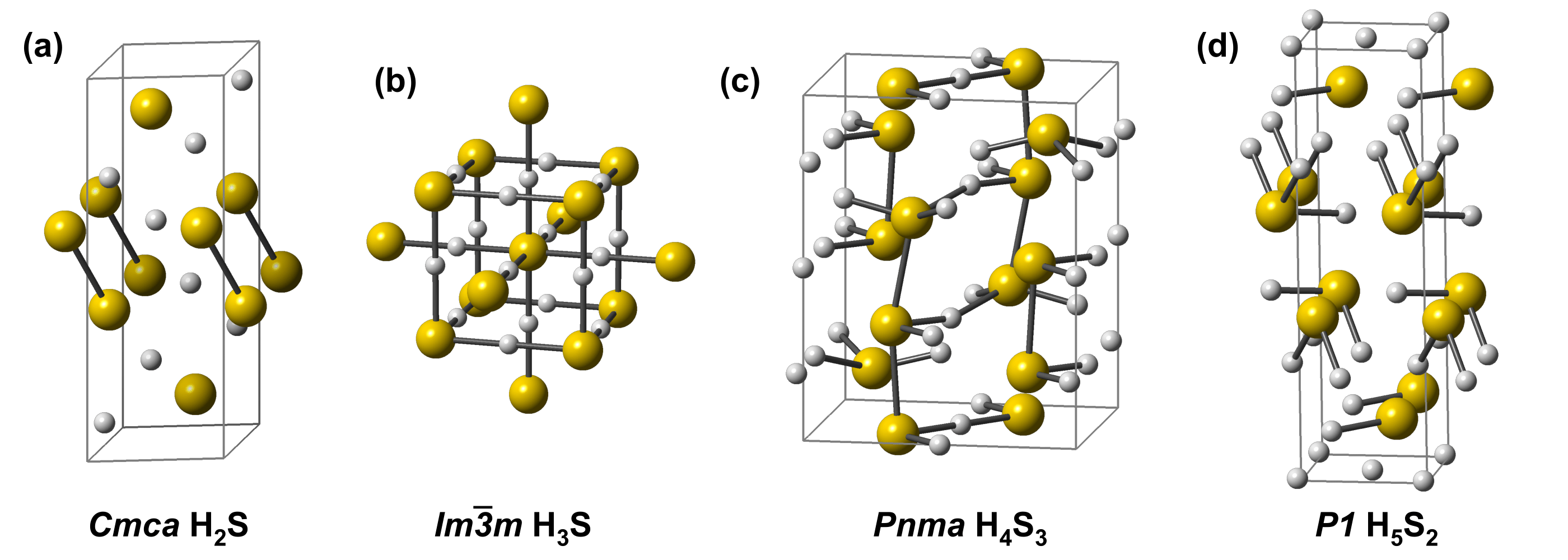}
\end{center}
\caption{Structures of superconducting phases predicted for compounds containing hydrogen and sulfur with various stoichiometries: (a) $Cmca$-H$_2$S \cite{Li:2014}, (b) $Im\bar{3}m$-H$_3$S \cite{Duan:2014}, (c) $Pnma$-H$_4$S$_3$ \cite{Li:2016-S} and (d) $P1$-H$_5$S$_2$ \cite{Ishikawa:2016}.}
\label{fig:Group16}
\end{figure}

\noindent\textbf{Sulfur} \\ 
The discovery of high temperature superconductivity in the high pressure hydrogen/sulfur phase diagram is a striking example of how a feedback loop between experiment and theory can lead to the synthesis of remarkable materials. A recent mini-review provides an excellent synopsis of the work carried out on this system to date \cite{Yao-S-review:2018}. Hydrogen sulfide, H$_2$S, has been theoretically and experimentally studied extensively under pressure \cite{Shimizu:1992,Endo:1994,Shimizu:1995,Loveday:1997,Fujihisa:1998,Endo:1998,Endo:1996,Shimizu:1997,Sakashita:1997, Rousseau:1999, Sasaki:1991,Rousseau:2000a,Wang:2007-S,Wang:2010-S,Li:2014,Durajski:2015-S}. For example, \emph{Ab~Initio} molecular dynamics computations \cite{Rousseau:2000a,Wang:2007-S,Wang:2010-S} and CSP techniques \cite{Li:2014} have been employed to propose structural candidates for various phases. Li et al.\ showed that H$_2$S is thermodynamically stable with respect to the elemental phases up to 200~GPa, and proposed candidate structures for the non-metallic phases IV and V \cite{Li:2014}. Moreover, they calculated a $T_c$ of 80~K at 160~GPa for the $Cmca$ phase illustrated in Fig.\ \ref{fig:Group16}(a) \cite{Li:2014}.

Strobel and co-workers carried out experiments that showed compound formation between H$_2$S and H$_2$ at 3.5~GPa, and evidence of a clathrate-like structure by 17~GPa \cite{Strobel:2011a}.  This work inspired a computational study by Duan and co-workers on the high pressure behavior of systems with the (H$_2$S)$_2$H$_2$ (or H$_3$S) stoichiometry \cite{Duan:2014}. DFT calculations predicted that metallic $R3m$ and $Im\bar{3}m$ symmetry H$_3$S phases became preferred above 111 and 180~GPa, respectively. The EPC of both structures was found to be particularly high, and the $T_c$ was estimated as being 155-166~K at 130~GPa and 191-204~K at 200~GPa. The cubic $Im\bar{3}m$ structure is illustrated in Fig.\ \ref{fig:Group16}(b).

The theoretical work of Li and co-workers \cite{Li:2014}, on the other hand, inspired Drozdov et al.\ to study the superconducting behavior of H$_2$S under pressure \cite{Drozdov:2015a}. The measured $T_c$s for samples prepared at  $T\le$100~K were in good agreement with the values computed for H$_2$S \cite{Li:2014}. The much higher $T_c$ observed for the sample prepared above room temperature was consistent with the predictions for H$_3$S made by Duan's group \cite{Duan:2014}, leading to the suggestion that at these conditions the hydrogen sulfide decomposed into elemental sulfur and H$_3$S, and the H$_3$S phase gave rise to the remarkable superconductivity. 
XRD has shown that the sulfur positions in the superconducting phase are consistent with the theoretically predicted $Im\bar{3}m$ and $R3m$-H$_3$S structures \cite{Einaga:2016}. The Meissner effect confirmed the record breaking $T_c$ \cite{Troyan:2016-S}, and a recent optical reflectivity study suggested that this material is a conventional superconductor where the superconductivity is due to electron-phonon interaction \cite{Capitani:2017a}. The direct synthesis of H$_3$S starting from H$_2$ and S followed by characterization via XRD and Raman spectroscopy led to the conclusion that the $Im\bar{3}m$-H$_3$S phase forms above 140~GPa \cite{Goncharov:2017}. Another experimental study, however, provided evidence for the synthesis of a $Cccm$-H$_3$S phase up to 160~GPa, but not for the proposed superconducting $Im\bar{3}m$ and $R3m$ symmetry phases \cite{Guigue:2017}. 

A plethora of theoretical calculations have been carried out to identify the stoichiometry and structure of the superconducting phase, analyze the factors contributing to $T_c$, and investigate the effect of anharmonicity, the isotope effect, and the quantum nature of the proton \cite{Flores:2016a,Papaconstantopoulos:2015,Bernstein:2015,Duan:2015-S,Errea:2015a,Akashi:2015-S,Errea:2016,Quan:2016a,Sano:2016a,Ortenzi:2016a,Gorkov:2016a,Goncharov:2016a,Bussmann:2017-S,Durajski:2016-S-P,Jarlborg:2016a,Szczkesniak:2017-S,Azadi:2017,Durajski:2015-S,Arita:2017a}. A majority of these focused on the $Im\bar{3}m$ symmetry H$_3$S phase. However, a few recent studies have questioned if this phase is responsible for the observed superconductivity. For example, it was shown that the decomposition of $Im\bar{3}m$-H$_3$S into H$_2$ plus an SH$_2$ phase, which can be thought of as an (SH$^-$)(H$_3$S$^+$) perovskite structure, is favorable \cite{Gordon:2016}. \emph{Ab initio molecular dynamics} (AIMD) simulations at 200~GPa and 200~K revealed that this perovskite structure can segregate into cubic and tetragonal regions that form a modulated structure whose diffraction pattern matches well with the one obtained experimentally \cite{Majumdar:S-2017}. Another study proposed that an infinite number of metastable long-period modulated crystals formed from the intergrowth of H$_2$S and H$_3$S with a composition of H$_x$S$_{1-x}$ with $2/3<x<3/4$ (so-called ``Magn\'eli'' Phases) could account for the experimentally observed dependence of the $T_c$ versus pressure \cite{Akashi:2016a}. Interestingly, the modulated structures predicted via AIMD resemble the Magn\'eli phases.

In addition, CSP searches have shown that the H$_2$S$_3$, H$_3$S$_2$, HS$_2$, H$_4$S$_3$ \cite{Li:2016-S}, H$_5$S$_8$, H$_3$S$_5$ \cite{Goncharov:2016a} and H$_5$S$_2$ \cite{Ishikawa:2016} stoichiometries could potentially form under pressure. The superconducting $Pnma$-H$_4$S$_3$ and $P1$-H$_5$S$_2$ structures are illustrated in Fig.\ \ref{fig:Group16}(c) and Fig.\ \ref{fig:Group16}(d), respectively. Moreover, a recent study concluded that H$_2$S can be kinetically protected up to very high pressures and only this phase can account for the superconductivity observed by Drozdov and co-workers in the samples prepared at low temperatures \cite{Li:2016-S}.

Theoretical work assuming the $Im\bar{3}m$-H$_3$S structure showed that the $T_c$ could further be increased by substitution of sulfur with more electronegative elements \cite{Heil:2015a}. Another study predicted a maximum superconducting critical temperature of 280~K at 250~GPa for H$_3$S$_{0.925}$P$_{0.075}$ \cite{Ge:2016-S}. High temperature superconductivity has also been proposed for other compressed sulfides including Li$_2$S, Na$_2$S and K$_2$S \cite{Hirsch:2015-S}. \\

\noindent\textbf{Selenium} \\ 
Following the discovery of high $T_{c}$ superconductivity in condensed sulfur hydride, the isoelectronic selenium analog was investigated for its potential superconductivity under high pressure. At 200~GPa and 300~GPa $C2/m$-HSe$_2$ and $P4/nmm$-HSe were found to be the lowest points on the convex hull, respectively \cite{Zhang:2015b}. An $Im\bar{3}m$-H$_3$Se structure that was isostructural to the proposed superconducting H$_3$S phase also lay on the convex hull at both of these pressures. The $T_c$ of H$_3$Se and HSe were estimated as being in the range of 100~K and 40~K, respectively. Another study that was published at about the same time also predicted the high pressure stability of the $Im\bar{3}m$-H$_3$Se phase, and SCDFT (density functional theory for superconductors) calculations yielded a $T_c$ of 130~K at 200~GPa \cite{Flores:2016a}. Its decreased $T_c$ as compared with the sulfur containing compound is a result of a smaller EPC, which likely originates from the better screening of the hydrogen vibrations by the larger ionic size of selenium. 

In a recent study H$_2$Se was synthesized from the elemental phases at 0.4~GPa and 473~K \cite{Pace:2017a}. At 12~GPa a transition to a structure resembling phase IV of H$_2$S was found to occur. Moreover, a host-guest $I4/mcm$-(H$_2$Se)$_2$H$_2$ structure, which was analogous to the one observed in (H$_2$S)$_2$H$_2$, formed above 4.2~GPa. Both of these hydrides of selenium decomposed above 24~GPa at 300~K \cite{Zhang:2018-Se}. In addition to H$_2$Se, a recent study synthesized a $Cccm$-H$_3$Se phase at 4.6~GPa and 300~K. At 170~K $Cmcm$-H$_3$Se was found to persist up to 39.5~GPa. Raman measurements and visual observations suggested that metalization occurred above 23~GPa. \\

\noindent\textbf{Tellurium} \\ 
Computational studies of structures with the H$_x$Te$_y$ ($x=1-8, y=1-3$) stoichiometries have been carried out up to 300~GPa. CSP techniques found that $P6/mmm$-H$_4$Te, $C2/m$-H$_5$Te$_2$ and $P6_3/mmc$-HTe were stable with respect to the elements by 200~GPa \cite{Zhong:2016}. In addition to these a $C2/c$-HTe$_3$ phase was present on the 300~GPa convex hull. Quasi-molecular H$_2$ units were found within H$_4$Te and H$_5$Te. The latter phase also possessed linear H$_3$ motifs. The largest $T_c$ calculated was 104~K for H$_4$Te at 170~GPa. The estimated $T_c$ for H$_5$Te was 58~K at 200~GPa. It was speculated that the structures and stoichiometries of the stable tellurium hydrides differed from those of sulfur and selenium because tellurium has a larger atomic core and smaller electronegativity as compared to its lighter brethren. The main contributions to the EPC for H$_4$Te and H$_5$Te arose from the intermediate-frequency hydrogen-based wagging and bending modes, as opposed to the higher frequency H-stretching modes that were found to be so important in hydrides containing sulfur and selenium.\\ 

\noindent\textbf{Polonium} \\ 
CSP techniques have been employed to find the most stable structures with the PoH$_n$, $n=1-6$, stoichiometries up to 300~GPa \cite{Liu:2015d}. The first hydride to become stable with respect to the elements was PoH$_2$ in the $Cmcm$ spacegroup at 100~GPa. At higher pressures the following stable phases were identified: $P6_3/mmc$-PoH, $Pnma$-PoH$_2$, $C2/c$-PoH$_4$ and $C2/m$-PoH$_6$. All of these phases were good metals, and with the exception of PoH they all contained H$_2$ units. Whereas the $T_c$ of PoH$_4$ was estimated as being 41-47~K at 200~GPa, the $T_c$ of all of the other phases was $<$5~K.

%%%%%%%%%%%%%%%%%%%%%%%%%%%%%%%%%%%%%%%%%%%%%%%%%%%%%%%%%%%%%%%%%%%%%%%%%%%%%%%%%%%%%%%%%%%%%%%%%%%%%%%%%%%%%%%%%%%%%%%%%%%%
%%%%%%%%%%%%%%%%%%%%%%%%%%                       Halogens                              %%%%%%%%%%%%%%%%%%%%%%%%%%%%%%%%%%%%%
%%%%%%%%%%%%%%%%%%%%%%%%%%%%%%%%%%%%%%%%%%%%%%%%%%%%%%%%%%%%%%%%%%%%%%%%%%%%%%%%%%%%%%%%%%%%%%%%%%%%%%%%%%%%%%%%%%%%%%%%%%%%

\section{Group 17: Halogen Hydrides} 

\noindent\textbf{Fluorine, Chlorine, Bromine} \\
The isomorphic low temperature phases of HF, HCl and HBr contain planar zigzag chains of hydrogen-bonded molecules held together by vdW forces. At atmospheric pressures HF crystallizes in a $Cmc2_1$ structure with four formula units in the cell \cite{Johnson:1975a}. Because the hydrogen bonds in the heavier halogen hydrides are weaker, they assume orientationally disordered molecular phases at high temperatures. At room temperature HCl and HBr adopt a structure isomorphic to  $Cmc2_1$-HF \cite{Sandor:1967a,Ikram:1993a}. HF (6~GPa) \cite{Pinnick:1989a}, HBr (32-39~GPa) \cite{Katoh:1999a} and HCl (51~GPa) \cite{Aoki:1999a} undergo a transformation to a $Cmcm$ phase wherein all of the H-X bonds are symmetric at the pressures given in the parentheses. Whereas the symmetric HCl phase is stable, the HBr phase is not \cite{Ikeda:1999a} and a decomposition reaction that yields Br$_2$ molecules occurs. 

Second order M\o ller-Plesset perturbation theory (MP2) calculations have been performed on solid HF and DF up to 20~GPa \cite{Sode:2012a}. Moreover, DFT based CSP techniques have been used to study the structural evolution of HF, HCl and HBr up to 200~GPa \cite{Zhang:2010a}. The known $Cmc2_1$ and $Cmcm$ phases were found, and it was predicted that HF would transform to a $Pnma$ phase at 143~GPa, whereas HCl and HBr were found to assume a $P\bar{1}$ structure above 108~GPa and 59~GPa, respectively. These phases all contained symmetric H-X bonds. Whereas HF was found to be a large band gap insulator at the highest pressures studied, HCl and HBr were semi-metals. A study that followed soft phonon modes calculated with DFT predicted that a $Cmcm \rightarrow P2_1/m$ phase transition would occur at 134-196~GPa for HBr and above 233~GPa for HCl  \cite{Duan:2010}. The $T_c$ was estimated to be 27-34~K for HBr at 160~GPa and 9-14~K for HCl at 280~GPa. Five years later CSP techniques were yet again employed to search for the most stable phases of HBr \cite{Chen:2015a,Lu:2015a} and HCl \cite{Chen:2015a} under pressure. Both studies showed that based on the enthalpy alone, HBr was not likely to decompose into the elemental phases under pressure. An $I\bar{4}2d$-HBr \cite{Lu:2015a} phase that was nearly isoenthalpic with $P\bar{1}$-HBr up to 125~GPa was discovered. Above 120-125~GPa both studies predicted that a $C2/m$ symmetry HBr phase \cite{Lu:2015a,Chen:2015a} is the most stable. The $C2/m$ symmetry HBr phases found in the two studies are different and their estimated $T_c$ values are: 25~K at 150~GPa \cite{Lu:2015a} and $<$1~K at 120~GPa \cite{Chen:2015a}. A $C2/m$ symmetry structure was also found to be the most stable HCl phase above 250~GPa, and at this pressure its $T_c$ was estimated as being 20~K \cite{Chen:2015a}. 

\begin{figure}[h!]
\begin{center}
\includegraphics[width=0.7\columnwidth]{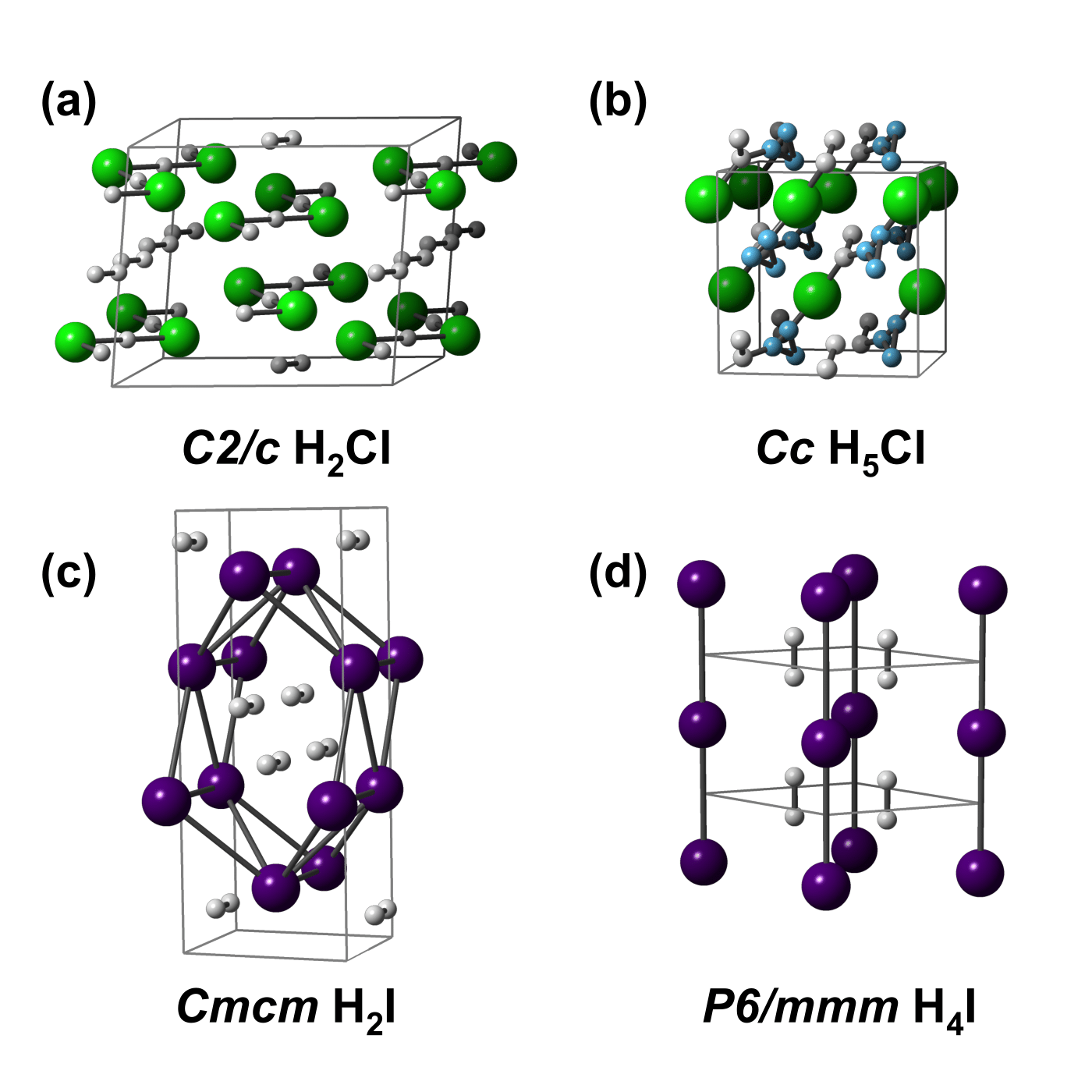}
\end{center}
\caption{Halogen polyhydrides that are predicted to be stable under pressure. (a) $C2/c$-H$_2$Cl contains hydrogen atoms and H$_2$ units \cite{Wang:2015-17}, (b) $Cc$-H$_5$Cl contains triangular H$_3^+$ units (colored in light blue) and H$_2$ molecules \cite{Wang:2015-17}, (c) $Cmcm$-H$_2$I and (d) $P6/mmm$-H$_4$I contain H$_2$ units \cite{Shamp:2015,Duan:2015-I}. }
\label{fig:halogen}
\end{figure}

More recently, CSP has been employed to study polyhydrides of the halogens under pressure. H$_2$Cl and H$_5$Cl were predicted to become stable with respect to decomposition into H$_2$ and HCl between 60-300~GPa, and they lay on the 100, 200 and 300~GPa convex hulls \cite{Wang:2015-17}. The $C2/c$-H$_2$Cl phase illustrated in Fig.\ \ref{fig:halogen}(a) contained one-dimensional HCl chains interposed with H$_2$ molecules.  As shown in Fig.\ \ref{fig:halogen}(b), the three-center-two-electron H$_3^+$ motif, which was first found in interstellar space, was present within $Cc$-H$_5$Cl. By 300~GPa it approached the ideal equilateral triangle configuration. In addition to these phases a later CSP study found that H$_3$Cl lies on the 20, 50, and 100~GPa convex hulls, and H$_7$Cl comprises the 100~GPa hull \cite{Duan:2015-F-Cl}. None of the aforementioned phases were superconducting. Two years later CSP added one more stoichiometry, H$_4$Cl$_7$, to the 100-400~GPa convex hulls \cite{Zeng:2017-Cl}. Moreover, the ZPE was found to stabilize a few H$_n$Cl ($n=2,3,5$) phases that had not been predicted previously. One example was a metallic $R\bar{3}m$-H$_2$Cl phase that was superconducting below 44~K at 400~GPa. 

A CSP study showed that the only hydride of fluorine that lies on the convex hull under pressure is HF \cite{Duan:2015-F-Cl}.
Finally, H$_n$Br phases with $n=2-5,7$ were shown to be thermodynamically stable above 40~GPa \cite{Duan:arxiv-Br}. At 240~GPa $Cmcm$-H$_2$Br.  and $P6_3/mmc$-H$_4$Br were predicted to be superconducting below 12~K and 2.4~K, respectively.\\

\noindent\textbf{Iodine} \\
The behavior of the monohydride of iodine differs from its lighter brethren. At atmospheric conditions HI assumes a planar distorted hydrogen-bonded diamond lattice, but its detailed structure is unknown \cite{Ikram:1993a}. Pressure induces an insulator to metal transition in HI below 50~GPa \cite{Straaten:1986a}, but the impurities and side-products observed above 70~GPa are thought to be indicative of decomposition \cite{Straaten:1988a}. Indeed, CSP calculations showed that already at 1~atm HI is metastable with respect to the elemental phases \cite{Shamp:2015}. This same study found that an insulating $P\bar{1}$-H$_5$I phase, which is comprised of  (HI)$^{\delta+}$ and H$_2^{\delta-}$ molecules, is stable between 30-90~GPa. Moreover, the metallic $Cmcm$-H$_2$I and  $P6/mmm$-H$_4$I structures shown in Fig.\ \ref{fig:halogen}(c) and Fig.\ \ref{fig:halogen}(d) were found on the 100, 150 and 200~GPa convex hulls. These phases contained monoatomic iodine lattices and H$_2^{\delta-}$ units. At 100~GPa they were estimated to become superconducting below 7.8~K and 17.5~K, respectively. These two phases were also found by Duan and coworkers who in addition predicted a $R\bar{3}m$ symmetry H$_2$I phase wherein the H$_2$ units had dissociated that had a $T_c$ of 33~K at 240~GPa \cite{Duan:2015-I}. 

Recently, a novel HI(H$_2$)$_{13}$ molecular compound, which was stable between 9 and 130~GPa, was synthesized using I$_2$ and H$_2$ as starting materials \cite{Binns:2018a}. AIMD simulations showed that this phase adopts $Fm\bar{3}c$ symmetry. Compression of HI and H$_2$ led to the formation of an $I4/mcm$ symmetry H$_2$(HI)$_2$ phase that was stable between 3.5 to 12.5~GPa instead. Superconductivity was not found in these phases, however it was noted that the H$_2$ content of HI(H$_2$)$_{13}$ is high; 93 mol~\%, and 17.7 weight~\% . \\

\noindent\textbf{Astatine} \\
To the best of our knowledge, the hydrides of astatine have not yet been studied theoretically nor experimentally.

%%%%%%%%%%%%%%%%%%%%%%%%%%%%%%%%%%%%%%%%%%%%%%%%%%%%%%%%%%%%%%%%%%%%%%%%%%%%%%%%%%%%%%%%%%%%%%%%%%%%%%%%%%%%%%%%%%%%%%%%%%%%
%%%%%%%%%%%%%%%%%%%%%%%%%%                       Aerogens                              %%%%%%%%%%%%%%%%%%%%%%%%%%%%%%%%%%%%%
%%%%%%%%%%%%%%%%%%%%%%%%%%%%%%%%%%%%%%%%%%%%%%%%%%%%%%%%%%%%%%%%%%%%%%%%%%%%%%%%%%%%%%%%%%%%%%%%%%%%%%%%%%%%%%%%%%%%%%%%%%%%

\section{Group 18: Aerogen Hydrides}

\noindent\textbf{Helium, Neon} \\ 
Because the Jovian planets are primarily composed of helium and hydrogen, the astrophysical community has been intensely interested in the behavior of these elements upon mixing at high pressures. The calculated Gibbs free energies of mixing of He and Ne over a wide range of density, temperature and composition have led to the conclusion that at conditions resembling those of the interior of Saturn the two elements are likely to separate \cite{Morales:2009a,Morales:2013}. Such phase segregation has been observed in H$_2$-He mixtures in a DAC up to 8~GPa at 300~K \cite{Loubeyre:1985}. Moreover, experiments have shown that when the Ne concentration is between 0.25-99.5\%, Ne and H$_2$ are immiscible \cite{Barrett:1966a}. To the best of our knowledge no solid binary compounds of these two elements have been made under pressure (but a binary Ne(He$_2$)$_2$ compound has been experimentally observed and theoretically studied \cite{Loubeyre:1993,Cazorla:2009}).  \\

\begin{figure}[h!]
\begin{center}
\includegraphics[width=0.9\columnwidth]{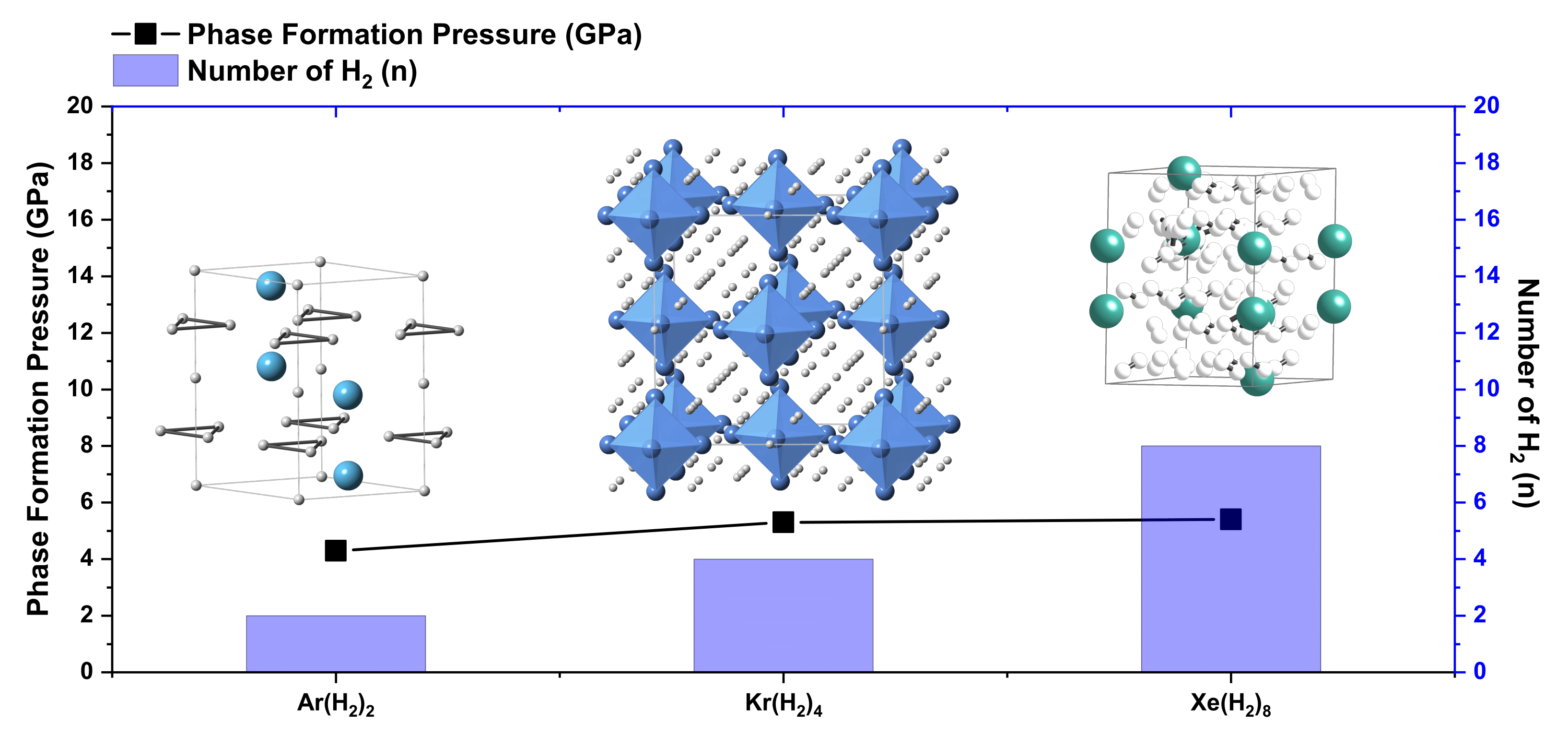}
\end{center}
\caption{Supercells of the vdW compounds: Ar(H$_2$)$_2$ \cite{Loubeyre:1994a, Ji-Ar:2017a}, Kr(H$_2$)$_4$ \cite{Kleppe:2014a}, and Xe(H$_2$)$_8$ \cite{Somayazulu:2010a, Somayazulu:2015a}. The black line denotes the pressures at which these phases were synthesized. The purple bars correspond to the number of H$_2$ molecules in the phase. The larger the aerogen, the larger the number of H$_2$ molecules that can be accommodated.}
\label{fig:Group18}
\end{figure}

\noindent\textbf{Argon, Krypton, Xenon} \\ 
An interesting, yet intuitive, trend emerges for the heavier noble gas hydrides under pressure; namely that the maximum hydrogen content in the binary compounds is proportional to the size of the aerogen atom. Specifically, the following stoichiometries, illustrated in Fig.\ \ref{fig:Group18}, have been identified in experiment: Ar(H$_{2}$)$_{2}$ \cite{Loubeyre:1994a, Ji-Ar:2017a}, Kr(H$_{2}$)$_{4}$ \cite{Kleppe:2014a}, and Xe(H$_{2}$)$_{8}$ \cite{Somayazulu:2010a, Somayazulu:2015a}. 

In the mid 1990s experiments showed evidence for the formation of an Ar(H$_2$)$_2$ compound at 4.3~GPa that adopted a structure isomorphous with the MgZn$_2$ Laves phase \cite{Loubeyre:1994a}. Raman measurements indicated that this structure was stable up to 175~GPa, at which point the H$_2$ molecules started to dissociate and undergo metalization. IR experiments up to 220~GPa  questioned this conclusion, since they showed the persistence of molecular hydrogen within this phase \cite{Hemley:2000}. 
A number of theoretical studies have been carried out in an attempt to explain this discrepancy. Tight-binding calculations predicted that band gap closure would occur above 400~GPa in Ar(H$_2$)$_2$ \cite{Chacham:1995a}. On the other hand, AIMD simulations suggested that an MgZn$_2\rightarrow$AlB$_2$ structural transition would occur around 250~GPa, with concomitant metalization \cite{Bernard:1997a}. More recent GGA calculations found that band gap closure occurs at 420~GPa within the AlB$_2$ structure \cite{Matsumoto:2007a}. Another AIMD simulation concluded that the structures Ar(H$_2$)$_2$ adopts are temperature dependent, with the MgCu$_2$ phase being more stable than the MgZn$_2$ and AlB$_2$ phases below 215~GPa at 0~K \cite{Cazorla:2010a}. Moreover, across a broad pressure range the MgZn$_2$ geometry was found to be favored above $\sim$60-100~K. CSP techniques have also been applied to find the most stable structures up to 300~GPa \cite{Yao:2011b}. Below 66~GPa the MgCu$_2$ structure was found to be only slightly more stable than the MgZn$_2$ alternative. However, a hitherto unconsidered CeCu$_2$ structure was clearly the lowest enthalpy candidate above 66~GPa. Band gap closure in this phase occurred at a pressure higher than that necessary to metallize hydrogen at the same level of theory. This was explained by noting that at a given pressure the intermolecular H-H distances in Ar(H$_2$)$_2$ are larger than those in pure hydrogen because of the presence of the noble gas, thereby decreasing the orbital overlap. 
This issue was finally resolved in a 2017 study that employed synchrotron XRD, as well as Raman and optical spectroscopy \cite{Ji-Ar:2017a}. It was shown that Ar(H$_2$)$_2$ retains the MgZn$_2$ structure with molecular H$_2$ units up to 358~GPa, at which point it had a 2~eV band gap.

Only one study has been carried out on the hydrides of krypton. A Kr(H$_2$)$_4$ phase was synthesized above 5.3~GPa \cite{Kleppe:2014a}. Its Kr sublattice possessed the $Fm\bar{3}m$ spacegroup, with rotationally disordered H$_2$ molecules occupying the interstitial sites up to 50~GPa. 

Experiments in a DAC up to 255~GPa showed crystallographic and spectroscopic evidence for the formation of Xe-H$_{2}$ compounds \cite{Somayazulu:2010a}. At 4.8~GPa a vdW compound that can be viewed as a superstructure based on the hcp lattice of solid hydrogen modulated by layers containing xenon dimers was formed. Its stoichiometry  was estimated as being Xe(H$_2$)$_7$ below, and Xe(H$_2$)$_8$ above 5.4~GPa. Increasing the pressure increased the chemical interaction within and between the xenon dimers, and it also weakened the H-H bond. This $R3$ symmetry phase remained semiconducting to the highest pressures studied. Because only the positions of the Xe atoms could be determined experimentally, CSP techniques were employed to find the preferred positions of the hydrogen atoms in Xe(H$_2$)$_7$ and Xe(H$_2$)$_8$\cite{Kaewmaraya:2011a}. A Bader charge analysis showed that at 263~GPa the Xe atoms lost on average 0.5$e$ to the hydrogen atoms. GW calculations predicted that Xe(H$_2$)$_8$ would metallize around 250~GPa.  Subsequent experiments were able to refine the crystal structures of the hydrides of xenon at various pressures \cite{Somayazulu:2015a}. Between 4.8-7.1~GPa the structure was indexed as having the $P\bar{3}m1$ spacegroup, however the site occupancy, and potentially stoichiometry, changed as a function of pressure. Its structural evolution at higher pressures was consistent with the previously reported $R3$ phase. Decompression at low temperature to ambient pressures illustrated that Xe-H$_2$ phases can be stable up to 90~K. 

CSP techniques have also been used to predict the most stable XeH$_n$ ($n=1-8$) structures up to 300~GPa \cite{Yan:2015a}. Only XeH$_2$ and XeH$_4$ were found to be thermodynamically stable, and all other stoichiometries were metastable. $Cmca$-XeH$_2$ was stable with respect to the elemental phases already at 1~GPa, and it was the lowest point on the convex hull up to 300~GPa. A Bader analysis revealed pressure induced charge transfer from Xe to H atoms. XeH$_4$ assumed $Amm2$ symmetry below, and $Cm$ symmetry above 100~GPa. With the exception of XeH, all of the phases contained H$_2$ molecules. Whereas hybrid functionals showed that XeH$_2$ metalized near 300~GPa, XeH was found to be metallic already at atmospheric pressures. The $T_c$s of XeH at 100~K and XeH$_2$ at 400~GPa were both estimated to be near 30~K, and they decreased with increasing pressure.  \\

\noindent\textbf{Radon} \\ 
To the best of our knowledge, the hydrides of radon have not yet been studied theoretically nor experimentally.

%%%%%%%%%%%%%%%%%%%%%%%%%%%%%%%%%%%%%%%%%%%%%%%%%%%%%%%%%%%%%%%%%%%%%%%%%%%%%%%%%%%%%%%%%%%%%%%%%%%%%%%%%%%%%%%%%%%%%%%%%%%%
%%%%%%%%%%%%%%%%%%%%%%%%%%                       Conclusions                           %%%%%%%%%%%%%%%%%%%%%%%%%%%%%%%%%%%%%
%%%%%%%%%%%%%%%%%%%%%%%%%%%%%%%%%%%%%%%%%%%%%%%%%%%%%%%%%%%%%%%%%%%%%%%%%%%%%%%%%%%%%%%%%%%%%%%%%%%%%%%%%%%%%%%%%%%%%%%%%%%%

\section{Conclusions}
In just over a decade the phase diagrams of most binary hydrides as a function of stoichiometry and pressure have been investigated via first-principles calculations. Many of these studies have been carried out using state-of-the-art crystal structure prediction techniques. Because experiments at very high pressures can be difficult to carry out, at first only a handful of high hydride phases were synthesized. However, intense efforts have resulted in the synthesis of a number of hydrides with unique structures and stoichiometries under pressure including those of lithium \cite{Pepin:2015a}, sodium \cite{Struzhkin:2016}, silicon \cite{Strobel:2009a,Wang:2009-Si}, phosphorus \cite{Drozdov:2015-P}, sulfur \cite{Drozdov:2015a}, argon \cite{Loubeyre:1994a, Ji-Ar:2017a}, iron \cite{Pepin:2014,Pepin:2017a}, selenium \cite{Pace:2017a,Zhang:2018-Se}, krypton \cite{Kleppe:2014a}, niobium \cite{Liu:2017}, rhodium \cite{Li:2011a}, iodine \cite{Binns:2018a}, xenon \cite{Somayazulu:2010a, Somayazulu:2015a}, lanthanum \cite{Geballe:2018a}, tungsten \cite{Strobel:2009a,Scheler:2013b}, iridium \cite{Scheler:2013c} and platinum \cite{Scheler:2011a}. Superconductivity has been measured in some of these (phosphorus, sulfur, platinum), and predicted in others, as shown in Table \ref{tab:Tc}. %(iron \cite{Majumdar:2017a,Kvashnin:2018a}, lanthanum \cite{Liu:2017-La-Y,Peng:Sc-2017}). 
In particular, the discovery of conventional phonon-mediated superconductivity in compressed hydrogen sulfide at 203~K and 150~GPa \cite{Drozdov:2015a}, which occurred via a synergistic feedback loop between theory and experiment \cite{Yao-S-review:2018}, was a spectacular breakthrough in the field. Theoretical investigations have attempted to understand the factors responsible for high temperature superconductivity in the compressed high hydrides \cite{Peng:Sc-2017,Tanaka:2017a}, and  materials that are superconducting at room temperature (but very high pressures) \cite{Liu:2017-La-Y,Peng:Sc-2017} have even been predicted. Superhydride research is blossoming, and it is therefore likely that it will lead to the discovery of quite interesting materials.

\pagebreak

%\begin{longtable}{c c c c c c c c c}
\begin{longtable}{ccccccccc}
\caption{A compilation of the highest computationally estimated $T_c$ (K) values for hydrides of the main group and transition metal elements that are available, along with the stoichiometry, space-group and pressure at which these values were obtained.} \\
%\begin{center}
%\begin{tabular}{c c c c c c c c c}
\hline
Group & System       & Pressure (GPa) & T$_c$ (K) &  $\mu ^*$   & Space Group \\
\hline
1     & LiH$_6$      & 300            & 82 $^{d,}$ \cite{Xie:2014a}   & 0.13     &  $R\bar{3}m$ \\
      & KH$_6$       & 230            & 59-70 $^{d,}$ \cite{Zhou:2012a} & 0.13-0.10  &  $C2/c$ \\
\hline
2     & BeH$_2$      & 365            & 97 $^{d,}$ \cite{Yu:2014} & 0.10 &  $P4/nmm$ \\
      & MgH$_6$      & 400 & 271 $^{e,}$ \cite{Feng:2015a} & 0.12 & $Im\bar{3}m$   \\
      & CaH$_6$      & 150 & 220-235 $^{b,}$ \cite{Wang:2012} &0.13-0.10 & $Im\bar{3}m$   \\
      & SrH$_6$      & 250 & 156 $^{b}$ & 0.10 & $R\bar{3}m$   \\
      & BaH$_6$      & 100 & 30-38 $^{d,}$ \cite{Hooper:2012b} & 0.13-0.10 & $P4/mmm$   \\
\hline
3     & ScH$_9$      & 300 & 233 $^{b,}$ \cite{Zurek:2018b} & 0.10 &  $I4_{1}md$ \\
      & YH$_{10}$    & 250 &  305-326 $^{b,}$ \cite{Liu:2017-La-Y} & 0.13-0.10 &  $Fm\bar{3}m$  \\
      & LaH$_{10}$   & 200 & 288 $^{b,}$ \cite{Peng:Sc-2017} & 0.10 &  $Fm\bar{3}m$  \\
      & CeH$_{10}$   & 200 & 50-55 $^{b,e,}$ \cite{Peng:Sc-2017} & 0.13-0.10 &  $Fm\bar{3}m$  \\
\hline
4      & TiH$_2$     & 1~atm & 7 $^{c,}$ \cite{Shanavas:2016a} & 0.10 &  $Fm\bar{3}m$  \\
      & ZrH          & 120 & 11 $^{d,}$ \cite{Li:2017} & 0.10 &  $Cmcm$  \\
      & HfH$_2$      & 260 & 11-13 $^{d,}$ \cite{Liu:2015} & 0.13-0.10 &  $P2_{1}/m$  \\
\hline
5     & VH$_{2}$     & 60 & 4 $^{d,}$ \cite{Chen:2014} & 0.10 & $Pnma$ \\
      & NbH$_{4}$    & 300 & 50 $^{b,}$ \cite{Durajski:2014b} & 0.10 & $I4/mmm$ \\
      & TaH$_{6}$    & 300 & 124-136 $^{d,}$ \cite{Zhuang:2017a} & 0.13-0.10 & $Fdd2$ \\
\hline
6     & CrH$_{3}$    & 81 & 37 $^{d,}$ \cite{Yu:2015} & 0.10 & $P6_{3}/mmc$ \\
\hline
7     & TcH$_{2}$    & 200 & 7-11 $^{d,}$ \cite{Li:2016} & 0.13-0.10 & $I4/mmm$ \\
\hline
8     & FeH$_{5}$    & 130 & 51 $^{d,}$ \cite{Majumdar:2017a} & 0.10 & $I4/mmm$ \\
      & RuH$_{3}$    & 100 & 4 $^{d,}$ \cite{Liu:2015c} & 0.10 & $Pm\bar{3}m$ \\
      & OsH & 100    & 2 $^{d,}$ \cite{Liu:2015b} & 0.10 & $Fm\bar{3}m$ \\
\hline
9     & RhH          &  4  & $\sim$2.5 $^{d,~e,}$ \cite{Kim:2011a} & 0.13 & $Fm\bar{3}m$  \\
      & IrH          &  80  & 7 $^{d,}$ \cite{Kim:2011a} & 0.13 & $Fm\bar{3}m$  \\
\hline
10    & PdH/PdD/PdT  & 1~atm & 47/34/30 $^{b,}$ \cite{Errea:2013} & 0.085 & $Fm\bar{3}m$   \\
      & PtH          & 77  & 25 $^{d,}$ \cite{Kim:2011a} & 0.13 & $Fm\bar{3}m$  \\
\hline     
11    & AuH          &  220  & 21 $^{d,}$ \cite{Kim:2011a}  & 0.13  & $Fm\bar{3}m$ \\
\hline
13    & B$_{2}$H$_{6}$       & 360 & 90-125 $^{d,}$ \cite{Abe:2011a} & 0.20-0.13 & $Pbcn$ \\
      & AlH$_{3}$(H$_2$)     & 250 & 132-146 $^{d,}$ \cite{Hou:2015-Al} & 0.13-0.10 & $P2_{1}/m$ \\
      & GaH$_{3}$            & 120 & 90-123 $^{b,}$ \cite{Szczesniak:2014} & 0.20-0.10 & $Pm\bar{3}n$ \\
      & InH$_{3}$            & 200 & 34-41 $^{d,}$ \cite{Liu:2015a} & 0.13-0.10 & $R\bar{3}$ \\
\hline
14      & Si$_{2}$H$_{6}$    & 275 & 139-153 $^{d,}$ \cite{Jin:2010-Si} & 0.13-0.10 & $Pm\bar{3}m$ \\
      & GeH$_{3}$            & 180 & 140 $^{d,}$ \cite{Abe:2013-Ge} & 0.13 & $Pm\bar{3}n$ \\
      & SnH$_{14}$           & 300 & 86-97 $^{d,}$ \cite{Esfahani:2016} & 0.13-0.10 & $C2/m$ \\
      & PbH$_{4}$(H$_2$)$_2$ & 230 & 107 $^{d,}$ \cite{Cheng:2015} & 0.10 & $C2/m$ \\
\hline
15      & PH$_{2}$ & 270 & 87 $^{a,~e,}$ \cite{Flores:2016-P} & -- & $I4/mmm$ \\
      & AsH$_{8}$ & 450 & 151 $^{d,}$ \cite{Fu:2016a} & 0.10 & $C2/c$ \\
      & SbH$_{4}$ & 150 & 95-106 $^{d,}$ \cite{YanbinMa:2015b} & 0.13-0.10 & $P6_{3}/mmc$ \\     
      & BiH$_{5}$ & 300 & 105-119 $^{d,}$ \cite{YanbinMa:2015a} & 0.13-0.10 & $C2/m$ \\
\hline
16    & H$_{3}$S & 200 & 191-204 $^{d,}$ \cite{Duan:2014} & 0.13-0.10 & $Im\bar{3}m$ \\
      & H$_{3}$Se & 200 & 131 $^{a,}$ \cite{Flores:2016a} & -- & $Im\bar{3}m$ \\
      & H$_{4}$Te & 170 & 95-104 $^{d,}$ \cite{Zhong:2016} & 0.13-0.10 & $P6/mmm$ \\
      & PoH$_{4}$ & 250 & 46-54 $^{d,}$ \cite{Liu:2015d} & 0.13-0.10 & $C2/c$ \\
\hline
17    & H$_{2}$Cl & 400 & 44-45 $^{d,}$ \cite{Zeng:2017-Cl} & 0.13-0.10 & $R\bar{3}m$ \\
      & HBr & 200 & 44-51 $^{d,}$ \cite{Duan:2010} & 0.13-0.10 & $P2_{1}/m$ \\
      & H$_2$I & 240 & 24-33 $^{d,}$ \cite{Duan:2015-I} & 0.13-0.10 & $R\bar{3}m$ \\
\hline
18    & XeH & 100 & $\sim$29 $^{e,}$  \cite{Yan:2015a} & 0.12 & $Immm$ \\
\hline
\hline
%\end{longtable}
\label{tab:Tc}
\end{longtable}
\noindent $^a$ $T_c$ was predicted using SCDFT. \\
$^b$ $T_c$ was calculated by solving the Eliashberg equations numerically. \\
$^c$ $T_c$ was calculated using the simplified Allen-Dynes formula. \\
$^d$ $T_c$ was calculated using the Allen-Dynes modified McMillan equation. \\ 
$^e$ These values were estimated from plots found in the original papers.
%\end{table}

%
%%%%%%%%%%%%%%%%%%%%%%%%%%%%%%%%%%%%%%%%%%%%%%%%%%%%%%%%%%%%%%%%%%%%%%%%%%%%%%%%%%%%%%%%%%%%%%%%%%%%%%%%%%%%%%%%%%%%%%%%%%%%
%%%%%%%%%%%%%%%%%%%%%%%%%%                           ACKNOWLEDGEMENTS                  %%%%%%%%%%%%%%%%%%%%%%%%%%%%%%%%%%%%%
%%%%%%%%%%%%%%%%%%%%%%%%%%%%%%%%%%%%%%%%%%%%%%%%%%%%%%%%%%%%%%%%%%%%%%%%%%%%%%%%%%%%%%%%%%%%%%%%%%%%%%%%%%%%%%%%%%%%%%%%%%%%
%
\pagebreak
\section{Acknowledgements}
We acknowledge the NSF (DMR-1505817) for financial, and the Center for Computational Research (CCR) at SUNY Buffalo for computational support. T.B.\ acknowledges financial support from the Department of Energy National Nuclear Security Administration under Award Number DE-NA0002006. 
%
%%%%%%%%%%%%%%%%%%%%%%%%%%%%%%%%%%%%%%%%%%%%%%%%%%%%%%%%%%%%%%%%%%%%%%%%%%%%%%%%%%%%%%%%%%%%%%%%%%%%%%%%%%%%%%%%%%%%%%%%%%%%
%%%%%%%%%%%%%%%%%%%%%%%%%%                             BIBLIOGRAPHY                    %%%%%%%%%%%%%%%%%%%%%%%%%%%%%%%%%%%%%
%%%%%%%%%%%%%%%%%%%%%%%%%%%%%%%%%%%%%%%%%%%%%%%%%%%%%%%%%%%%%%%%%%%%%%%%%%%%%%%%%%%%%%%%%%%%%%%%%%%%%%%%%%%%%%%%%%%%%%%%%%%%
%
\pagebreak
%\bibliography{./Review}
\providecommand{\latin}[1]{#1}
\makeatletter
\providecommand{\doi}
  {\begingroup\let\do\@makeother\dospecials
  \catcode`\{=1 \catcode`\}=2 \doi@aux}
\providecommand{\doi@aux}[1]{\endgroup\texttt{#1}}
\makeatother
\providecommand*\mcitethebibliography{\thebibliography}
\csname @ifundefined\endcsname{endmcitethebibliography}
  {\let\endmcitethebibliography\endthebibliography}{}

\end{document}